\documentclass[]{gji}
\usepackage{timet,color}
\usepackage[urlcolor=blue,citecolor=black,linkcolor=black]{hyperref}
\usepackage{url} 
\usepackage{graphicx}
\usepackage{url}
\usepackage{lineno}
\usepackage{mathtools}
\usepackage{soul}
\usepackage{amsmath, amssymb}
\usepackage{blindtext}
\usepackage{indentfirst}
\usepackage{wrapfig}
\usepackage{gensymb} 
\usepackage{natbib}
\newcounter{supfig}
\DeclareUnicodeCharacter{1EA1}{\d{a}}
\title[Full-waveform earthquake source inversion using simulation-based inference]
  {Full-waveform earthquake source inversion using simulation-based inference}
\author[A. A. Saoulis et al.]
  {A. A. Saoulis$^{1,2}$\thanks{a.saoulis@ucl.ac.uk}, 
    D. Piras$^{3,4}$,
   A. Spurio Mancini$^{1,5}$, 
   B. Joachimi$^1$, 
  A. M. G. Ferreira$^2$
 \\
   $^1$ Department of Physics \& Astronomy, University College London, Gower Street, London, WC1E 6BT, United Kingdom \\
   $^2$ Department of Earth Sciences, University College London, 5 Gower Place, London, WC1E 6BS, United Kingdom  \\
   $^3$ Centre Universitaire d’Informatique, Université de Genève, 7 route de Drize, 1227 Genève, Switzerland\\
   $^4$ Département de Physique Théorique, Université de Genève, 24 quai Ernest Ansermet, 1211 Genève 4, Switzerland\\
    $^5$ Department of Physics, Royal Holloway, University of London, Egham Hill, Egham, TW20 0EX, United Kingdom
  }

\begin{document}

\label{firstpage}

\maketitle

\begin{summary}
This paper presents a novel framework for full-waveform seismic source inversion using simulation-based inference (SBI). Traditional probabilistic approaches often rely on simplifying assumptions about data errors, which we show can lead to inaccurate uncertainty quantification. SBI addresses this limitation by learning an empirical probabilistic relationship between the parameters and data, without making assumptions about the data errors. This is achieved through the use of specialised machine learning models, known as neural density estimators, which can then be integrated into the Bayesian inference framework. We apply the SBI framework to point-source moment tensor inversions as well as joint moment tensor and time-location inversions. We construct a range of synthetic examples to explore the quality of the SBI solutions, as well as to compare the SBI results with standard Gaussian likelihood-based Bayesian inversions. We then demonstrate that under real seismic noise, common Gaussian likelihood assumptions for treating full-waveform data yield overconfident posterior distributions that underestimate the moment tensor component uncertainties by up to a factor of 3. We contrast this with SBI, which produces well-calibrated posteriors that generally agree with the true seismic source parameters, and offers an order-of-magnitude reduction in the number of simulations required to perform inference compared to standard Monte Carlo sampling techniques. Finally, we apply our methodology to a pair of moderate magnitude earthquakes in the North Atlantic. We utilise seismic waveforms recorded by the recent UPFLOW ocean bottom seismometer array as well as by regional land stations in the Azores, comparing full moment tensor and source-time location posteriors between SBI and a Gaussian likelihood approach. We find that our adaptation of SBI can be directly applied to real earthquake sources to efficiently produce high quality posterior distributions that significantly improve upon Gaussian likelihood approaches.

\end{summary}

\begin{keywords}
 Earthquake source observations -- Waveform inversion -- Bayesian inference -- Inverse theory -- Machine learning
\end{keywords}

\section{Introduction}

Seismic source mechanism characterisation is of vital importance for a variety of geoscience applications. Active tectonics and crustal deformation studies \citep[e.g.,][]{grimison1986azores,jackson1988relationship, molnar1989fault, demets1990current,demets2010geologically}, seismic tomography \citep{dahlen2000frechet,hung2000frechet,ritsema2011s40rts,bozdaug2016global}, seismic hazard assessment (for example via peak ground acceleration calculations; e.g.,   \citealt{olsen2000site,komatitsch2004simulations, denolle2014strong}) and earthquake physics analyses \citep[e.g.,][]{wiens2001seismological, oglesby2012fault,prieto2012earthquake}  often require accurate earthquake source models. Moreover, advances in these disciplines are contingent not only on reliable source mechanism determination, but also on robust and informative uncertainty quantification over the source parameters of interest \citep{weston2011global, ferreira2011global, blom2023mitigating}. 

Over the last half-century, seismic source mechanisms have become increasingly better characterised. Breakthroughs in decomposing earthquake sources through normal mode summation \citep{gilbert1975application,backus1976moment} led to the first waveform inversion methods \citep{Dziewonski1981,Kanamori1981} becoming rapidly popularised, and in many cases improving over earlier first-motion polarity techniques. Catalogues such as the Global Centroid Moment Tensor (GCMT) catalogue \citep{Ekstrom_2012} and the U.S. Geological Survey catalogue \citep{usgs-2024} now systematically compute moment tensor solutions semi-automatically using waveforms recorded by the global seismic network. 

However, there are still some key remaining challenges in seismic source inversions. For example, the study of earthquakes in mid-ocean settings is typically limited by large distances to seismic stations, which provide useful but restricted information about the source process \citep{lopez2015extended}. Local/regional data from marine seismology experiments can help address this issue, but such expensive experiments are usually limited to short time periods and the data collected are often noisy \citep[e.g.,][]{stahler2016performance, tsekhmistrenko2024performance}. Crucially, robust uncertainty quantification is still lacking in many earthquake source studies, which limits meaningful interpretations \citep{weber2006probabilistic}. Over the past twenty years, the Bayesian treatment of seismic source inversion has received increasing interest \citep{lomax2000probabilistic,tarantola2005inverse,vackavr2017bayesian,vasyura2020bayesian}. Seismic source inversion can be complicated by inter-parameter trade-offs (meaning different parameter combinations fit the data equally well) and strong non-uniqueness \citep{kawakatsu1996observability,Dufumier97, ford2010network}, and previous work has demonstrated that classical statistical analysis of standard errors can significantly underestimate uncertainty in source parameters \citep{Duputel2012,valentine2012assessing}. A number of studies have demonstrated that thorough Bayesian treatment of seismic events uncovers uncertainties and parameter trade-offs, often revealing greater complexity than simple maximum likelihood solutions \citep{Duputel2012, pugh2016bayesian, hu2023seismic}. \citet{HMCSimute23}, for example, demonstrated how inferred focal mechanisms and their distributions are sensitive to the seismic frequencies considered in the inversion, and that a Bayesian analysis of seismic source inversion can produce slightly multimodal posteriors. In addition, recent work has demonstrated that full-waveform probabilistic inversions can be highly sensitive to the noise model parametrisation \citep{Duputel2012,Musta2016, mustac2018variability}.

This study utilises novel machine learning (ML) tools to demonstrate that standard Bayesian inference techniques, which require a functional model of the data errors encoded in the likelihood function, have substantial limitations when applied to full-waveform data with real noise. We propose a novel framework to perform full-waveform moment tensor inversions with robust, well-calibrated uncertainty estimates through simulation-based inference (SBI), sometimes referred to as ``likelihood-free inference'' \citep{lueckmann2019likelihood, Cranmer2020}. SBI uses ML to build empirical models of the noise background, given observations or simulations of noise, avoiding the need to make highly simplifying assumptions. SBI has gained significant interest across a range of scientific fields, such as in cosmology \citep{Alsing2019,dax2021real,spurio2022cosmopower} and planetary science \citep{vasist2023neural}, by synthesising the flexibility of ML-based probabilistic modelling and the statistical rigour of Bayesian inference. Its recent popularity has been catalysed by improvements to the specialised neural network architectures used for estimating probability densities, known in general as neural density estimators (NDEs) \citep{dinh2017density, Papamakarios_MAF_2017, Alsing2019}. 

To the best of our knowledge this study is the first application of SBI in seismology. We therefore begin this paper by presenting the motivation and prior work in SBI, introducing modern ML techniques for probability distribution modelling, and presenting a data compression technique to treat the full-waveform source inversion problem with SBI. We then present our methodology and apply it to a range of synthetic waveform inversions. Subsequently we use an ocean bottom seismology dataset recently collected in the Atlantic region \citep[the UPFLOW dataset;][]{Ferreira2024,tsekhmistrenko2024performance} to study two oceanic earthquakes. This allows us to probe the advantages and pitfalls of using SBI in earthquake source inversions using noisy seismic data.

\section{Related Work and Motivation}

Seismologists have already demonstrated the advantages of using ML to perform inference over model parameters in inverse problems \citep[see e.g.][for a review]{valentine2023emerging}. Recent work has utilised ML to train rapid-executing emulators of the forward model, which can then be used within a Bayesian framework for increased inference speed \citep[e.g.,][]{Mancini2021,smith2022hyposvi,Piras2023}. Alternatively, often motivated by inference efficiency and speed, ML techniques have been used to learn posterior distributions over model parameters directly, utilising both simulated and human-annotated data \citep[e.g.,][]{Münchmeyer21, nooshiri2022multibranch, zhang2022loc}. Learning to model the likelihood or posterior directly from data allows practitioners to forego any forward model and perform inference with a ``frozen'' representation of the inverse problem; this concept is often referred to as amortization in the SBI literature \citep{papamakarios2016fast, Cranmer2020, Hermans2021}. Some previous works in seismic source characterisation used Mixture Density Networks, a class of NDE, for this task. For example, \citet{Münchmeyer21} used large, publicly available earthquake catalogues to train an end-to-end full-waveform ML model to estimate source location and magnitude posteriors. In addition, \citet{kaufl_framework_2013} used simulated static displacement data to model the posterior distribution over the source parameters directly.

Our foremost motivation for applying SBI to seismic inversion is the highly variable and non-Gaussian nature of noise in full-waveform data \citep[e.g.,][]{ nakata2019seismic}. There is a wide range of causes of seismic noise, from anthropogenic activities \citep[e.g.,][]{mcnamara2004ambient} to microseismic sources related to ocean waves \citep[e.g.,][]{longuet1950theory,kedar2008origin}, coastal hum \citep[e.g.,][]{mcnamara2004ambient,ardhuin2015ocean}, as well as ocean currents which can strongly affect ocean bottom seismometer (OBS) data in particular \citep{corela2022obs}. Rather than making strong assumptions about the noise model, SBI uses ML to build an empirical likelihood function, either explicitly or implicitly (by modelling the posterior), given many noisy simulated observations. In this sense, it learns to encode the full noise model of the forward process in its probabilistic modelling of the problem. This approach therefore sidesteps the classic limitation of approximating the complex seismic noise as Gaussian, as is very often done in seismology \citep[e.g.,][]{ferreira2006long,Ekstrom_2012, Fichtner2018}. To this end, this work can be viewed as exploring whether and by how much this common assumption impacts the posterior distribution produced by standard Bayesian analyses in full-waveform modelling.

Some previous studies explored the effect of non-Gaussian noise or errors in Bayesian inversions in seismology. \citet{tilmann2020another} explored the impact of the Gaussian error assumption on phase-picks, and concluded that a hybrid error model of a uniform distribution summed with a Gaussian distribution is required to perform successful inversions with realistic pick distributions. A comprehensive treatment was provided in \citet{Stahler2016_a,Stahler2016_b}, where the distribution of errors for their given inversion scheme and forward model was computed with reference to a ``high-quality catalogue'', and was thereafter used to parametrise a simple empirical likelihood function. However, the reliance of this approach on a manually curated catalogue makes extension to new domains challenging. Moreover, unknown errors and biases in such a catalogue could corrupt the quality of the empirical likelihood function. Indeed, the limitations of the schema presented in \citet{Stahler2016_a,Stahler2016_b}  provide motivation for extending the notion of an empirical likelihood to the more flexible SBI approach of direct simulation, where the true source parameters are known.

Another key motivation of SBI is that it can drastically reduce the number of evaluations of the forward model to generate robust, well-calibrated posterior distributions. Rather than relying on classical Markov Chain Monte Carlo (MCMC) techniques to traverse the model space, which generally requires a large number of simulations to converge, SBI can use a one-off dataset of simulations to learn an amortised representation of the inverse problem. Once trained, SBI emulates the forward model; the ML model learns to interpolate the probability density function between realised simulations in the model space, improving sample efficiency and thus reducing the number model evaluations required by several orders of magnitude \citep[e.g.,][]{Alsing2019, Cranmer2020}. In a parallel line of work, Hamiltonian Monte Carlo (HMC) has recently received attention for substantially improving sampling efficiency in seismic source inversion problems \citep{Fichtner2018,masfara2022efficient,HMCSimute23}. These applications demonstrate that HMC is a very promising method for accelerating seismic source inversion, though its parameters can be tricky to tune and, pertinent to this work, it still requires approximations on the form of the likelihood.

\section{Simulation-based Inference}

Simulation-based inference has emerged across a variety of the physical sciences as a way of incorporating complex, non-Gaussian noise models into the Bayesian inversion framework \citep{gutmann2016bayesian,Alsing2019}. Modern SBI leverages recent advancements in ML to replace the need for an analytic likelihood function with an empirical surrogate \citep{Cranmer2020}. It therefore represents a potential tool to treat the highly non-Gaussian noise in seismic waveforms more accurately. 

\subsection{Bayesian Inference}

The Bayesian formalism provides a natural procedure for performing inference over a set of parameters $\mathbf{m}$ given noisy observations $\mathbf{D}$:
\begin{equation}
p(\mathbf{m} \mid \mathbf{D}) = \frac{p(\mathbf{D} \mid \mathbf{m}) p(\mathbf{m})}{p(\mathbf{D})}.
\label{eq:bayes}
\end{equation}
Here, $p(\mathbf{m})$ is the prior, $p(\mathbf{D} \! \mid \! \mathbf{m})$ is the likelihood, and $p(\mathbf{D})$ is referred to as the evidence or marginal likelihood. The posterior, $p(\mathbf{m} \! \mid \! \mathbf{D})$, is the key quantity that represents a complete solution to the inference problem \citep{tarantola2005inverse}. Analytic computation of the posterior distribution in Eq. \ref{eq:bayes} becomes intractable beyond a family of simple distributions.

Classically, Bayesian inference has relied on a series of techniques that allow for drawing samples from the posterior distribution, $\mathbf{m}_i \sim p(\mathbf{m}\mid\mathbf{D})$, which can serve as an adequate surrogate for an analytic expression of the posterior. These range from simpler, more direct methods such as importance sampling and rejection sampling, as well as MCMC methods which surged to prominence for their robustness and elegance for drawing samples from the posterior. Thorough reviews are provided in e.g. \citet{neal1993probabilistic,gilks1995markov,tarantola2005inverse}. These methods, however, generally rely on the availability of functional forms of the prior and likelihood, such as a uniform distribution over the prior and a Gaussian distribution over the likelihood (though non-functional priors can be treated through random-walk strategies such as \textit{extended} Metropolis sampling\citealp[;][]{mosegaard1995monte}). Seismology has historically relied on assumptions about the noise generating processes to produce a simplified but adequately informative likelihood that can be incorporated into these sampling methods. 

The likelihood represents the probability of seeing a set of observations given specific model parameters, $p(\mathbf{D} \! \mid \! \mathbf{m})$. The forward model $g(\mathbf{m})$ is an essential component of defining the likelihood, which in seismological modelling is generally taken as a deterministic mapping. The probabilistic aspect of the likelihood is implicitly defined by the measurement uncertainties and seismic noise. At this stage, most seismological applications assume particular analytic forms for the likelihood function; for instance, by assuming the noise model is Gaussian:
\begin{equation}
p(\mathbf{D} \mid \mathbf{m}) \propto \exp  \left[- \frac{1}{2} \left[\mathbf{D} - g(\mathbf{m})\right]^T \mathbf{C}^{-1} \left[\mathbf{D} - g(\mathbf{m})\right] \right],
\label{eq:latent_variables}
\end{equation}
where the Gaussian covariance of the noise is given by $\mathbf{C}$. For any problems with complex sources of noise, this approximation can potentially lead to significant errors in inferring the posterior.

Note, however, that we can directly draw realisations from the likelihood function $\mathbf{D}_i \sim  p(\mathbf{D} \!\mid \!\mathbf{m})$ by combining the forward model with real observations of noise. For seismic source inversions, this can be done additively due to the independence of noise from the model parameters. The key step in SBI is to use these realisations to build an empirical model of the likelihood function (or related quantities, such as the posterior.)



\subsection{Empirical Probability Density Modelling}

\begin{figure}
\includegraphics[width=0.48\textwidth]{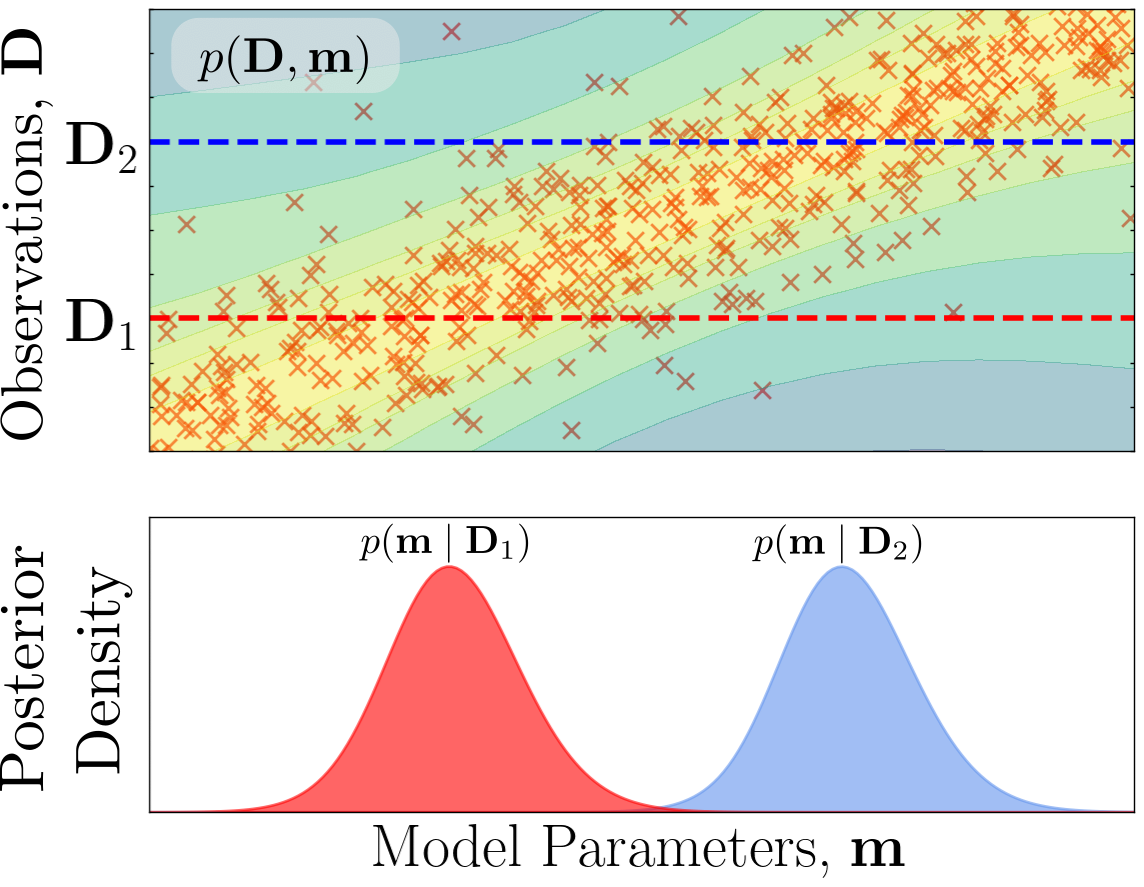}
\caption{Upper panel: realistic observations $(\mathbf{m}_i, \mathbf{D}_i)$ (red crosses) define a joint distribution (contoured background). Lower panel: the posterior distribution can be modelled empirically from this dataset, allowing for direct evaluation $\hat{p}(\mathbf{m} \mid \mathbf{D}_{1,2})$ for a given observation.}
\label{fig:modelling_distributions}
\end{figure}

By combining seismological forward modelling (for our purposes, given in Section \ref{sec:forward_modelling}) with noise observations, we can build a `realistic' forward model to emulate observations from the likelihood $\mathbf{D}_i \sim  p(\mathbf{D} \! \mid \! \mathbf{m}_i)$.  The framework builds a dataset of simulated noisy observations along with the associated model parameters using this forward model i.e. $\mathcal{D} = \{\mathbf{D}_i, \mathbf{m}_i\}$. Then, some form of statistical learning is leveraged to model a quantity that can be used to generate samples from the posterior distribution. Some examples include:
\begin{equation}
    \psi(\mathbf{D}, \mathbf{m} \mid \mathbf{w}) =
    \begin{dcases*} 
        \hat{p}(\mathbf{D}, \mathbf{m}), & \text{an empirical joint distribution;} \\
        \hat{p}(\mathbf{D}\mid \mathbf{m}), & \text{an empirical likelihood;} \\
        \hat{p}(\mathbf{m}\mid \mathbf{D}), & \text{an empirical posterior,}

    \end{dcases*}
    \label{eq:empirical_distributions}
\end{equation}
where $\psi$ can be any machine learning model, e.g. a neural network, with parameters $\mathbf{w}$. The first option is to model the joint distribution, after which the posterior density can be evaluated at a given observation: $p(\mathbf{m}\mid \mathbf{D_\text{obs}}) \simeq \hat{p}(\mathbf{m}, \mathbf{D} = \mathbf{D}_\text{obs})$. The second option of modelling the likelihood can be used in conjunction with classical sampling techniques such as MCMC to draw samples from the posterior. The final option, used in this work, is to model the posterior distribution directly, visualised in Fig. \ref{fig:modelling_distributions}. Here, simulated observations are sampled from a user-specified prior and the likelihood $\mathbf{D}_i \sim p(\mathbf{D} \mid \mathbf{m} ) p (\mathbf{m})$, and ML is used to model the posterior density $\hat{p}(\mathbf{m} \mid \mathbf{D})$, implicitly learning an empirical likelihood \citep{papamakarios2016fast}.  Eq. \ref{eq:empirical_distributions} is by no means an exhaustive list; recent work in SBI has demonstrated the value of modelling an empirical likelihood-to-evidence ratio, $\hat{p}(\mathbf{D}\mid \mathbf{m})/ \hat{p}(\mathbf{D})$, for example \citep{Hermans2020}. 
There are several factors to consider when selecting which quantity to model from Eq. \ref{eq:empirical_distributions}, and a number of studies have systematically compared these approaches \citep{Papamakarios2019,lueckmann2021benchmarking}. The likelihood approach is a common choice since it tends to have a simpler, generally uni-modal form (compared to generally more multimodal joint or posterior distributions) \citep{Alsing2019}, and it allows different priors to be explored without any need for retraining. This work models the posterior distribution $p(\mathbf{m} \! \mid \! \mathbf{D})$ directly; we explore relatively simple posteriors corresponding to a fixed prior, and this approach is desirable since posterior samples can be generated near instantly once $\psi$ is trained.

\subsection{Neural density estimators (NDEs)}
\label{sec:ndes}

The task of inference then relies on finding and training a model $\psi$ capable of modelling complex probability distributions. Much of the success of SBI hinges on the advancements made in the flexibility and efficiency of NDEs over the last decade. Neural density estimation is a conceptually simple task at its core; given a dataset of observations of $\{\mathbf{x}_i\}$ where $i$ runs over all examples in the dataset, a NDE must be trained to estimate the density $p(\mathbf{x})$ across $\mathbf{x}$. Conditional probabilities can be similarly treated, with a dataset of $\{\mathbf{x}_i, \mathbf{y}_i\}$ allowing the empirical modelling of e.g. density $p(\mathbf{y} \! \mid \! \mathbf{x})$. In practice, the challenge in performing this modelling is to design a flexible density estimator with a tractable (ideally easy to compute) density function. 

Practical advancements in explicit probability density estimation \citep{bishop1994mixture,Rezende_15,Papamakarios_MAF_2017} have enabled the direct use of ML for Bayesian inference. Generally, neural networks are trained to parametrise either mixtures of density models \citep[Mixture Density Networks,][]{bishop1994mixture} or a series of transformations to convert a simple base distribution into the target probability distribution \citep{Rezende_15}. This latter approach is known as \textit{normalising flows} and has seen intense interest from the ML community, with rapid progress leading to improvements in the efficiency and flexibility of these models \citep{germain2015made,uria2016neural}. The Masked Autoregressive Flow (MAF) \citep{Papamakarios_MAF_2017} leverages a number of these improvements and has emerged as a highly successful NDE, with several proven applications modelling complex and diverse probability distributions in SBI \citep{Alsing2019}. \citet{Papamakarios_review21} provides a thorough review of normalising flows and the various incarnations of this method in trainable NDEs. 

\begin{figure*}
    \centering
    \includegraphics[width=\textwidth]{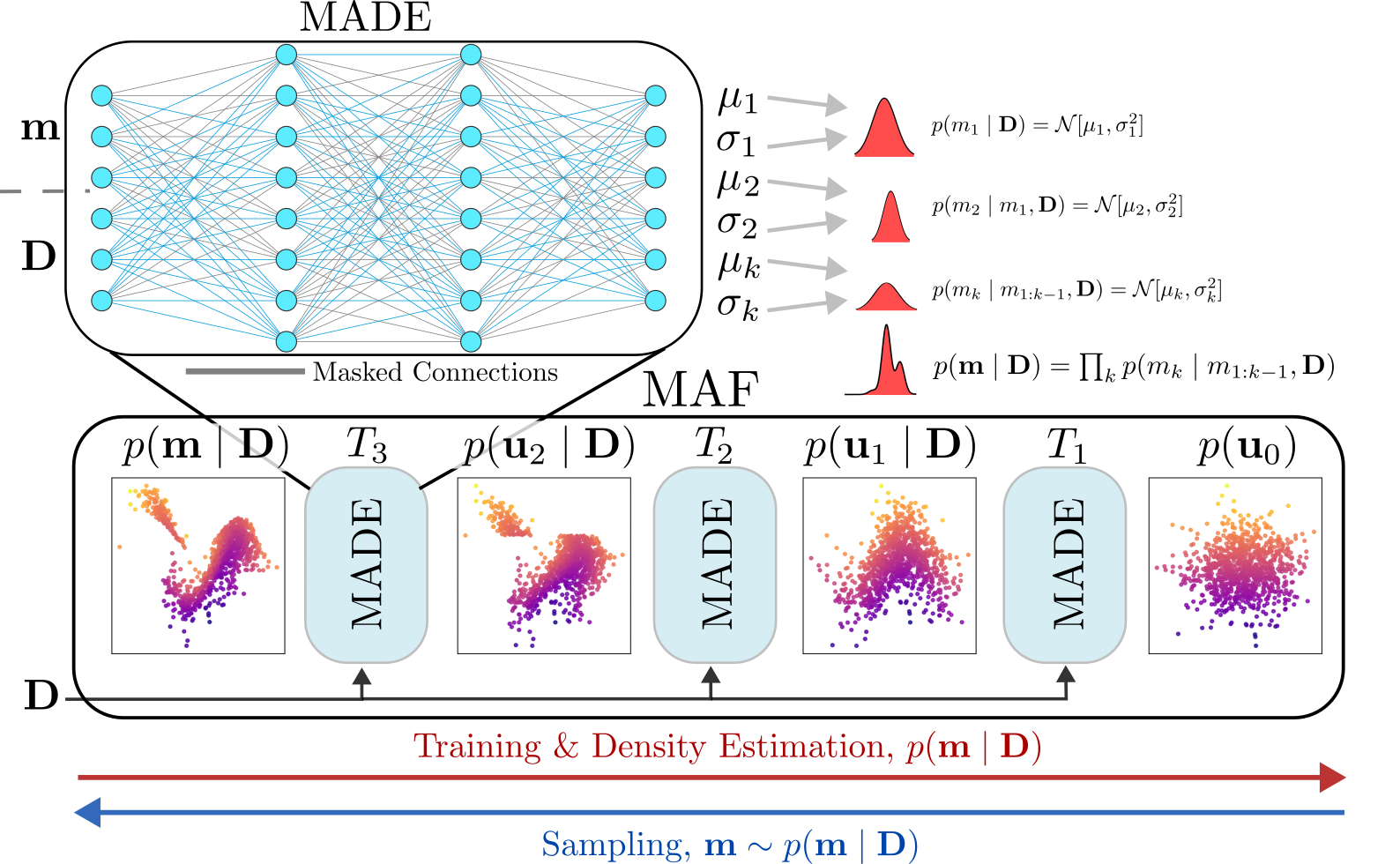}
    \caption{Schematics of the Masked Autoencoder for Distribution Estimation (MADE) and Masked Autoregressive Flow (MAF). MAFs are built by chaining together MADEs, each representing a transformation $T_j$, and permuting the order of conditioning to ensure suitably flexible density estimation. Each MADE $\varphi$ models a chain of conditional Gaussian distributions, parametrised by $\{\mu_k, \sigma_k\}$ where $k$ indexes the element of $\mathbf{x}$, to produce estimates of the density. Training can be thought of as learning to transform observations of the posterior $p(\mathbf{m} \mid \mathbf{D})$ to a simple base distribution $p(\mathbf{u}_0)$. Density estimation (left-to-right) repeatedly applies the change of variables formula in Eq. \ref{eq:change_of_variables} by computing $\text{det}J_T(\mathbf{u}_j)$ for each intermediate transformed variable $\mathbf{u}_j$. Sampling (right-to-left) successively applies transformation $T_j$ to samples from the base distribution $\mathbf{u}_0 \sim p(\mathbf{u}_0)$. See the text in Section \ref{sec:ndes} for more details. }
    \label{fig:MADE_MAF}
\end{figure*}

Normalising flows are a class of models that are trained to transform a base distribution $p(\mathbf{u})$, which is generally a simple distribution such as a Gaussian, into a target distribution $p(\mathbf{x})$. The transformation $T$ operates on the base distribution, $\mathbf{x} = T(\mathbf{u})$, where $T$ must satisfy certain conditions, such as invertibility and differentiability. This then allows the probability density $p(\mathbf{x})$ to be computed by the change of variables formula:
\begin{equation}
p_\mathbf{x}(\mathbf{x}) = p_\mathbf{u}(\mathbf{u}) | \text{det} J_T(\mathbf{u})|^{-1},
\label{eq:change_of_variables}
\end{equation}
where $J_T(\mathbf{u})$ is the Jacobian of T, i.e., the matrix of partial derivatives with respect to components of $\mathbf{u}$. One type of normalising flow are autoregressive models, where $\mathbf{x}$ is decomposed into its elements $x_k$. One can then proceed by learning to model a succession of conditional distributions, thereby constructing the full probability distribution by applying the chain rule:
\begin{equation}
p(\mathbf{x}) = \prod_k p(x_k \mid x_{1:k-1}).
\label{eq:autoregressive}
\end{equation}
Each conditional in $p(x_k \! \mid \! x_{1:k-1})$ can then be modelled by parametrising a simple base distribution $p(u_k; \lambda_k)$, where $\lambda_k$ represents a set of parameters output by a neural network. Autoregressive models produce a triangular Jacobian $J_T(\mathbf{u})$ by construction, allowing for simple computation of $\text{det}|J_T(\mathbf{u})|$ in Eq. \ref{eq:change_of_variables} and inversion of the transformations. The Masked Autoencoder for Distribution Estimation (MADE) \citep{germain2015made}, see Fig. \ref{fig:MADE_MAF}, ensures the autoregressive property is satisfied by carefully masking portions of the neural network $\varphi$, allowing the conditionals of autoregressive models to be evaluated in a single forward pass of $\varphi$ \citep{germain2015made,Papamakarios_MAF_2017}. This masking enables efficient density estimation and parallel evaluation on specialised computing architectures. The MADE can utilise a simple, empirically parametrised 1D Gaussian distribution for each conditional in Eq. \ref{eq:autoregressive}; a feedforward neural network is trained to estimate each mean and variance:
\begin{equation}
p(x_k \mid x_{1:k-1}) = \mathcal{N}(\varphi_{\mu_k}(x_{1:k-1}), \varphi_{\sigma^2_k}(x_{1:k-1})).
\end{equation}

These flows also have the useful property that they can be composed together, in effect learning to parametrise successive transformations, allowing for significant improvements in the flexibility of the model. MAFs, which we utilise in this work, are constructed by stacking together several MADE networks, indexed by $j$, and varying the autoregression order from  Eq. \ref{eq:autoregressive} between each MADE. In this work, MAFs are trained to model the posterior $p(\mathbf{m} \mid \mathbf{D})$, so each transformation $T_j$ parametrised by a MADE is also conditioned on the observation $\mathbf{D}$. This architecture is depicted in Fig. \ref{fig:MADE_MAF}. 

Once the network architecture has been selected, the NDE $\hat{p}(\mathbf{x})$ can be trained to model the observed density $p(\mathbf{x})$. The mismatch between the distributions can be quantified using the Kullback–Leibler (KL) divergence, leading to the following objective function:
\begin{equation}
\begin{aligned}
\mathcal{L} &= \mathrm{KL}\big(p(\mathbf{x}) \, \| \, \hat{p}(\mathbf{x})\big) = \mathbb{E}_{p(\mathbf{x})} \left[ \log p(\mathbf{x}) - \log \hat{p}(\mathbf{x}) \right] \\
&= \text{const.} -\mathbb{E}_{p(\mathbf{x})} \left[ \log \hat{p}(\mathbf{x}) \right] ,
\end{aligned}
\end{equation}
where the first term can be ignored as it is a constant that does not depend on the network parameters \citep[see e.g.,][]{Papamakarios_review21}. For a dataset $\mathcal{D} = \{\mathbf{m}_i, \mathbf{D}_i\}$, posterior estimation can therefore be achieved through the simple objective $\mathcal{L}=-\mathbb{E}_{(\mathbf{m}_i, \mathbf{D}_i) \sim \mathcal{D}} \left[ \log \hat{p}(\mathbf{m}_i \mid \mathbf{D}_i) \right]$, i.e. the network should aim to maximise the probability density it outputs over the observed data. This straightforward formulation can be combined with standard stochastic gradient descent techniques common to most ML applications to train the NDE.

\subsection{Data Compression}
\label{sec:data_compression}

One key limitation of the class of NDEs used for SBI is that they scale very poorly to high-dimensional inputs. The reason for this is as much an advantage as a detriment; these NDEs rely on fully-connected layers to parametrise the transformations they use to reconstruct the target probability distribution. This means that they assume very little structure on the inputs. On the one hand, this makes them highly flexible learners, since they can model near arbitrary relationships between all the variables; on the other hand, this necessitates that the input variables fed into these NDEs are all meaningful. If this latter condition is not satisfied, the NDEs will struggle to learn complex relationships between variables that are generally latent in high-dimensional data. 

\begin{figure*}
\includegraphics[width=\textwidth]{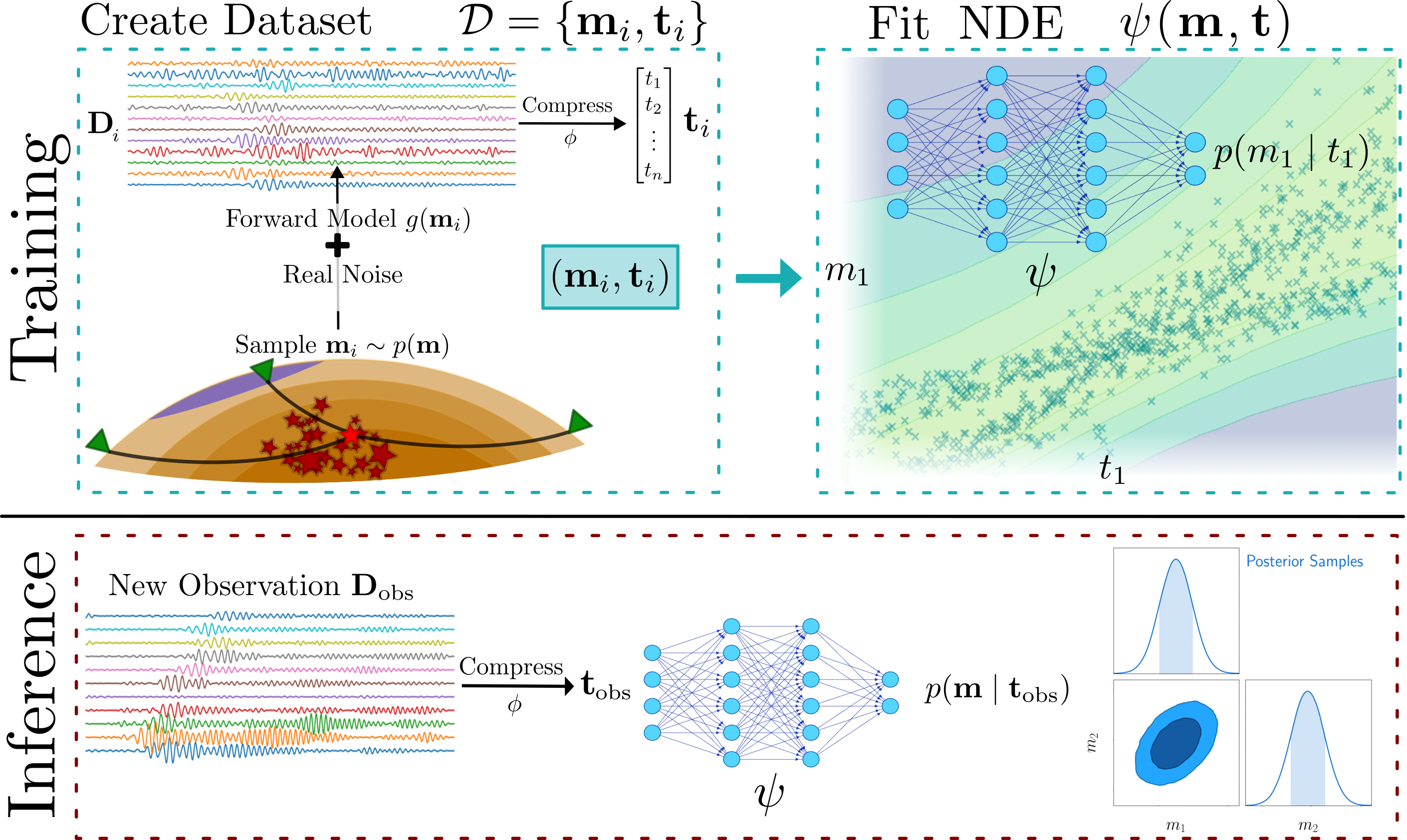}
\caption{Diagram of the SBI workflow for earthquake source inversion. Top: seismic sources are sampled from a prior $p(\mathbf{m})$, simulated, and combined with real noise data. This emulates sampling from a realistic likelihood, $\mathbf{D}_i \sim p(\mathbf{D} \mid \mathbf{m} )$. The resulting observations $\mathbf{D}_i$ are compressed with compression method $\phi$ to create a dataset of summary statistics $\mathbf{t}_i$ and model parameters $\mathbf{m}_i$. A NDE $\psi$ is then trained on this data to model the associated probability density function. Once trained, inference (bottom) can be performed without access to the forward model by compressing a new observation $\mathbf{D}_\text{obs}$ and passing the compressed observation $\mathbf{t}_\text{obs}$ directly through the NDE. See the text in Section \ref{sec:data_compression} for more details.}
\label{fig:data_generation}
\end{figure*}

In order to manage this incongruity between the requirements of  NDEs and the large data dimensions generally manipulated in scientific inference problems, usually an initial compression step is performed to convert data vectors $\mathbf{D} \in \mathbb{R}^N$ into a set of \textit{informative} summary statistics $\mathbf{t} \in \mathbb{R}^M$, where $M \ll N$ . Then, the SBI workflow can be broadly summarised in four conceptual steps:
\begin{enumerate}
    \item Sample from model prior $p(\textbf{m})$ and likelihood $p(\mathbf{D} \mid \mathbf{m})$ to generate a set of simulation pairs $\{\mathbf{m}_i, \mathbf{D}_i\}$.
    \item Select a compression algorithm $\phi : \mathbf{D} \rightarrow \textbf{t}$ and compress the resulting data vectors to generate the corresponding dataset of pairs $\mathcal{D} = \{\mathbf{m}_i, \mathbf{t}_i\}$.
    \item Train the NDE $\psi$ on the dataset $\mathcal{D}$ to model one of the relevant probability densities in Eq. \ref{eq:empirical_distributions}, such as the posterior distribution $p(\mathbf{m} \mid \mathbf{t})$. 
    \item Perform inference by compressing observations  $\mathbf{t}_\text{obs} = \phi(\mathbf{D}_\text{obs})$ and sampling from the (empirical) posterior using $\psi (\mathbf{m}, \mathbf{t}_\text{obs})$.
\end{enumerate}
This procedure is summarised in Fig. \ref{fig:data_generation} in the context of inferring earthquake source parameters from seismic waveform data using neural posterior estimation.

When selecting the compression technique $\phi$, the primary concern is to preserve as much discriminative power of summary statistics $\mathbf{t}$ with respect to the model parameters $\mathbf{m}$ as possible. Much prior work across the field of SBI has focused on how to formalise this requirement. 

\subsection{Optimal Score Compression}

Optimal score compression, which compresses high-dimensional observations down to the dimensionality of the model parameters, provides one approach to satisfy this requirement \citep{Alsing2017,Alsing2018}. This technique produces summary statistics that saturate the \textit{information inequality}, which provides a lower bound on the variance of a set of statistics $\mathbf{t}$: 
\begin{linenomath*}
\begin{equation}
\text{Var}_{\mathbf{m}}[t_k] \geq (\mathbf{A}^T\mathbf{F}^{-1}\mathbf{A})_{kk},
\label{eq:information_inequality}
\end{equation}
\end{linenomath*}
where the Fisher information matrix is defined as $\mathbf{F} = -E_{\mathbf{m}}[\nabla \nabla^T \mathcal{L}]$ for log-likelihood $\mathcal{L}$, and $\mathbf{A} = \nabla E_{\mathbf{m}}[\mathbf{t}(\mathbf{m})]$. Here, $E_\mathbf{m}$ represents an expectation over the model parameters. Eq. \ref{eq:information_inequality} states that the variance of the statistic over the model parameters can be bounded from below (where equality holds for an ``optimal'' statistic). Stated informally, to maximally constrain 
 model parameters $\mathbf{m}$ when inferring them from an observation $\mathbf{t}$ we should select a compression scheme such that the variance over $\mathbf{t}$ is minimised for a given $\mathbf{m}$.


With this objective in mind, it is sufficient to simply find a compression scheme that satisfies this lower bound. Optimal score compression makes use of the fact that for a local linearisation of the log-likelihood $\mathcal{L}$, the derivative of this expanded likelihood saturates the lower bound of Eq. \ref{eq:information_inequality}. Assuming a Gaussian likelihood, we have:
\begin{linenomath*}
\begin{equation}
\mathcal{L} = - \frac{1}{2} \left[\mathbf{D} - g(\mathbf{m})\right]^T \mathbf{C}^{-1} \left[\mathbf{D} - g(\mathbf{m})\right] - \frac{1}{2} \ln{|\mathbf{C}|},
\label{eq:gaussian_log_likelihood}
\end{equation}
\end{linenomath*}
where $g(\cdot)$ denotes the forward model and $\mathbf{C}$ is the Gaussian covariance. It is important to note that this is distinct from assuming a Gaussian noise model for inference; here, the likelihood is used to create a summary statistic of the data observations.\footnote{This is still an incorrect assumption, but crucially it will lead to sub-optimal compression (and therefore reduced constraining power per Eq. \ref{eq:information_inequality}), rather than incorrect posteriors caused by using a Gaussian likelihood for inference. Improved compression schemes that do not rely on this assumption are an open area of research \citep{prelogovic2024informative, lanzieri2024optimal}.} Taylor expanding the log-likelihood to first order about the expansion point $\mathbf{m}_*$ yields:
\begin{linenomath*}
\begin{equation}
\mathcal{L} = \mathcal{L}_* + \delta \mathbf{m}^T \nabla_{\mathbf{m_*}} \mathcal{L}_*.
\label{eq:first_order_expansion}
\end{equation}
\end{linenomath*}
In this linearised setting, \citet{Alsing2018} showed that the \textit{score} function $\mathbf{t} = \nabla_{\mathbf{m_*}} \mathcal{L}_*$ yields the desired estimator that satisfies Eq. \ref{eq:information_inequality}. Since for seismic source inversion the noise covariance matrix is independent of the model parameters, i.e. $\nabla_{\mathbf{m}}\mathbf{C} = 0$, the score has a simple form for a Gaussian likelihood:
\begin{linenomath*}
\begin{equation}
\mathbf{t} = \mathbf{G}^T_{\mathbf{m_*}} \mathbf{C}^{-1}_* (\mathbf{D} - \boldsymbol{\mu}_*),
\label{eq:optimal_score}
\end{equation}
\end{linenomath*}
where $\boldsymbol{\mu}_*$ is the synthetic data at the expansion point  $\mathbf{m}_*$, and $\mathbf{G}_{\mathbf{m_*}} = \nabla_{\mathbf{m_*}} g(\mathbf{m})$ is the gradient of this data vector with respect to the model parameters. Eq. \ref{eq:optimal_score} thus defines the compression operator $\phi$ for the optimal score compression scheme. 

\citet{Alsing2018} provided an alternative form of the summary statistic that satisfies Eq. \ref{eq:information_inequality} and has a simpler interpretation:
\begin{linenomath*}
\begin{equation}
\mathbf{t} = \mathbf{m}_* + \mathbf{F}^{-1}_* \mathbf{G}^T_{\mathbf{m_*}} \mathbf{C}^{-1} (\mathbf{D} - \boldsymbol{\mu}_*), 
\label{eq:optimal_score_mle}
\end{equation}
\end{linenomath*}
where the Fisher information $\mathbf{F}_* = \mathbf{G}^T_{\mathbf{m_*}} \mathbf{C}^{-1} \mathbf{G}_{\mathbf{m_*}}$ for the Gaussian likelihood in Eq. \ref{eq:gaussian_log_likelihood}. This form of the summary statistic $\mathbf{t}$ corresponds to a local maximum likelihood estimate (MLE) of the model parameters, $\widehat{\mathbf{m}}$, given the observation $\mathbf{D}$; equivalently, it is an estimate of the model parameters after a single step of least-squares from expansion point $\mathbf{m}_*$. This work uses the latter expression (Eq. \ref{eq:optimal_score_mle}) throughout as its familiar form allows for easier interpretation, but inversions using Eq. \ref{eq:optimal_score} would yield identical results. 

There are alternative compression schemes that could be used. One possibility is to rely on ML models to learn to preserve as much information as possible about the parameters of interest given a set of observations $\mathbf{D}$. This is an active area of research with a wide range of strategies for training deep neural networks to compress high dimensional observations \citep{fluri2021cosmological, lu2022simultaneously, sharma2024comparative}, see e.g. \citet{lanzieri2024optimal} for a review. One approach, for example, aims to maximise the mutual information between model parameters $\mathbf{m}$ and a learnt set of summary statistics $\mathbf{t}$ using deep learning \citep{charnock2018automatic, Jeffrey2020, prelogovic2024informative}. We leave exploring such alternatives for seismic source inversion to future work. 

\subsection{Assessing Posterior Quality}
\label{sec:posterior_quality}

In order to compare the results of SBI with the standard, likelihood-based approaches commonly used across seismology, quantitative checks must be performed on the resulting posterior distributions. If the true posterior $p(\mathbf{m} \mid \mathbf{D})$ is Gaussian and known analytically, a simple check to perform is to compute the reduced $\chi^2$ of samples drawn from each inference method. Values of the reduced $\chi^2$ that diverge from $1$ can indicate mismodelling of the posterior distribution. Section \ref{sec:artificial_gaussian_noise} considers a simplified toy inversion that performs exactly this test to explore differences between SBI and the likelihood-based approach where the true posterior is known.

For realistic problems where the posterior is analytically intractable, there is a wide range of tests that can be performed to quantify the quality of posterior samples, particularly in synthetic examples where the true model parameters are known. Such self-consistency checks aim to quantify the level of agreement between inferred posteriors and the known model parameters. \citet{lueckmann2021benchmarking} provided a thorough investigation of such tests and found that standard techniques such as posterior predictive checks suffer from significant failure modes. On the other hand, empirical coverage tests emerged as a particularly robust technique for measuring the posterior quality. Coverage tests perform many artificial inversions and check that the true model parameters fall within a given probability interval the expected proportion of the time. The frequency that the true model parameters  are ``covered'' by a given probability interval can then be used to identify whether the inferred posteriors are overconfident or under-confident (uncertainty intervals too narrow or broad, respectively). Empirical coverage tests are explained in more detail in Section S1. 

Unfortunately, computing empirical coverage can become very computationally expensive in high-dimensions as many methods rely on explicit density modelling to compute the credible region around the ground truth for quantile estimation \citep[see e.g.][]{Hermans2021}). \citet{lemos2023sampling} provided an alternative approach, named Tests of Accuracy with Random Points (TARP), to estimate the coverage much more efficiently. This approach computes the fraction of samples that occur between a random reference point and the true model parameters, allowing one to estimate the observed distribution of credibility levels over a number of inversions and check if they match the expected coverage. A visualisation of this procedure, and how it detects incorrect posteriors, is given in Figs S\ref{SI:1}--S\ref{SI:3}. This study will utilise TARP to check whether Gaussian-likelihood inversions and SBI produce well-calibrated posterior distributions. 

\section{Application to synthetic and real waveform source inversions}

In the remainder of this paper we will apply SBI to synthetic and real data earthquake point source inversions for: (i)  six independent moment tensor components, fixing the source location; and (ii) space-time location (centroid latitude, longitude, depth and origin time) and the seismic moment tensor (i.e., making a total of 10 source parameters). We will compare SBI with classical Bayesian inversion methods and will carry out synthetic inversions considering both Gaussian and real data noise. Inversion results will be examined in the moment tensor component basis, as well as in a physically interpretable reparametrisation using strike, dip, rake, the non-double-couple component $\epsilon$ \citep[following the convention from][]{giardini1984systematic}, and moment magnitude $M_W$. Finally, the SBI approach will be applied to two moderate magnitude earthquakes in the Azores and Madeira regions combining seismic waveforms recorded by stations in the region's ocean islands and by the recent large-scale UPFLOW ocean bottom seismology experiment in the region \citep{tsekhmistrenko2024performance}.

\subsection{Earthquakes Studied and Data Used}

\begin{figure*}
    \centering
    \includegraphics[width=\textwidth]{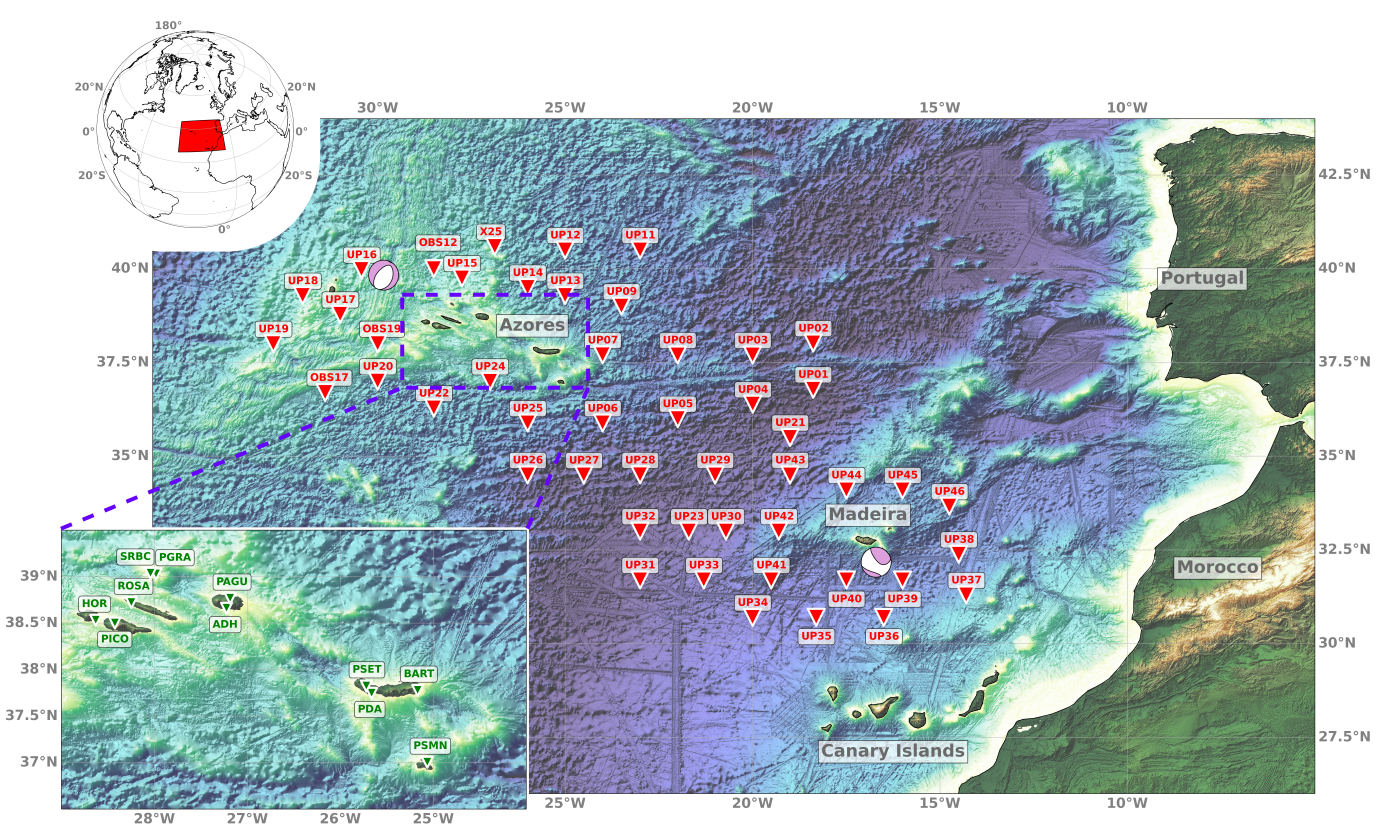}
    \caption{\textit{Main panel}: UPFLOW ocean bottom seismometer array located in the Azores-Madeira-Canaries region in the mid-Atlantic. This consists of 50 ocean-bottom seismometers that were deployed in July-August 2021, denoted by the labelled markers. 49 OBSs were recovered in August-September 2022, with UP14 being lost. Of these, a subset of 43 stations could be used after data quality checks \citep{tsekhmistrenko2024performance}. GCMT focal mechanism solutions are plotted as beachballs for two regional moderate magnitude earthquakes that we study in this work. \textit{Lower left}: A zoom-in on the Azores islands, with a number of labelled permanent land stations operated by Instituto Português do Mar e da Atmosfera (IPMA) that are used in this study.}
    \label{fig:UPFLOW_map}
\end{figure*}

The UPFLOW passive seismology experiment deployed 50 and recovered 49 OBSs in a $\sim \!1,000 \! \times\! 2,000$ $\text{km}^2$  area in the Azores-Madeira-Canary islands region with station spacing $\sim 110 - 160$ km between June 2021 and August 2022. The OBS array is shown in Fig. \ref{fig:UPFLOW_map}. Most of the instruments had three component broadband seismic sensors and hydrophones, and recorded seismic data with quality as high or even higher than in previous OBS experiments \citep{tsekhmistrenko2024performance}. Yet, as expected for OBS data, high noise levels are observed, with noise observations for most stations typically being close to the new high noise model \citep{tsekhmistrenko2024performance}. Such high and non-Gaussian noise levels make UPFLOW's data particularly suitable for testing our SBI approach. All UPFLOW data have clock drift estimated and corrected for \citep{tsekhmistrenko2024performance, cabieces2024clock}. While future work will use data corrected for tilt and compliance, these corrections should not affect this study's results because the data's dominant period is  $T\sim25$ s. This is at the limit for which tilt and compliance corrections become substantial for the UPFLOW dataset \citep{tsekhmistrenko2024performance}.

We study two regional moderate magnitude earthquakes. These events, located near the Azores and Madeira, have respective GCMT event codes 202201130646A and 202202160432A. We also use several permanent land stations in the Azores region operated by Instituto Português do Mar e da Atmosfera (IPMA). The earthquakes and permanent land stations are also shown in Fig. \ref{fig:UPFLOW_map}.

\subsection{Forward Modelling}
\label{sec:forward_modelling}

For a given earth model, a Green's function database can be computed that represents the solution to the wave-equation across the entire domain. This means that one can pay a single, one-off computational cost to compute a database and thereafter generate seismograms for all possible source-receiver configurations. This study uses \texttt{Instaseis} \citep{instaseis_paper}, a tool which queries a pre-computed strain database created by \texttt{AxiSEM} \citep{nissen2014axisem,axisem} to produce full displacement seismograms across a receiver array in just a few seconds. 

Following the convention of \citet{aki2002quantitative}, the displacement field $\mathbf{u}$ at position $\mathbf{x}$ and time $t$ for a point source  can be simplified to:
\begin{equation}
    u_n(\mathbf{x}, t)= \sum_{p,q} \int \mathbf{M}_{pq} (t-\tau) \frac{\partial}{\partial \xi_q}\mathbf{G}_{np}(\mathbf{x}, \tau; \mathbf{\xi}) d\tau,
\end{equation}
where $\mathbf{M}_{pq} (t-\tau)$ are the time-varying moment tensor components, which represent all six possible moment couples about the point source, $\mathbf{G}$ is the Green's function solution to the wave equation, and $\mathbf{\xi}$ is the position of the source. Here indices $\{n,p,q\}$ run over the three spatial dimensions. We may then use spatial reciprocity, a property satisfied by solutions of the wave equation, to re-express the displacement field in terms of receiver-side Green's function $\mathbf{G}_{np}(\mathbf{\xi}, \tau; \mathbf{x})$
\begin{equation}
    u_n(\mathbf{x}, t)= \sum_{p,q} \int \mathbf{M}_{pq} (t-\tau) \frac{\partial}{\partial \xi_q}\mathbf{G}_{np}(\mathbf{\xi}, \tau; \mathbf{x}) d\tau .
    \label{eq:eq_reciprocity}
\end{equation}
This step allows us to express the displacement field $\mathbf{u}$ in terms of a database of Green's function derivatives for all possible receiver locations at the surface $\mathbf{x}$ by computing and storing for efficient query $\partial \mathbf{G}_{np}(\mathbf{\xi}, \tau; \mathbf{x}) / \partial \xi_q$.

The time-varying component of the moment tensor describes how the moment couples vary over the duration of the earthquake rupture. This can often play an important role in modelling realistic earthquakes, particularly for large magnitude earthquakes where ruptures last long periods of time. In this study, which focuses on moderate magnitude regional earthquakes, it is assumed that the slip duration is small relative to the shortest period being simulated, i.e. $\tau_\text{slip} < T_\text{sim}$ meaning that the slip duration can be neglected. Mathematically it can be treated as a Dirac delta in Eq. \ref{eq:eq_reciprocity}, i.e. $\mathbf{M}_{pq} (t-\tau) = \mathbf{M}_{pq} \delta(t - \tau)$, where $\mathbf{M}_{pq}$ is now constant in time, such that the displacement field can be expressed simply as:
\begin{equation}
    u_n(\mathbf{x}, t)= \sum_{p,q}  \mathbf{M}_{pq} \frac{\partial}{\partial \xi_q}\mathbf{G}_{np}(\mathbf{\xi}, t; \mathbf{x}).
    \label{eq:final_displacement}
\end{equation}
Throughout the rest of this work, this equation will serve as the forward model to generate synthetic seismograms to perform inference. When considering displacement seismograms in multiple spatial directions and at $N$ separate receiver locations, all measurements can be combined such that $ \mathbf{D} = \{u^1_x,u^1_y,u^1_z, \ldots, u^N_x, u^N_y, u^N_z \}$. 

Note also that for a fixed source location $\mathbf{\xi}$, the displacement field $\mathbf{u}$ at each receiver is expressly linear with respect to the moment tensor components. This property is utilised to explore a simplified toy problem in the Section \ref{sec:artificial_gaussian_noise}, since it allows comparison of SBI with analytic results associated with linear Gaussian inference problems \citep{tarantola2005inverse, Alsing2018}.

This study utilises two crustal 1D earth models: one computed beneath station ROSA in the Azores islands \citep{Ferreira2020}, and a Madeira model provided by IPMA. Below the crustal model, LITHO1.0 \citep{pasyanos2014litho1} is used until its maximum depth, after which the Preliminary reference Earth model (PREM) 1D \citep{dziewonski1981preliminary} is used. Green's functions were computed using AxiSEM-1D \citep{axisem} down to a dominant mesh period of 10 s. Once the databases were computed and processed, Instaseis was used to generate synthetic vertical and horizontal component seismograms at each station. $15$-minute windows starting from one minute before the centroid time are used for each station. Event and noise observations have the instrument response removed, before being processed in exactly the same way as the synthetics. All real and synthetic waveforms are filtered using a 4-corner band-pass filter between $0.02$ and $0.04$ Hz (T $\sim 25 - 50$ s). 

These relatively long period data are used to mitigate against modelling errors, both due to the simplistic 1D Earth models being used, as well as unmodelled ocean-layer ringing effects observed in the OBS data. In our setting, particularly when treating waveforms with high signal-to-noise ratios, we expect large modelling errors that dominate contributions to the uncertainty and bias the inference results. This is relevant for the moderate magnitude events studied in this manuscript, and we present a discussion of this issue in Section \ref{sec:model_errors}.

\subsection{Neural density estimators (NDEs) and Bayesian inversion: practical implementation}

This study uses the \texttt{sbi} Python package \citep{tejero-cantero2020sbi} for training and sampling from NDEs. Throughout this work, the same MAF architecture is trained (only the input and output dimension is varied according to the problem). This uses the default parameters from the \texttt{sbi} package; there are 5 MADE blocks which use a feedforward neural network with two layers, each with 50 hidden nodes, to parametrise the conditional Gaussian distributions. $\tanh$ activations are used throughout. The MAF is always trained using the same procedure, again using the \texttt{sbi} package defaults: batch size of 50 samples, learning rate of $5\times10^{-4}$ with an Adam optimiser \citep{kingma2014adam}, and early stopping that ends training once validation loss has not improved for 20 epochs. A random $10\%$ fraction of the dataset $\mathcal{D}$ is selected as the validation set. Five-point stencils are used to compute the derivatives required for optimal score compression and iterative least squares throughout. 

To provide comparisons with the SBI approach, a range of Gaussian likelihood-based inversions are performed using MCMC. For the synthetic scenarios below, a simple Metropolis-Hastings sampling strategy is used where the step-size is manually tuned for each problem. For the real event inversions, an ensemble affine sampler approach following \citet{goodman2010ensemble} is used, implemented using the Python library \texttt{emcee} employing the default RedBlue move sampling strategy \citep{foreman2013emcee}. This is significantly more sample-efficient than more standard sampling approaches such as Metropolis-Hastings, particularly for the higher-dimensional inversions required for seismic sources. However, it does introduce a large synchronisation overhead compared to the embarrassingly parallel simulation strategy required for SBI. All experiments are performed utilising $20$ cores on a desktop class machine.

\subsection{Parametrising the Covariance}
\label{sec:parametrise_covariance}
In the sections that follow, comparisons will be drawn between the inferred posterior distributions using the SBI approach, as well as the standard Gaussian likelihood-based MCMC approach. Naturally, the Gaussian likelihood-based approach requires the specification of a noise covariance matrix. Also note, however, that optimal score compression requires an estimate of the covariance matrix. The key difference here is that using a misspecified Gaussian likelihood for inference can lead to biased, miscalibrated posteriors, whereas in principle in the optimal score compression case it should only lead to lossy compression, or broader uncertainty estimates than optimal.

This study uses the $15$-minute time-window prior to an event to estimate the variances $\hat{\sigma}^2$ for each station and component. This can be used to compute a simplified diagonal event covariance matrix ($\mathbf{C}_\text{E}$), with constant blocks associated with each of $n$ stations and $l$ components: 
\begin{equation}
(\hat{\mathbf{C}}_\text{diag}^{nl})_{ij} = (\hat{\sigma}^{nl})^2 \delta_{ij}.
\end{equation}

\citet{Duputel2012} demonstrated that for oversampled data (where the sampling rate is much higher than the lowest frequency content), it is important to include autocorrelation terms in the covariance $(\hat{\mathbf{C}}_\text{E}^{nl})_{ij}$. We follow their approach of a per station and component block diagonal covariance, with an exponentially decaying autocorrelation:
\begin{equation}
(\hat{\mathbf{C}}_\text{exp}^{nl})_{ij} = (\hat{\sigma}^{nl})^2 \exp{[-|\Delta t_{ij}|/t_0]},
\end{equation}
where the autocorrelation timescale $t_0$ is simply the shortest period in the filtering range, and $\Delta t_{ij}$ is the timelag between elements. 

In order to emulate realistic noise, a dataset of randomly selected 15-minute time windows of noise across the network is collected and processed identically to the synthetics. For each of these windows, a per-window covariance estimate ${\hat{\mathbf{C}}_N}$ is computed using the previous 15 minutes. However, note that for a given event inversion the event noise $\mathbf{C}_\text{E}$ may substantially differ from the sampled noise ${\hat{\mathbf{C}}_N}$, which could lead to different uncertainties during inference. We therefore use ${\hat{\mathbf{C}}_N}$ to linearly rescale each noise window to roughly match the event's covariance diagonal $\mathbf{C}_\text{E} \sim (\hat{\sigma}^{nl})^2$. These rescaled noise windows are added to synthetics during the training stage to produce realistic noise matched to the event noise level, as well as emulating errors in estimating the event covariance $\mathbf{C}_\text{E}$. This method allows us to draw observations from a realistic likelihood, $\mathbf{D}_i \sim p(\mathbf{D} \! \mid \! \mathbf{m}; \hat{\mathbf{C}}_\text{E})$.


Note that this approach implicitly makes an assumption of stationarity in order to make time-domain modelling feasible. This follows standard practice in seismology and is an adequate first-order estimate.
 


\subsection{Adapting Score Compression to Non-Linear Domains}
\label{sec:non_linear_method}

We also apply the SBI approach to full 10-parameter source inversions, where inference is performed over the earthquake source time-location (4 parameters) and the 6 independent moment tensor components. It is well-known   that such inversions are mildly non-linear in the vicinity of the true location, and so can be solved iteratively provided a good first estimate of the time-location is available \citep[see e.g.][]{Kanamori1981}.

For such realistic problems that are not expressly linear, selection of the expansion point $\mathbf{m}_*$ is key for successful application of optimal score compression. Since the compression is only locally optimal, it stands to reason that accurate inversions require that $\mathbf{m}_*$ be located in or near the bulk of the posterior mass. However, this requirement poses a problem in specifying the prior; an ``uninformative'' prior should cover the entire parameter space, including regions where locally computed gradients $\mathbf{G}_{\mathbf{m_*}}$ break down. 

For the problem of seismic source inversion, specifying broad priors over the moment tensor components (which can typically be allowed to vary over several orders of magnitude) substantially degrades the performance of the compression. This problem is exacerbated by the fact that the magnitude of the gradients for seismic displacement data vectors, $\mathbf{G}_{\mathbf{m_*}}$, is proportional to the moment tensor components. In practice, experiments in preparation for this work found that the rapid degradation of $\mathbf{t}$ away from the expansion point $\mathbf{m}_*$ completely corrupts the posterior $p(\mathbf{m} \mid \mathbf{t})$, such that it cannot be used as an approximation of the true posterior $p(\mathbf{m} \mid \mathbf{D})$. As such, a compromise is required that constrains the prior to regions of the model space where the linearisation of optimal score compression still roughly holds, while still ensuring the bulk of the posterior mass is supported by this prior.

This work proceeds by first computing the maximum likelihood solution $\mathbf{m_*}$ using a weighted iterative least squares using the approximate Gaussian likelihood. Once the solution for $\mathbf{m_*}$ has converged, the relevant quantities $\boldsymbol{\mu}_*$ and Fisher information $\mathbf{F}$ can be computed. Finally, the Gaussian posterior covariance estimate $\mathbf{\hat{C}} = \mathbf{F}^{-1}$ \citep[see e.g.,][]{tarantola2005inverse}, which gives an estimate of the per-parameter standard deviation $\boldsymbol{\sigma}^2 = \text{diag}(\mathbf{\hat{C}})$, can be used to construct a prior over model parameters $p(\mathbf{m})$. In order to avoid an overly informative prior that may bias the final result, instead of using the Gaussian posterior directly, a uniform hypercube centered on $\mathbf{m_*}$ is used:
\begin{equation}
    p_\text{Fisher}(\mathbf{m}; N) = \mathcal{U}\left[\mathbf{m_*} - N \boldsymbol{\sigma} , \mathbf{m_*} + N\boldsymbol{\sigma}\right],
\label{eq:fisher_prior}
\end{equation}
where $N$ is a free parameter to trade-off the limitations of an overly constrained prior ($N$ too small) against too broad a prior where the compression breaks down ($N$ too large). This study found values of $N\in[5,10]$ to be adequate. Once the prior distribution has been constrained, dataset generation, training, and inference proceed as before. This procedure is demonstrated in Fig. S\ref{SI:8}, and some of its limitations are explored through artificial inversion coverage tests in Fig. S\ref{SI:9}.

\section{Results}

\subsection{Synthetic Moment Tensor Inversion using Gaussian Noise}
\label{sec:artificial_gaussian_noise}

To begin with, a simple toy example of moment tensor inversion with fixed source location is explored. First, Gaussian noise is considered; this is particularly appealing since the linearity of the problem with respect to moment tensor components makes inference analytically tractable \citep{tarantola2005inverse}. In addition to this, optimal score compression saturates the information inequality globally since the first order expansion in Eq. \ref{eq:first_order_expansion} is exact. This means that any departure between the learned posterior $\psi(\mathbf{D}, \mathbf{m}) = \hat{p}(\mathbf{m} \mid \mathbf{D})$ and the true posterior is as a result of modelling errors in the NDE $\psi$. A component of the modelling error should vary as a function of the dataset size $\mathcal{D}$, or equivalently the number of simulations.

\begin{figure*}
    \centering
    \includegraphics[width=0.8\textwidth]{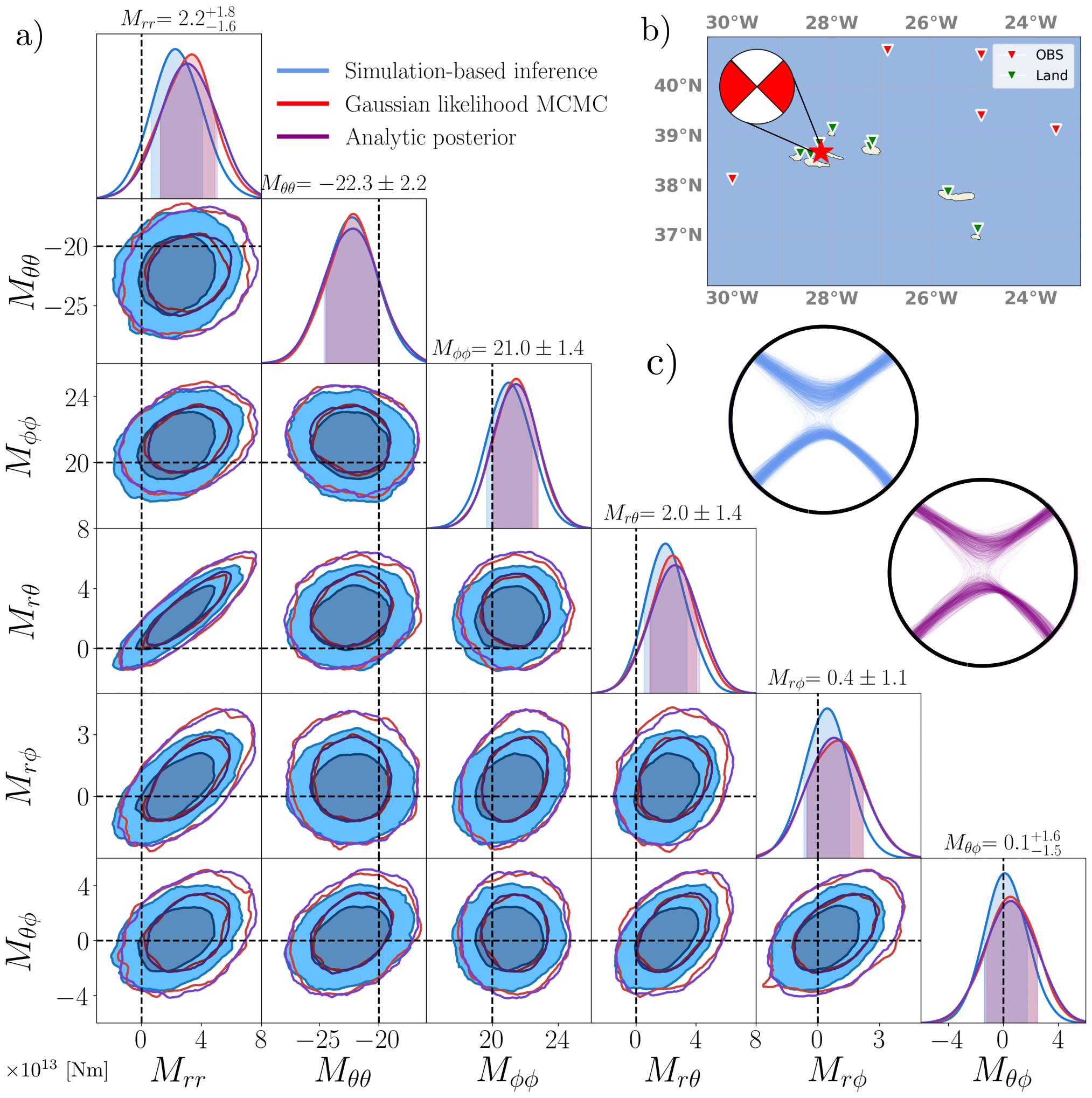}
    \caption{Solutions for the artificial strike-slip event with known Gaussian noise considered in this study. Panel a) shows the analytic posterior distribution (purple), MCMC posterior samples produced using a correctly specified Gaussian likelihood (red), and the SBI approach using a MAF trained on $4,000$ simulations (blue). Inner and outer contour rings correspond to $[68,95]\%$ credibility contours. The dashed lines represent the true solution. The SBI maximum \textit{a posteriori} solutions and 1D uncertainty estimates are given at the top of each column. Moment tensor component units are given in the bottom left. Panel b) shows the simplified station configuration used for the artificial examples, as well as the location and focal mechanism of the artificial event. Panel c) compares the beachball solutions of the SBI sample ensemble against the analytic posterior, showing very little physical difference between the results.}
    \label{fig:mt_gaussian_inversion}
\end{figure*}

An artificial pure strike-slip event is constructed with the following moment tensor solution: $M_{np} \equiv [M_{rr}, M_{\theta \theta}, M_{\phi \phi}, M_{r \theta}, M_{r \phi}, M_{\theta \phi}] = [0, -20, 20, 0, 0, 0]\times 10^{13} ~ \text{Nm}$. Vertical and horizontal component waveforms are simulated with a known Gaussian noise model for a simplified 13 station configuration around the Azores islands, shown in Fig. \ref{fig:mt_gaussian_inversion}.

The procedure summarised in Fig. \ref{fig:data_generation} is followed, proceeding by uniformly sampling moment tensor components within the hypercube $M_{np} \sim \mathcal{U}[-3, 3] \times 10^{14} ~ \text{Nm}$. For each sampled moment tensor $\mathbf{M}$, per station per component Gaussian noise is added to the data vector $\mathbf{D}$.  Fig. \ref{fig:mt_gaussian_inversion} shows the full inversion results of this synthetic example, with comparisons between the SBI approach, a long-run MCMC chain with the known Gaussian likelihood, and the analytically known posterior. Both posteriors closely match the true posterior, though the SBI approach does not fully overlap with the analytic result. This minor offset is caused by modelling errors in the neural density estimator (NDE); the NDE has been trained to model the posterior over a very large prior region, so small inaccuracies are unsurprising. It should be noted that while some parameters are not perfectly recovered, this is directly as a result of the noise in the data. Noise-free synthetic tests lead to a perfect coincidence between the posterior maximum and the ground truth, as expected.

\begin{figure}
  \centering
  \includegraphics[width=0.48\textwidth]{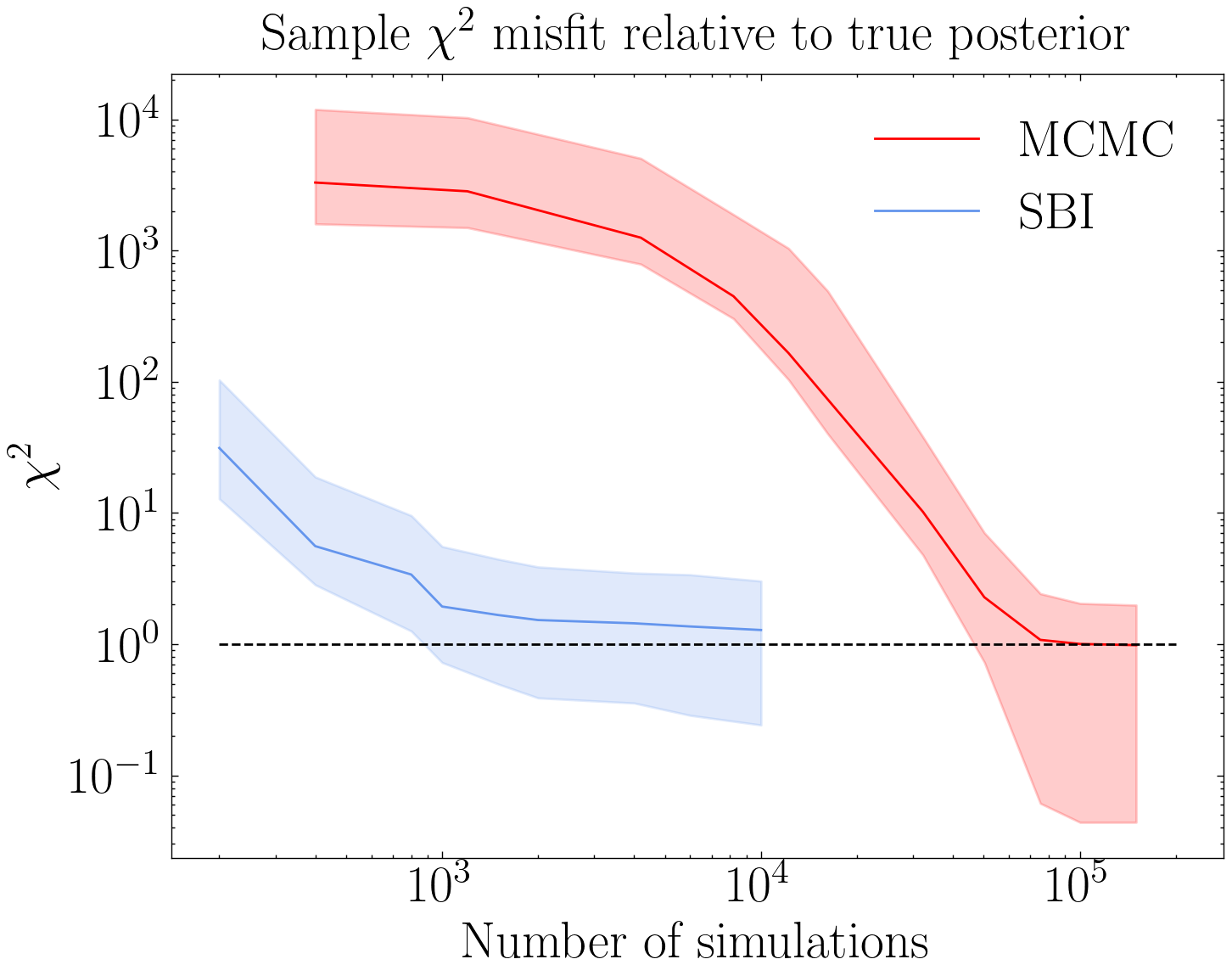}
  \caption{Mean reduced $\chi^2$ misfits of the samples drawn from each method over $50$ independent inversions; shaded regions indicate $10\% - 90\%$ quantiles of the sample quality over all event inversions, repeated three times. For the SBI approach, the number of simulations is equivalent to the dataset size $\mathcal{D}$. $\chi^2$ of the MCMC chain is calculated using a burn-in fraction of $0.5$. SBI yields accurate estimates of the posterior with an order of magnitude reduction in simulations.}
  \label{fig:nde_misfit}

\end{figure}

We then generate $50$ independent events within the same prior, and explore how inference performance varies as a function of the number of simulations for each technique. We retrain the same NDE architecture for a range of dataset sizes and perform inference on each event to check the inferred posterior quality. Fig. \ref{fig:nde_misfit} compares how the sample ensemble of each inversion approach improves with the number of simulations. These results indicate at least an order of magnitude reduction in the number of samples required to generate realistic posterior contours using the SBI approach.  Misfits are calculated using the analytically known posterior $ p(\mathbf{m \mid D}) = \mathcal{N}[\mathbf{m}_\text{MLE}, \mathbf{F}^{-1}$], such that $\chi^2 = (\mathbf{m}_\text{MLE} - \mathbf{m}_\text{samples})^T \mathbf{F} (\mathbf{m}_\text{MLE} - \mathbf{m}_\text{samples}) / n_{\text{dof}}$, where $n_{\text{dof}} \!= \! \text{dim} \: \mathbf{m}$ \citep{tarantola2005inverse}.

\subsection{Synthetic Moment Tensor Inversion with Real Noise}
\label{sec:artificial_real_noise}
Next, real data noise is introduced into the 6-parameter moment tensor inversion. The process from above is repeated, but this time real noise collected from each individual station is added to the relevant synthetics. This requires the curation of a small database of pre-processed noise traces (around 2500 windows of concurrent traces across the array, collected over 14 days of data), which have been bandpass filtered, down-sampled down to the desired frequency of 1 Hz, and have had the instrument response removed. For this application, we collect 15 minute noise windows, and rescale them to the event noise $\hat{\mathbf{C}}_E$ using the previous 15 minutes, following the approach explained in Section \ref{sec:parametrise_covariance}. Importantly, this integrates over errors induced in estimating the event covariance matrix using the previous time-window. These noise samples are then added to each synthetic data-vector $\mathbf{D}$ before compression, such that the distortion to the distributions over $\mathbf{t}$ are reflected in the training stage of SBI.  

\begin{figure*}
    \centering
    \includegraphics[width=\textwidth]{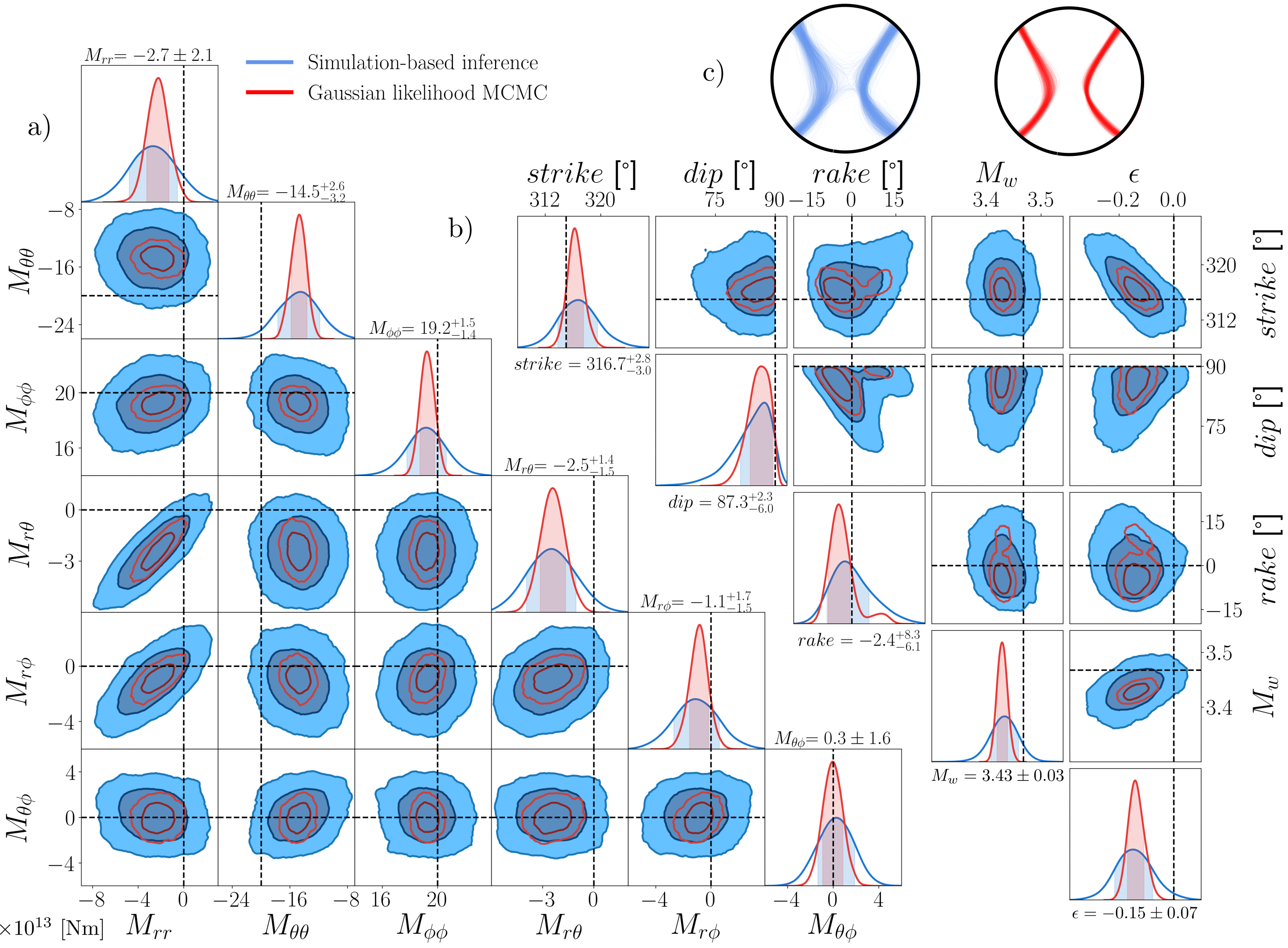}
    \caption{Inference results for the artificial strike-slip event considered in this study with real noise, comparing MCMC samples using an approximate Gaussian likelihood parametrised by $\mathbf{C}_{\text{exp}}$, and SBI using $4,000$ simulations. Panel a) shows the solutions in the moment tensor parametrisation, and b) in the physically interpretable parametrisation of strike, dip, rake, magnitude $M_W$ and non-double-couple component $\epsilon$. The Gaussian likelihood approach yields significantly narrower contours than the SBI approach, leading to inconsistency between the inferred posterior and the true solution. Panel c) shows a fuzzy beachball plot of the two sample chains, indicating that the SBI posterior is consistent with a pure double-couple strike-slip event. }
    \label{fig:mt_real_inversion}
\end{figure*}

The same artificial earthquake is used as in the previous section. The inversion results are shown in Fig. \ref{fig:mt_real_inversion}, both in the original moment tensor parametrisation as well as in the more physically interpretable basis of a best double-couple earthquake parametrised by its fault strike, dip and rake \citep{Dziewonski1981}. Both posteriors have a similar maximum \textit{a posteriori} (MAP) point because both rely on the same Gaussian covariance matrix $\hat{\mathbf{C}}$ to compute the maximum likelihood estimate MLE (since the MLE is not too biased, and the prior is uniform). However, the likelihood-based MCMC approach yields significantly overconfident contours, while the SBI posterior consistently covers the true synthetic moment tensor parameters. To some degree, the overconfidence is caused by misspecification of the Gaussian likelihood - a combination of errors in estimating the per-station, per-component variances, and unmodelled off-diagonal covariances. A broader problem is the fact that the noise is non-stationary and non-Gaussian, and therefore cannot be fully modelled by standard Gaussian noise approximations. 

\begin{figure*}
    \centering
    \includegraphics[width=\textwidth]{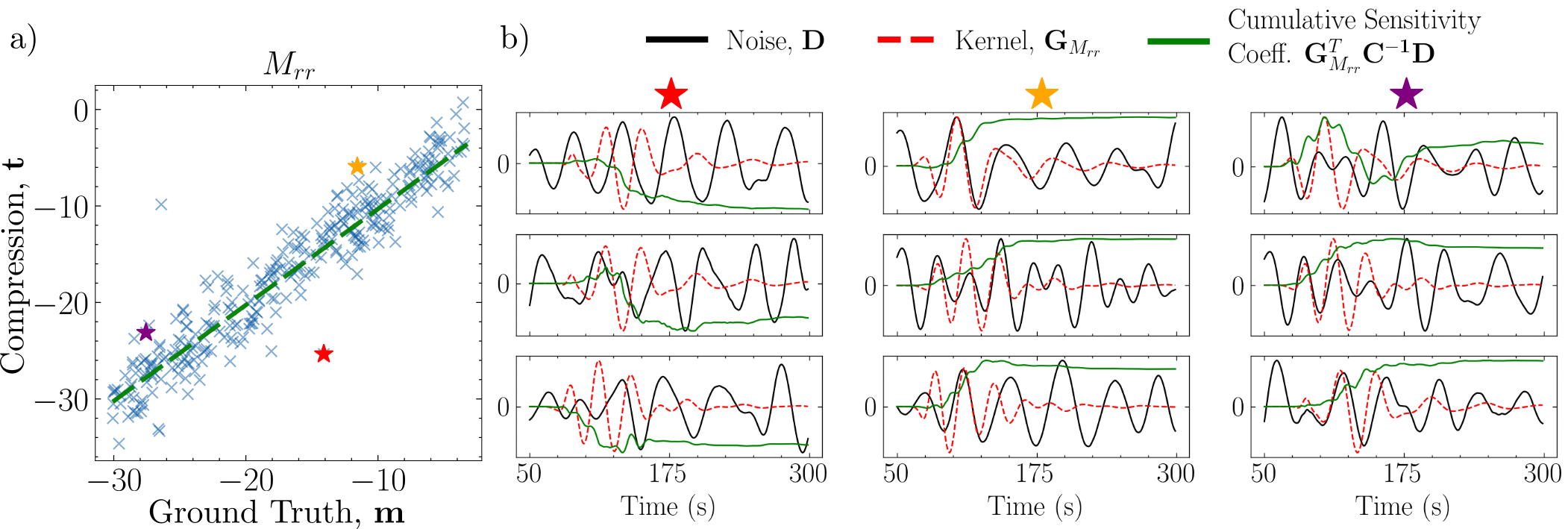}
    \caption{Panel a) displays a 1D slice of dataset $\{\mathbf{m}_i, \mathbf{t}_i\}$ for $M_{rr}$ against the corresponding compression estimates $t_{M_{rr}}$, which are MLEs of $M_{rr}$ under our compression scheme. The dashed green line indicates perfect recovery of the model parameters, while each scatter point is a compressed noisy observation. Panel b) plots the noise (black) that caused poor compression for 3 separate events (each coloured star and column corresponds to a different event). It compares the noise (black) to the sensitivity kernel $\mathbf{G}_{M_{rr}} = \partial_{M_{rr}} g(\mathbf{m})$ (dashed red). The product of these terms will lead to noise-induced errors in the compression scheme, whose cumulative sum over time is shown (green). In each of the poorly compressed events, the noise \mbox{(anti-)correlates} with the sensitivity kernels, driving large compression errors that are not properly accounted for by the chosen covariance parametrisation. These errors appear to be driven by the strong non-stationarity of the noise amplitude as well as varying frequency-dependence.}
    \label{fig:real_noise_examples}
\end{figure*}

We hypothesise that the primary driver of the overconfidence is the strongly non-stationary, frequency-dependent noise that may occasionally coincide with key regions of the signal. This is investigated in Fig. \ref{fig:real_noise_examples}. We observed that occasions where the noise is strongly (anti)-correlated with the sensitivity kernels $\mathbf{G}_{\mathbf{m_*}}$ (at levels inconsistent with the estimated Gaussian covariance) significantly skew the maximum likelihood solution. Since this non-stationarity and frequency dependence is unmodelled by the Gaussian likelihood, such an approximation will incorrectly overestimate the confidence. On the other hand, the SBI approach should account for these biases through its empirical modelling of the error distribution in $\mathbf{t}$ under the effect of real noise in $\mathbf{D}$, leading to better calibrated posteriors.

\begin{figure}
  \centering
  
    \includegraphics[width=0.48\textwidth]{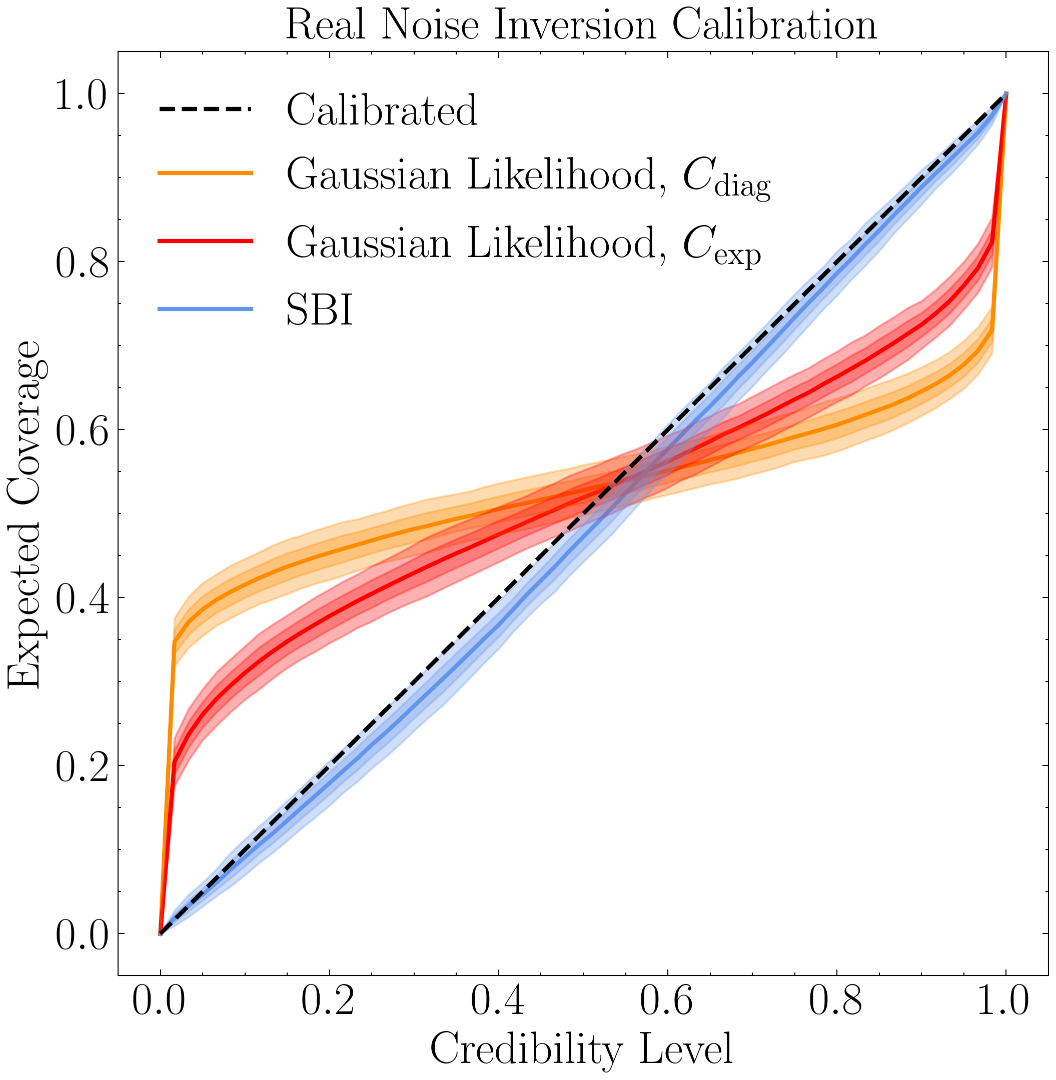}
  \caption{Expected coverage plots for both inversion techniques, generated by performing $600$ independent inversions, with $\pm[1,2]\sigma$ uncertainty regions estimated by bootstrapping. We find that the exponential tapering parametrisation $C_\text{exp}$ (red) significantly improves posterior quality over the na\"ive diagonal covariance $C_\text{diag}$ (orange), but still yields highly overconfident uncertainties. The SBI approach (blue) used 10,000 simulations and the $C_\text{exp}$ covariance parametrisation for compression. Its coverage plot is very close to the perfect calibration scenario (black dashed line).}
  \label{fig:coverage}

\end{figure}

For a systematic evaluation, an estimate of the empirical coverage of the inversion techniques approach can be performed using the TARP approach explained in Section \ref{sec:posterior_quality}. This is presented in Fig. \ref{fig:coverage}, where 600 independent events generated within $M_{np} \sim \mathcal{U}[-3, 3] \times 10^{14}~\text{Nm}$ are inverted using both techniques to estimate their calibration in a real noise settings. For computational efficiency, a Metropolis-Hastings sampling approach with 20 independent chains is used for the likelihood-based inversions. Each chain contains 20,000 samples, with a burn-in fraction of 0.5 and thinning factor of 5 to reduce sample correlation. The proposal step-size in the Metropolis-Hastings algorithm was tuned for the problem, and convergence was manually verified for a number of chains. 

Fig. \ref{fig:coverage} indicates substantial issues with the Gaussian likelihood-based approaches. The expected coverage significantly diverges from the ideal distribution at the tails; this indicates that a very high proportion of true model parameters fall well outside the inferred posteriors. This shows that Gaussian likelihood MCMC yields highly overconfident posteriors, which significantly underestimate the uncertainties in the model parameters. This finding is consistent with earlier work which found that an improved covariance parametrisation led to broader posterior uncertainty estimates \citep{Musta2016}. On the other hand, the SBI approach yields very well-calibrated posteriors, with a slight indication of a mild bias, similar to that observed in the Gaussian noise inversions in Fig. \ref{fig:mt_gaussian_inversion}. Figs S\ref{SI:4} and S\ref{SI:5} show that Gaussian likelihoods persistently underestimate uncertainties across both land and OBS seismometer arrays, while SBI remains well-calibrated. Fig. S\ref{SI:6} explores the degree of overconfidence with further coverage testing, and demonstrates that the Gaussian likelihood approach appears to overly-constrain each parameter by up to a factor of 3.

Note, however, that while the SBI approach is well-calibrated, meaning that it produces posteriors that are statistically consistent with the true model parameters, this does not mean SBI recovers the true posterior. Since the compression step is suboptimal (the Gaussian covariance $\hat{\mathbf{C}}$ used to compress the data is incorrect), the true posterior will be tighter than the contours our SBI method produces. Tighter posteriors could be obtained with improved parametrisations of the covariance to compress the data for SBI; we leave this for exploration in future work.

To summarise, this section has shown that the standard Gaussian likelihood approximation, alongside common parametrisations of the covariance matrix $\hat{\mathbf{C}}$, yields highly overconfident posteriors that often fail to cover the true model parameters when applied to realistic seismic observations. On the other hand, the SBI approach produces robust and trustworthy posteriors that are broadly consistent with the model parameters being inferred.

\subsection{Real Source Inversions}

The non-linear inversion strategy from Section \ref{sec:non_linear_method} is applied to two moderate magnitude events in the region, both of which have GCMT catalogue solutions shown in Fig. \ref{fig:UPFLOW_map}. The first of these was located slightly north of the Azores islands, while the second occurred just south of Madeira. For both events, all the Azores land stations and UPFLOW OBS (only vertical components to ensure the use of high-quality data) are initially included in the inversion. Observations are then compared with initial synthetics, allowing stations and components to be excluded on the basis of any clear modelling errors by visual inspection. Note that no data are excluded purely on the basis of a low signal-to-noise ratio, and no windowing is performed, since these effects should not affect the inversion (again, provided they do not obscure any modelling errors). The resulting subset of accepted stations and components is then fixed and used throughout all dataset generation and inference. 

For each event, the GCMT solution is used as a starting location, and a modified Levenberg–Marquardt algorithm \citep{fletcher1971modified} is used to perform a damped iterative least-squares inversion until reaching convergence on a local best-fitting solution $\mathbf{m}_*$. The prior is constrained as in Eq. \ref{eq:fisher_prior}, and inference is performed using both SBI and Gaussian likelihood-based MCMC. All results are presented in Figures \ref{fig:azores_4.8_real} and \ref{fig:madeira_4.8_real}, with $x, y, \Delta t$ given as shifts relative to the GCMT reference solution. 

\subsubsection{North Azores $M_W \: 4.8$, 13/01/2022}

\begin{figure*}
    \centering
    \includegraphics[width=\textwidth]{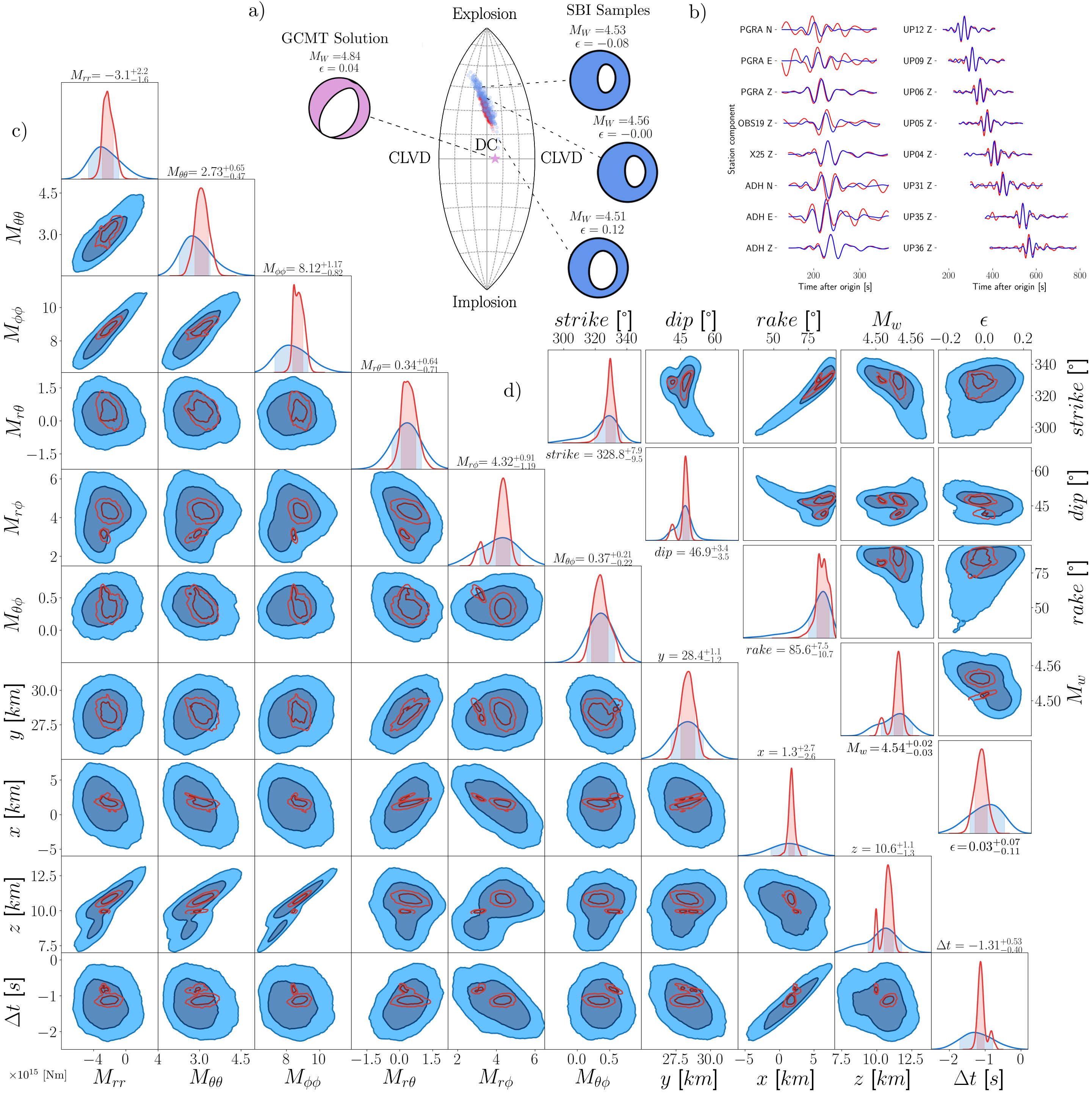}
    \caption{10-parameter inference solutions for the Azores islands $M_W \sim 4.8$ event that occurred on 13/01/2022 using 46 stations, comparing Gaussian likelihood inversion results (red) with the SBI approach (blue) trained using 10,000 simulations. Panel a) shows a lune plot \citep{tape2012geometric, Tape2015} comparing both solution ensembles and the GCMT solution. The focal mechanism contributions are represented on the eigenvalue lune, which conveys the source type of the moment tensor. Note that GCMT solutions impose a zero volumetric component, so the solutions are not directly comparable. b) Compares the observations (red) against best fitting solution $g(\mathbf{m}_*)$ (blue); c) shows the full 10-parameter trade-off plots for the SBI and Gaussian likelihood inversions; and d) shows the same posterior solutions parametrised by fault strike, dip, rake, magnitude $M_W$ and non-double-couple component $\epsilon$.  }
    \label{fig:azores_4.8_real}
\end{figure*}

\begin{figure*}
    \centering
    \includegraphics[width=\textwidth]{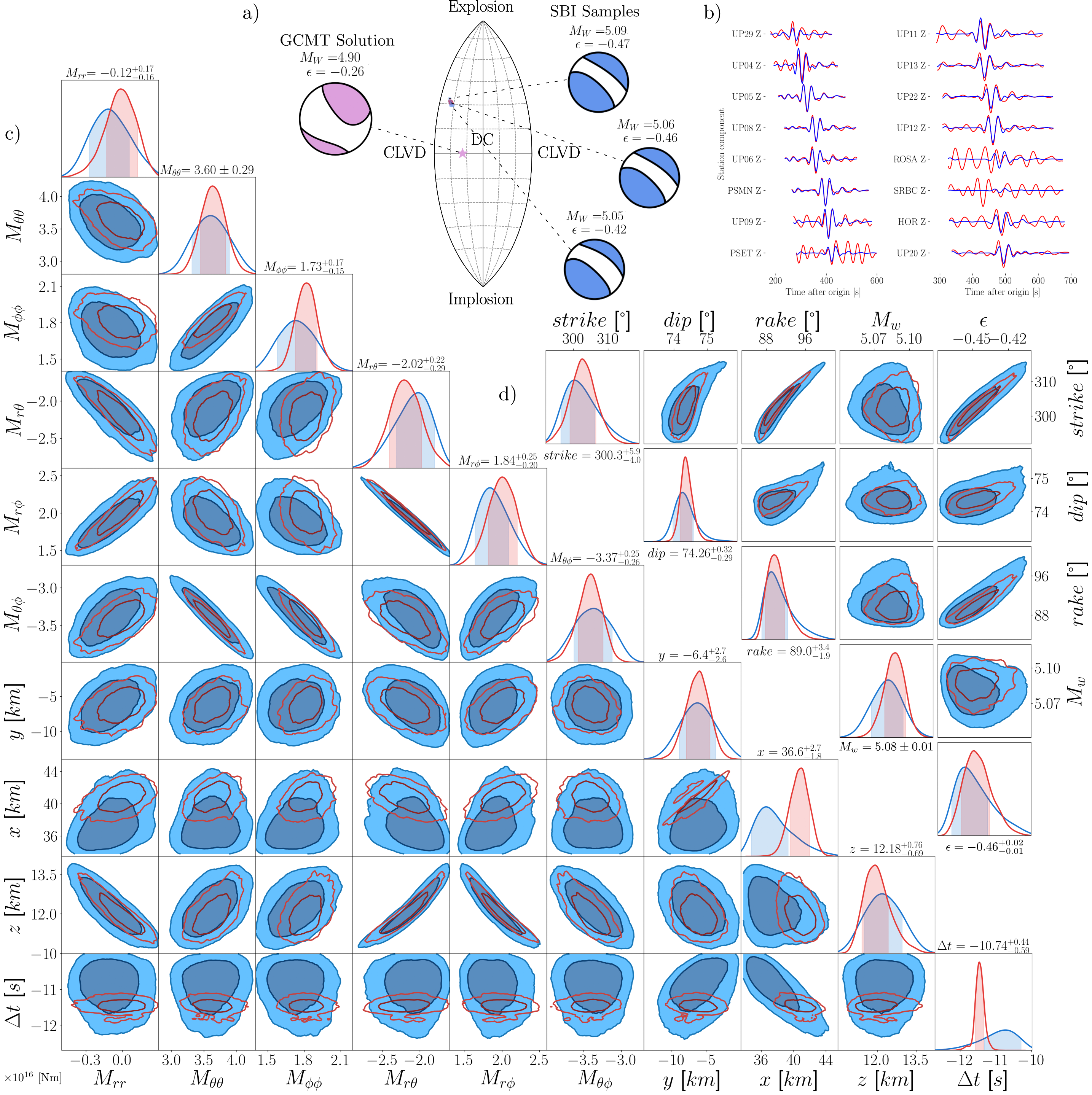}
    \caption{10-parameter inference solutions for the Madeira  $M_W \sim 4.9$ event that occured on 16/02/2022 using 19 stations, comparing Gaussian likelihood inversion results with the SBI approach trained using 10,000 simulations. Panels are the same as Fig. \ref{fig:azores_4.8_real}. This inversion utilises a more tightly constrained prior $p_\text{Fisher}(\mathbf{m} ; N=4)$.}
    \label{fig:madeira_4.8_real}
\end{figure*}

Data from across the seismic array is collected and processed identically as in Sections \ref{sec:artificial_gaussian_noise} and \ref{sec:artificial_real_noise} for the $M_W \: 4.8$ event, north of the Azores islands. We ran iterative least squares, using the GCMT solution as the initial time-location, until convergence on a maximum likelihood solution $\mathbf{m}_*$, and constrained with $p_\text{Fisher}(\mathbf{m} ; N=8)$ as in Eq. \ref{eq:fisher_prior}. Inference is run using 10,000 simulations for SBI and 160,000 samples with the Gaussian likelihood MCMC approach. Excluding a few minutes to perform iterative least squares, the SBI inversion took $24$ minutes ($16$ minutes for dataset generation, $8$ minutes training the MAF) and the MCMC approach took $400$ minutes. Results are presented in Fig. \ref{fig:azores_4.8_real}, with the GCMT moment tensor solution presented as reference (note that GCMT inversions constrain the volumetric component to zero). We also show a random subselection of data fits comparing the observations $\mathbf{D}_\text{obs}$ with the MLE solution $\mathbf{D} =  g(\mathbf{m}_*)$. These show good fits across most of the land station components, as well as a large number of the OBS vertical components across the UPFLOW array.

By virtue of the shared iterative least squares and prior constraining step, both sets of solutions have similar maximum likelihood solutions, but the inferred posteriors diverge significantly. As explored in Section \ref{sec:artificial_real_noise}, the Gaussian likelihood approach yields significantly narrower contours than those produced by the SBI approach. Good compression quality, demonstrated in Fig. S\ref{SI:9}, was found for this event (i.e. the prior constraining step in Eq. \ref{eq:fisher_prior} ensured low compression loss). This  provides strong evidence that the Gaussian likelihood yields highly overconfident posteriors in line with the synthetic inversion experiments carried out in the previous sections. The SBI posterior recovers some of the expected trade-offs, such as between the isotropic components $[M_{rr}, M_{\theta\theta}]$ and $ [M_{rr},M_{\phi\phi}]$. We find a large volumetric component across both ensembles; it is unclear to what degree this is a genuine feature of the event, or whether it is driven by insufficient data and modelling errors. Future work could investigate solution sensitivity to volumetric component by enforcing traceless priors during the inversions. Further interpretation, such as attributing the focal mechanism to tensional crack, is challenging without constraining information on subsurface structures, since the classical moment tensor model of \citet{aki2002quantitative} may not be applicable \citep{tape2013classical, alvizuri2016full}. We observe a wide range of trade-offs that differ between the SBI and Gaussian likelihood approach. These likely reflect genuine features of the posterior caused by the broader contours sampling different regions (and therefore trade-offs) in the domain, and by using an improved empirical likelihood that better captures the effects of seismic noise. Sub-optimal compression will perturb the SBI posterior, likely slightly biasing and blurring it over the domain, but in this instance simple checks ensure this is a minor effect (Fig. S\ref{SI:9}).  

Interestingly, both inversions produce significantly multi-modal posteriors which appear to be driven by the Moho discontinuity in the source-receiver region. This multi-modality is highly pronounced in the likelihood case, whereas SBI suggests that the two modes are connected and cannot be resolved completely separately due to the effects of seismic noise on the observations. One promising observation is that despite the Moho's effect, which will necessarily degrade the quality of the compression since local derivatives $\mathbf{G}_{\mathbf{m_*}}$ will be inaccurate, this does not prevent the SBI approach from modelling both modes of the distribution. During testing, we found that the SBI approach began to adequately resolve the two modes only after 4,000 simulations, before which the modes blurred together. On the whole, SBI produces a far greater range of samples, admitting solutions with the strike, dip, and rake varying by 20-30 degrees from the MAP, compared with 5-10 degrees utilising a Gaussian likelihood. Particularly striking are the long-tailed trade-offs between the fault angles that are obscured by a Gaussian likelihood approach. This example indicates that in settings with low signal-to-noise ratio, and therefore broader posterior distributions, our methodology produces solutions whose physical interpretation may significantly diverge from a Gaussian likelihood approach. 

\subsubsection{Madeira $M_W \: 4.9$, 16/02/2022}

Inference is performed on the $M_W \: 4.9$ event just south of Madeira. This event was more difficult to treat; most of the surrounding OBSs could not be utilised due to suspected near-field effects, and the OBSs nearer the Azores islands were at a sufficient distance to be impacted by significant modelling errors due to the highly heterogeneous regional earth structure. In order to stabilise the iterative least squares procedure, a prior around the GCMT source location was imposed:
\begin{equation}
p_\text{loc}(x, y) = \mathcal{N}\left( \begin{bmatrix} x_\text{GCMT} \\ y_\text{GCMT} \end{bmatrix} \degree, \begin{bmatrix} 0.01 & 0 \\ 0 & 0.01 \end{bmatrix} \circ^2 \right).
\end{equation}
This corresponds to a relatively tight prior with $\sigma \simeq 10$ km around the GCMT source location, in-line with GCMT source location errors uncovered in prior work \citep{weston2011global, weston2012systematic}. We incorporate this prior into the least squares procedure and the compression routine \citep[see][]{tarantola2005inverse,Alsing2019}. Finally, we use $p(\mathbf{m}) = p_\text{loc}(x, y) p_\text{Fisher}(\mathbf{m}; N)$ for both SBI (for dataset generation, implemented using truncated Gaussian sampling) and the Gaussian likelihood inversions.

The iterative least squares procedure was sensitive to the initial conditions, prior, and damping parameters, yielding various least squares solutions converging on false local minima or stationary points. This indicates significant model errors, as well as low sensitivity to some model parameters that made convergence challenging. Once an adequate minimum was found with low misfit values and well-fitting synthetics, poor compression performance was observed, with large compression errors caused by low model sensitivity compounding with non-linearity effects that degrade optimal score compression. This is explored in Fig. S\ref{SI:10}. We therefore selected $p_\text{Fisher}(\mathbf{m}; N=4)$ to mitigate against these problems. We proceeded with inference using 10,000 simulations for SBI and 200,000 samples with the Gaussian likelihood MCMC approach. Inference took $12$ minutes (including dataset generation, training, and sampling) and $240$ minutes, respectively. The results are presented in Fig. \ref{fig:madeira_4.8_real}.

Results are broadly similar to the moment tensor inversion in Fig. \ref{fig:azores_4.8_real}, with the Gaussian likelihood yielding slightly tighter posteriors than the SBI approach. Both sets of solutions show little variability, driven by high signal-to-noise ratios and poor sensitivities for this event; the lune plot in Fig. \ref{fig:madeira_4.8_real} indicates the two approaches yield very similar mechanisms. Again, both approaches resolve similar trade-offs between parameters. There are some differences in the posteriors; most notably, the MAP point is different for some parameters. This deviates from our earlier results, and is caused by the interplay between the non-uniformity introduced by the prior $p_\text{loc}(x, y)$ and the empirical model of the likelihood utilised by SBI, $\hat{p}(\mathbf{D} \mid \mathbf{m})$. For instance, the SBI MAP longitude shift $x$ is closer to the prior mean due to the much broader empirical likelihood the SBI approach produces. This underscores an important point about the advantages of SBI; when complex priors are imposed on the inference problem, a better model of the likelihood can even lead to significantly different MAP solutions in addition to different uncertainty intervals. 

For both events considered, we found similarities between our solutions and those in the GCMT catalogue, with focal mechanisms consistent with the regional plate boundaries. In the north Azores normal faulting aligned with the mid-Atlantic ridge predominates \citep{frietsch_robustness_2018, ousadou2024catalogue}, while for the Madeira event the high non-double-couple component and depth is consistent with existing catalogues that hint at volcanic processes \citep{matos2015upper,ipma2024}. However, in a strict sense the inferred posteriors did not agree with (``cover'') the GCMT solutions (particularly the magnitude of the events), with GCMT parameters well outside the $5 \sigma$ credibility region. This may be at least partly due to differences in the datasets used, notably the use of regional data in this study, whereas the GCMT solution is based on teleseismic data. However, the dominant cause for these discrepancies is likely to be substantial modelling errors, notably due to incorrect Earth structure models, which we did not address directly. Future studies of these events would benefit from regional high-resolution 3D tomographic modelling, particularly since amplitude modelling at OBSs requires careful treatment of bathymetry and water layer effects.

\section{Discussion}

We have found that across both synthetic and real source inversions, Gaussian-likelihood approaches with various covariance parametrisations are inherently limited for full-waveform moment tensor inversions. Failure to model autocorrelations caused by non-stationary, frequency-specific noise appears to lead to significantly overconfident posteriors. Some prior work has constructed more complex parametrisations of the covariance matrix $\hat{\mathbf{C}}$ by fitting tapered sinusoids to the autocorrelation structure \citep{Musta2016}. More generic empirical modelling of the Gaussian covariance is challenging due to the practical difficulties in estimating Gaussian auto-covariances of time series from limited samples \citep{wu2012covariance}. Transdimensional inference, which jointly inverts the noise and model parameters hierarchically, could also be explored for seismic source inversion for better estimates of the noise model \citep{sambridge2006trans, sambridge2013transdimensional}. However, all these approaches may be hamstrung by the strong non-stationarity of the frequency content and amplitude of seismic noise, as well as potential inherent non-Gaussianity. These factors suggest that wholesale empirical modelling of the likelihood, as performed by SBI, is highly desirable in the full-waveform inversion context. Additionally, further work could explore how these results vary under different frequency ranges. For instance, shorter period seismic waveforms close to the microseismic noise band may be noisier but more stationary than for the wave periods considered in this study \citep{peterson1993observations}. This could potentially lead to differences in the performance of standard Gaussian likelihood-based techniques. 

We successfully applied our SBI scheme to a 10-parameter source-location and moment tensor inversion for two real events near the Azores and Madeira, and explored the differences between SBI and a Gaussian likelihood approach. This demonstrated how our application can be naturally adapted to admit priors over the model parameters. In future work, constraints on the non-double-couple and volumetric component of the moment tensor, or physically realistic priors as in \citet{Tape2015, Stahler2016_a, Stahler2016_b} could be explored. Significant improvements to the likelihood model, as provided by SBI, can lead to differing MAP estimates and much broader uncertainties. These findings, especially the underestimation of uncertainties observed in Gaussian likelihood-based inversions, should be considered when interpreting previous results presenting moment tensor solutions that seem inconsistent with the regional tectonic setting or geological context.

\subsection{Limitations}
\subsubsection{Optimal Score Compression}
\label{sec:limitations_optimalscore}

Optimal score compression is a very efficient technique to compress high-dimensional seismic observations into low-dimensional summaries, making them amenable to simple ML-based density modelling. However, it poses significant limitations for full-waveform inversions; uninformative priors cannot be treated directly, restricting inference to quasi-linear regions of the model space. Additionally, while constraining the prior was adequate to apply optimal score compression in the mildly non-linear 10 parameter source inversion, higher-dimensional parameter spaces that include e.g. source-time functions may lead to increasing non-linearity and distant modes in the posterior that cannot be adequately modelled with this approach  \citep{vallee2011scardec, Stahler2016_a}.

These issues are inherent limitations of optimal score compression when applied to non-linearities in the observations. Often, practitioners perform two-step compression where the raw observations are transformed into a first set of more stable summaries, after which the linearising assumption in optimal score compression is adequate \citep{Alsing2019}. This approach may reduce the discriminative power of SBI but ensures its robustness. In seismology, this may involve utilising wavelet transforms and power spectra in conjunction with classical compression techniques such as principal component analysis \citep{vavryvcuk2017moment, yu2018moment, Piras2023}. Another approach could aim to adapt compression techniques to more time-shift robust techniques that utilise the commonly used decorrelation rather than $L2$ distance \citep{Stahler2016_a, pham2024gradient} or perhaps take inspiration from recent work using optimal transport in seismology \citep{engquist2016optimal,sambridge2022geophysical}. Alternatively, a more general approach could utilise end-to-end deep learning architectures inspired by e.g. \citet{Münchmeyer21, nooshiri2022multibranch} to directly learn how to compress these high-dimensional seismic observations using ML \citep{Jeffrey2020}. Such deep learning approaches could be integrated into an SBI framework to make inference robust to data artefacts and glitches, mitigating against the need for manual data inspection and selection \citep[as in e.g.,][]{mcbrearty2022earthquake, mcbrearty2023earthquake}.

\subsubsection{Addressing Modelling Errors}
\label{sec:model_errors}

One key limitation of this study is that it only partially addressed issues related with modelling errors (e.g., inaccurate earth structure used, missing crucial physical processes such as seismic wave propagation in the oceans, etc). Modelling errors can cause serious inaccuracies in seismic source inversion, potentially leading to highly overconfident, inaccurate posteriors. A wide range of practical efforts to address this issue have been utilised across the field. \citet{valentine2012assessing} explored ensembles of ``plausible'' solutions; \citet{Duputel2012,duputel2014accounting,vasyura2021accounting, pham2024towards} incorporated an additional ``corrective'' Gaussian terms in the likelihood that addresses auto-correlated errors arising from modelling errors (e.g. errors caused by phase-offsets), and demonstrate significantly improved robustness to model errors. However, since for instance waveform time shifts do not comply with a Gaussian distribution \citep{yagi2008importance, hu2023seismic}, these approaches represent simplifications and assumptions about the effect of model errors on the inference process. They could therefore produce incorrect uncertainty intervals that are difficult to interpret. The approach presented by \citet{Stahler2016_a,Stahler2016_b} remains robust and principled in dealing with modelling errors in the Bayesian framework, as they demonstrate that modelling the distribution over decorrelation errors between observed seismograms and synthetics prove more robust than Gaussian likelihood noise models. The success of this approach, however, hinges on the availability of an accurate, curated catalogue, which may be infeasible to acquire or whose biases may be propagated directly into the source parameter estimates. 

SBI represents a natural extension of these earlier works, allowing for a significantly more rigorous treatment of the effect of modelling errors that can guarantee high-quality, well-calibrated posteriors. By sampling from the prior earth model parameter uncertainties directly during the dataset generation step, these effects can effectively be baked into the empirical modelling of the likelihood function \citep{alsing2019nuisance}. This represents a more principled way to marginalise over these  `nuisance' parameters that represent significant sources of uncertainty which may not be adequately captured by Gaussian model error approximations \citep{vasyura2021accounting}, and whose direct inclusion in e.g. MCMC is infeasible due to the large increase in model dimension it would entail.  Not only would this allow seismologists to include the effects of a misspecified earth model, but more practical sources of error, such as unknown orientation or clock skew in OBS instruments \citep{zha2013determining, hable2018clock}, could also be naturally incorporated. 



\subsection{Future Applications}

The techniques introduced in this paper could be adapted to other aspects of seismology. Both full waveform and arrival time tomography suffer from non-Gaussian noise distributions \citep{tilmann2020another}, making SBI an appealing avenue to explore. Indeed, since the noise then becomes dependent on the model parameters, there is ample motivation to explore how this more complex source of stochasticity affects inferred posteriors. On the other hand, these domains tend to have very high-dimensional model spaces (in the case e.g. of large-scale tomography), or complex parametrisations that require adaptations like transdimensional sampling techniques \citep{bodin2009seismic, bodin2012transdimensionala}. Further research could be valuable in these applications where direct use of SBI as formulated here is not straightforward.

This work has utilised a simple formulation of SBI, where a single set of simulations is used to build an amortised posterior model \citep{papamakarios2016fast}. Active learning, where repeated rounds of inference use iterative estimates of the posterior to select which regions of the model space to probe, can result in further significant reductions in the number of forward model evaluations required to produce posterior estimates \citep{papamakarios2019sequential, brehmer2020mining, Cranmer2020}. These SBI techniques could therefore be used to enable robust probabilistic inversions in settings where each simulation is highly computationally expensive, such as short period 3D earth modelling \citep{sawade2022global}. SBI can also be utilised for model comparison, where estimates of the Bayesian evidence can be computed more efficiently than with classical sampling methods \citep{mancini2022bayesian}.

\section{Conclusion}

This study applied SBI to centroid moment  tensor inversions using seismic waveform data. Given that this is the first application of SBI in seismology, we presented a brief introduction of the framework, explored the motivations for our study due to the complex, non-Gaussian noise background in seismic observations, and presented a straightforward adaptation of SBI for full-waveform seismic source inversion. We demonstrated the strengths of SBI using a range of synthetic source inversion examples, and showed that SBI can faithfully reproduce the posterior with at least an order of magnitude fewer forward model evaluations, significantly accelerating the probabilistic inversion process relative to standard MCMC-based approaches.

The study then presented evidence that standard Gaussian likelihood assumptions about seismic noise significantly underestimate seismic source parameter uncertainties by up to a factor of $3$. On the other hand, SBI significantly improves upon the Gaussian likelihood approach by building an empirical model of the likelihood which is robust to the strong frequency dependence and non-stationarity of real seismological noise, allowing it to produce calibrated posteriors with more trustworthy uncertainty intervals. In addition, when more complex priors are considered for real event inversions, we showed that improved modelling of the likelihood provided by SBI can produce entirely different maximum \textit{a posteriori} estimates of the model parameters. 

This study has also uncovered some limitations of SBI as applied to full-waveform data, particularly resulting from the requirements of our compression scheme. Future work may need to draw upon the wide range of available improvements to extend SBI to more applications within seismology. Nonetheless, this manuscript has shown that SBI can be applied to full-waveform seismic source inversions to produce more accurate uncertainties than standard Gaussian likelihood approaches in a fraction of the time. 

\begin{dataavailability}

\texttt{Python} code used to produce the results of this study is available at \href{https://github.com/asaoulis/seismo-sbi}{https://github.com/asaoulis/seismo-sbi}.
This study utilised UPFLOW data, which are deposited in the GFZ EIDA node \citep{Ferreira2024} (network 8 J, data embargoed until May 2028).

\end{dataavailability}

\begin{acknowledgments}
We thank the UPFLOW project, funded by the European Research Council under the European Union's Horizon 2020 research and innovation program (grant agreement No 101001601). AAS was supported by the STFC UCL Centre for Doctoral Training in Data Intensive Science (grant ST/W00674X/1) and by departmental and industry contributions. AAS was also supported by the A. G. Leventis Foundation educational grant scheme. DP was supported by a Swiss National Science Foundation (SNSF) Professorship grant (No. 202671), and by the SNF Sinergia grant CRSII5-193826 ``AstroSignals: A New Window on the Universe, with the New Generation of Large Radio-Astronomy Facilities''.
\end{acknowledgments}

\newcommand{\newblock}{} 
\bibliographystyle{gji}
\bibliography{bibliography}

\begin{thebibliography}{143}
\expandafter\ifx\csname natexlab\endcsname\relax\def\natexlab#1{#1}\fi

\bibitem[Aki \& Richards(2002)]{aki2002quantitative}
Aki, K. \& Richards, P.~G., 2002.
\newblock {\it Quantitative seismology\/}.

\bibitem[Alsing \& Wandelt(2018)]{Alsing2018}
Alsing, J. \& Wandelt, B., 2018.
\newblock Generalized massive optimal data compression, {\it MNRAS\/}, {\bf 476}, 60--64.

\bibitem[Alsing \& Wandelt(2019)]{alsing2019nuisance}
Alsing, J. \& Wandelt, B., 2019.
\newblock Nuisance hardened data compression for fast likelihood-free inference, {\it Monthly Notices of the Royal Astronomical Society\/}, {\bf 488}(4), 5093--5103.

\bibitem[Alsing et~al.(2017)Alsing, Wandelt, \& Feeney]{Alsing2017}
Alsing, J., Wandelt, B., \& Feeney, S., 2017.
\newblock Massive optimal data compression and density estimation for scalable, likelihood-free inference in cosmology, {\it MNRAS\/}, {\bf 000}, 1--14.

\bibitem[Alsing et~al.(2019)Alsing, Charnock, Feeney, \& Wandelt]{Alsing2019}
Alsing, J., Charnock, T., Feeney, S., \& Wandelt, B., 2019.
\newblock Fast likelihood-free cosmology with neural density estimators and active learning.

\bibitem[Alvizuri \& Tape(2016)]{alvizuri2016full}
Alvizuri, C. \& Tape, C., 2016.
\newblock Full moment tensors for small events (m w< 3) at uturuncu volcano, bolivia, {\it Geophysical Journal International\/}, {\bf 206}(3), 1761--1783.

\bibitem[Ardhuin et~al.(2015)Ardhuin, Gualtieri, \& Stutzmann]{ardhuin2015ocean}
Ardhuin, F., Gualtieri, L., \& Stutzmann, E., 2015.
\newblock How ocean waves rock the earth: Two mechanisms explain microseisms with periods 3 to 300 s, {\it Geophysical Research Letters\/}, {\bf 42}(3), 765--772.

\bibitem[Backus \& Mulcahy(1976)]{backus1976moment}
Backus, G. \& Mulcahy, M., 1976.
\newblock Moment tensors and other phenomenological descriptions of seismic sources—i. continuous displacements, {\it Geophysical Journal International\/}, {\bf 46}(2), 341--361.

\bibitem[Bishop(1994)]{bishop1994mixture}
Bishop, C.~M., 1994.
\newblock Mixture density networks.

\bibitem[Blom et~al.(2023)Blom, Hardalupas, \& Rawlinson]{blom2023mitigating}
Blom, N., Hardalupas, P.-S., \& Rawlinson, N., 2023.
\newblock Mitigating the effect of errors in source parameters on seismic (waveform) tomography, {\it Geophysical Journal International\/}, {\bf 232}(2), 810--828.

\bibitem[Bodin \& Sambridge(2009)]{bodin2009seismic}
Bodin, T. \& Sambridge, M., 2009.
\newblock Seismic tomography with the reversible jump algorithm, {\it Geophysical Journal International\/}, {\bf 178}(3), 1411--1436.

\bibitem[Bodin et~al.(2012)Bodin, Sambridge, Tkal{\v{c}}i{\'c}, Arroucau, Gallagher, \& Rawlinson]{bodin2012transdimensionala}
Bodin, T., Sambridge, M., Tkal{\v{c}}i{\'c}, H., Arroucau, P., Gallagher, K., \& Rawlinson, N., 2012.
\newblock Transdimensional inversion of receiver functions and surface wave dispersion, {\it Journal of geophysical research: solid earth\/}, {\bf 117}(B2).

\bibitem[Bozda{\u{g}} et~al.(2016)Bozda{\u{g}}, Peter, Lefebvre, Komatitsch, Tromp, Hill, Podhorszki, \& Pugmire]{bozdaug2016global}
Bozda{\u{g}}, E., Peter, D., Lefebvre, M., Komatitsch, D., Tromp, J., Hill, J., Podhorszki, N., \& Pugmire, D., 2016.
\newblock Global adjoint tomography: first-generation model, {\it Geophysical Supplements to the Monthly Notices of the Royal Astronomical Society\/}, {\bf 207}(3), 1739--1766.

\bibitem[Brehmer et~al.(2020)Brehmer, Louppe, Pavez, \& Cranmer]{brehmer2020mining}
Brehmer, J., Louppe, G., Pavez, J., \& Cranmer, K., 2020.
\newblock Mining gold from implicit models to improve likelihood-free inference, {\it Proceedings of the National Academy of Sciences\/}, {\bf 117}(10), 5242--5249.

\bibitem[Cabieces et~al.(2024)Cabieces, Harris, Ferreira, Tsekhmistrenko, Hicks, Kr{\"u}ger, Geissler, Hannemann, \& Schmidt-Aursch]{cabieces2024clock}
Cabieces, R., Harris, K., Ferreira, A., Tsekhmistrenko, M., Hicks, S., Kr{\"u}ger, F., Geissler, W., Hannemann, K., \& Schmidt-Aursch, M., 2024.
\newblock Clock drift corrections for large aperture ocean bottom seismometer arrays: application to the upflow array in the mid-atlantic ocean, {\it Geophysical Journal International\/}, {\bf 239}(3), 1709--1728.

\bibitem[Charnock et~al.(2018)Charnock, Lavaux, \& Wandelt]{charnock2018automatic}
Charnock, T., Lavaux, G., \& Wandelt, B.~D., 2018.
\newblock Automatic physical inference with information maximizing neural networks, {\it Physical Review D\/}, {\bf 97}(8), 083004.

\bibitem[Corela et~al.(2022)Corela, Loureiro, Duarte, Matias, Rebelo, \& Bartolomeu]{corela2022obs}
Corela, C., Loureiro, A., Duarte, J.~L., Matias, L., Rebelo, T., \& Bartolomeu, T., 2022.
\newblock The obs noise due to deep ocean currents, {\it Natural Hazards and Earth System Sciences Discussions\/}, {\bf 2022}, 1--21.

\bibitem[Cranmer et~al.(2020)Cranmer, Brehmer, \& Louppe]{Cranmer2020}
Cranmer, K., Brehmer, J., \& Louppe, G., 2020.
\newblock The frontier of simulation-based inference, {\it Proceedings of the National Academy of Sciences of the United States of America\/}, {\bf 117}, 30055--30062.

\bibitem[Dahlen et~al.(2000)Dahlen, Hung, \& Nolet]{dahlen2000frechet}
Dahlen, F., Hung, S.-H., \& Nolet, G., 2000.
\newblock Fr{\'e}chet kernels for finite-frequency traveltimes—i. theory, {\it Geophysical Journal International\/}, {\bf 141}(1), 157--174.

\bibitem[Dax et~al.(2021)Dax, Green, Gair, Macke, Buonanno, \& Sch{\"o}lkopf]{dax2021real}
Dax, M., Green, S.~R., Gair, J., Macke, J.~H., Buonanno, A., \& Sch{\"o}lkopf, B., 2021.
\newblock Real-time gravitational wave science with neural posterior estimation, {\it Physical review letters\/}, {\bf 127}(24), 241103.

\bibitem[DeMets et~al.(1990)DeMets, Gordon, Argus, \& Stein]{demets1990current}
DeMets, C., Gordon, R.~G., Argus, D., \& Stein, S., 1990.
\newblock Current plate motions, {\it Geophysical journal international\/}, {\bf 101}(2), 425--478.

\bibitem[DeMets et~al.(2010)DeMets, Gordon, \& Argus]{demets2010geologically}
DeMets, C., Gordon, R.~G., \& Argus, D.~F., 2010.
\newblock Geologically current plate motions, {\it Geophysical journal international\/}, {\bf 181}(1), 1--80.

\bibitem[Denolle et~al.(2014)Denolle, Dunham, Prieto, \& Beroza]{denolle2014strong}
Denolle, M., Dunham, E., Prieto, G., \& Beroza, G., 2014.
\newblock Strong ground motion prediction using virtual earthquakes, {\it Science\/}, {\bf 343}(6169), 399--403.

\bibitem[Dinh et~al.(2017)Dinh, Sohl-Dickstein, \& Bengio]{dinh2017density}
Dinh, L., Sohl-Dickstein, J., \& Bengio, S., 2017.
\newblock Density estimation using real {NVP}, in {\em International Conference on Learning Representations\/}.

\bibitem[Dufumier \& Rivera(1997)]{Dufumier97}
Dufumier, H. \& Rivera, L., 1997.
\newblock {On the resolution of the isotropic component in moment tensor inversion}, {\it Geophysical Journal International\/}, {\bf 131}(3), 595--606.

\bibitem[Duputel et~al.(2012)Duputel, Rivera, Fukahata, \& Kanamori]{Duputel2012}
Duputel, Z., Rivera, L., Fukahata, Y., \& Kanamori, H., 2012.
\newblock Uncertainty estimations for seismic source inversions, {\it Geophysical Journal International Geophys. J. Int\/}, {\bf 190}, 1243--1256.

\bibitem[Duputel et~al.(2014)Duputel, Agram, Simons, Minson, \& Beck]{duputel2014accounting}
Duputel, Z., Agram, P.~S., Simons, M., Minson, S.~E., \& Beck, J.~L., 2014.
\newblock Accounting for prediction uncertainty when inferring subsurface fault slip, {\it Geophysical journal international\/}, {\bf 197}(1), 464--482.

\bibitem[Dziewonski \& Anderson(1981)]{dziewonski1981preliminary}
Dziewonski, A.~M. \& Anderson, D.~L., 1981.
\newblock Preliminary reference earth model, {\it Physics of the earth and planetary interiors\/}, {\bf 25}(4), 297--356.

\bibitem[Dziewonski et~al.(1981)Dziewonski, Chou, \& Woodhouse]{Dziewonski1981}
Dziewonski, A.~M., Chou, T.~A., \& Woodhouse, J.~H., 1981.
\newblock Determination of earthquake source parameters from waveform data for studies of global and regional seismicity, {\it J. Geophys. Res.; (United States)\/}, {\bf 86:B4}, 2825--2852.

\bibitem[Ekström et~al.(2012)Ekström, Nettles, \& Dziewoński]{Ekstrom_2012}
Ekström, G., Nettles, M., \& Dziewoński, A.~M., 2012.
\newblock The global cmt project 2004-2010: Centroid-moment tensors for 13,017 earthquakes, {\it Physics of the Earth and Planetary Interiors\/}, {\bf 200-201}, 1--9.

\bibitem[Engquist et~al.(2016)Engquist, Froese, \& Yang]{engquist2016optimal}
Engquist, B., Froese, B.~D., \& Yang, Y., 2016.
\newblock Optimal transport for seismic full waveform inversion, {\it arXiv preprint arXiv:1602.01540\/}.

\bibitem[Ferreira et~al.(2011)Ferreira, Weston, \& Funning]{ferreira2011global}
Ferreira, A., Weston, J., \& Funning, G., 2011.
\newblock Global compilation of interferometric synthetic aperture radar earthquake source models: 2. effects of 3-d earth structure, {\it Journal of Geophysical Research: Solid Earth\/}, {\bf 116}(B8).

\bibitem[Ferreira \& Woodhouse(2006)]{ferreira2006long}
Ferreira, A.~M. \& Woodhouse, J.~H., 2006.
\newblock Long-period seismic source inversions using global tomographic models, {\it Geophysical Journal International\/}, {\bf 166}(3), 1178--1192.

\bibitem[Ferreira et~al.(2020)Ferreira, Marignier, Attanayake, Frietsch, \& Berbellini]{Ferreira2020}
Ferreira, A.~M., Marignier, A., Attanayake, J., Frietsch, M., \& Berbellini, A., 2020.
\newblock Crustal structure of the azores archipelago from rayleigh wave ellipticity data, {\it Geophysical Journal International\/}, {\bf 221}, 1232--1247.

\bibitem[Ferreira(2024)]{Ferreira2024}
Ferreira, A. M.~G., 2024.
\newblock Upward mantle flow from novel seismic observations ({UPFLOW}), Other/Seismic Network [Dataset]. Available at: \url{https://geofon.gfz-potsdam.de/waveform/archive/network.php?ncode=8J&year=2021}.

\bibitem[Fichtner \& Simutė(2018)]{Fichtner2018}
Fichtner, A. \& Simutė, S., 2018.
\newblock Hamiltonian monte carlo inversion of seismic sources in complex media, {\it Journal of Geophysical Research: Solid Earth\/}, {\bf 123}, 2984--2999.

\bibitem[Fletcher(1971)]{fletcher1971modified}
Fletcher, R., 1971.
\newblock A modified marquardt subroutine for non-linear least squares.

\bibitem[Fluri et~al.(2021)Fluri, Kacprzak, Refregier, Lucchi, \& Hofmann]{fluri2021cosmological}
Fluri, J., Kacprzak, T., Refregier, A., Lucchi, A., \& Hofmann, T., 2021.
\newblock Cosmological parameter estimation and inference using deep summaries, {\it Physical Review D\/}, {\bf 104}(12), 123526.

\bibitem[Ford et~al.(2010)Ford, Dreger, \& Walter]{ford2010network}
Ford, S.~R., Dreger, D.~S., \& Walter, W.~R., 2010.
\newblock Network sensitivity solutions for regional moment-tensor inversions, {\it Bulletin of the Seismological Society of America\/}, {\bf 100}(5A), 1962--1970.

\bibitem[Foreman-Mackey et~al.(2013)Foreman-Mackey, Hogg, Lang, \& Goodman]{foreman2013emcee}
Foreman-Mackey, D., Hogg, D.~W., Lang, D., \& Goodman, J., 2013.
\newblock emcee: the mcmc hammer, {\it Publications of the Astronomical Society of the Pacific\/}, {\bf 125}(925), 306.

\bibitem[Frietsch et~al.(2018)Frietsch, Ferreira, Vales, \& Carrilho]{frietsch_robustness_2018}
Frietsch, M., Ferreira, A., Vales, D., \& Carrilho, F., 2018.
\newblock On the robustness of seismic moment tensor inversions for mid-ocean earthquakes: the {Azores} archipelago, {\it Geophysical Journal International\/}, {\bf 215}(1), 564--584, \_eprint: https://academic.oup.com/gji/article-pdf/215/1/564/25427935/ggy294.pdf.

\bibitem[Germain et~al.(2015)Germain, Gregor, Murray, \& Larochelle]{germain2015made}
Germain, M., Gregor, K., Murray, I., \& Larochelle, H., 2015.
\newblock Made: Masked autoencoder for distribution estimation, in {\em International conference on machine learning\/}, pp. 881--889, PMLR.

\bibitem[Giardini(1984)]{giardini1984systematic}
Giardini, D., 1984.
\newblock Systematic analysis of deep seismicity: 200 centroid-moment tensor solutions for earthquakes between 1977 and 1980, {\it Geophysical Journal International\/}, {\bf 77}(3), 883--914.

\bibitem[Gilbert \& Dziewonski(1975)]{gilbert1975application}
Gilbert, F. \& Dziewonski, A.~M., 1975.
\newblock An application of normal mode theory to the retrieval of structural parameters and source mechanisms from seismic spectra, {\it Philosophical Transactions of the Royal Society of London. Series A, Mathematical and Physical Sciences\/}, {\bf 278}(1280), 187--269.

\bibitem[Gilks et~al.(1995)Gilks, Richardson, \& Spiegelhalter]{gilks1995markov}
Gilks, W.~R., Richardson, S., \& Spiegelhalter, D., 1995.
\newblock {\it Markov chain Monte Carlo in practice\/}, CRC press.

\bibitem[Goodman \& Weare(2010)]{goodman2010ensemble}
Goodman, J. \& Weare, J., 2010.
\newblock Ensemble samplers with affine invariance, {\it Communications in applied mathematics and computational science\/}, {\bf 5}(1), 65--80.

\bibitem[Grimison \& Chen(1986)]{grimison1986azores}
Grimison, N.~L. \& Chen, W.-P., 1986.
\newblock The azores-gibraltar plate boundary: Focal mechanisms, depths of earthquakes, and their tectonic implications, {\it Journal of Geophysical Research: Solid Earth\/}, {\bf 91}(B2), 2029--2047.

\bibitem[Gutmann \& Corander(2016)]{gutmann2016bayesian}
Gutmann, M.~U. \& Corander, J., 2016.
\newblock Bayesian optimization for likelihood-free inference of simulator-based statistical models, {\it Journal of Machine Learning Research\/}.

\bibitem[Hable et~al.(2018)Hable, Sigloch, Barruol, St{\"a}hler, \& Hadziioannou]{hable2018clock}
Hable, S., Sigloch, K., Barruol, G., St{\"a}hler, S.~C., \& Hadziioannou, C., 2018.
\newblock Clock errors in land and ocean bottom seismograms: high-accuracy estimates from multiple-component noise cross-correlations, {\it Geophysical Journal International\/}, {\bf 214}(3), 2014--2034.

\bibitem[Hermans et~al.(2020)Hermans, Begy, \& Louppe]{Hermans2020}
Hermans, J., Begy, V., \& Louppe, G., 2020.
\newblock Likelihood-free mcmc with amortized approximate ratio estimators.

\bibitem[Hermans et~al.(2021)Hermans, Delaunoy, Rozet, Wehenkel, \& Louppe]{Hermans2021}
Hermans, J., Delaunoy, A., Rozet, F., Wehenkel, A., \& Louppe, G., 2021.
\newblock Averting a crisis in simulation-based inference, {\it stat\/}, {\bf 1050}, 14.

\bibitem[Hu et~al.(2023)Hu, Phạm, \& Tkal{\v{c}}i{\'c}]{hu2023seismic}
Hu, J., Phạm, T.-S., \& Tkal{\v{c}}i{\'c}, H., 2023.
\newblock Seismic moment tensor inversion with theory errors from 2-d earth structure: implications for the 2009--2017 dprk nuclear blasts, {\it Geophysical Journal International\/}, {\bf 235}(3), 2035--2054.

\bibitem[Hung et~al.(2000)Hung, Dahlen, \& Nolet]{hung2000frechet}
Hung, S.-H., Dahlen, F., \& Nolet, G., 2000.
\newblock Fr{\'e}chet kernels for finite-frequency traveltimes{\'e}ii. examples, {\it Geophysical Journal International\/}, {\bf 141}(1), 175--203.

\bibitem[IPMA(2024)]{ipma2024}
IPMA, 2024.
\newblock {Instituto Português do Mar e da Atmosfera (IPMA) Bulletin}.

\bibitem[Jackson \& McKenzie(1988)]{jackson1988relationship}
Jackson, J. \& McKenzie, D., 1988.
\newblock The relationship between plate motions and seismic moment tensors, and the rates of active deformation in the mediterranean and middle east, {\it Geophysical Journal International\/}, {\bf 93}(1), 45--73.

\bibitem[Jeffrey et~al.(2020)Jeffrey, Alsing, \& Lanusse]{Jeffrey2020}
Jeffrey, N., Alsing, J., \& Lanusse, F., 2020.
\newblock Likelihood-free inference with neural compression of des sv weak lensing map statistics, {\it MNRAS\/}, {\bf 000}, 1--18.

\bibitem[Kanamori \& Given(1981)]{Kanamori1981}
Kanamori, H. \& Given, J.~W., 1981.
\newblock Use of long-period surface waves for rapid determination of earthquake-source parameters, {\it Physics of the Earth and Planetary Interiors\/}, {\bf 27}, 8--31.

\bibitem[K{\"a}ufl et~al.(2013)K{\"a}ufl, Valentine, O'Toole, \& Trampert]{kaufl_framework_2013}
K{\"a}ufl, P., Valentine, A.~P., O'Toole, T.~B., \& Trampert, J., 2013.
\newblock A framework for fast probabilistic centroid-moment-tensor determination---inversion of regional static displacement measurements, {\it Geophysical Journal International\/}, {\bf 196}(3), 1676--1693, \_eprint: https://academic.oup.com/gji/article-pdf/196/3/1676/1566461/ggt473.pdf.

\bibitem[Kawakatsu(1996)]{kawakatsu1996observability}
Kawakatsu, H., 1996.
\newblock Observability of the isotropic component of a moment tensor, {\it Geophysical Journal International\/}, {\bf 126}(2), 525--544.

\bibitem[Kedar et~al.(2008)Kedar, Longuet-Higgins, Webb, Graham, Clayton, \& Jones]{kedar2008origin}
Kedar, S., Longuet-Higgins, M., Webb, F., Graham, N., Clayton, R., \& Jones, C., 2008.
\newblock The origin of deep ocean microseisms in the north atlantic ocean, {\it Proceedings of the Royal Society A: Mathematical, Physical and Engineering Sciences\/}, {\bf 464}(2091), 777--793.

\bibitem[Kingma \& Ba(2014)]{kingma2014adam}
Kingma, D.~P. \& Ba, J., 2014.
\newblock Adam: A method for stochastic optimization, {\it arXiv preprint arXiv:1412.6980\/}.

\bibitem[Komatitsch et~al.(2004)Komatitsch, Liu, Tromp, Suss, Stidham, \& Shaw]{komatitsch2004simulations}
Komatitsch, D., Liu, Q., Tromp, J., Suss, P., Stidham, C., \& Shaw, J.~H., 2004.
\newblock Simulations of ground motion in the los angeles basin based upon the spectral-element method, {\it Bulletin of the Seismological Society of America\/}, {\bf 94}(1), 187--206.

\bibitem[Lanzieri et~al.(2024)Lanzieri, Zeghal, Makinen, Boucaud, Starck, \& Lanusse]{lanzieri2024optimal}
Lanzieri, D., Zeghal, J., Makinen, T.~L., Boucaud, A., Starck, J.-L., \& Lanusse, F., 2024.
\newblock Optimal neural summarisation for full-field weak lensing cosmological implicit inference, {\it arXiv preprint arXiv:2407.10877\/}.

\bibitem[Lemos et~al.(2023)Lemos, Coogan, Hezaveh, \& Perreault-Levasseur]{lemos2023sampling}
Lemos, P., Coogan, A., Hezaveh, Y., \& Perreault-Levasseur, L., 2023.
\newblock Sampling-based accuracy testing of posterior estimators for general inference, {\it arXiv preprint arXiv:2302.03026\/}.

\bibitem[Lomax et~al.(2000)Lomax, Virieux, Volant, \& Berge-Thierry]{lomax2000probabilistic}
Lomax, A., Virieux, J., Volant, P., \& Berge-Thierry, C., 2000.
\newblock Probabilistic earthquake location in 3d and layered models: Introduction of a metropolis-gibbs method and comparison with linear locations, {\it Advances in seismic event location\/}, pp. 101--134.

\bibitem[Longuet-Higgins(1950)]{longuet1950theory}
Longuet-Higgins, M.~S., 1950.
\newblock A theory of the origin of microseisms, {\it Philosophical Transactions of the Royal Society of London. Series A, Mathematical and Physical Sciences\/}, {\bf 243}(857), 1--35.

\bibitem[L{\'o}pez-Comino et~al.(2015)L{\'o}pez-Comino, Stich, Ferreira, \& Morales]{lopez2015extended}
L{\'o}pez-Comino, J.~A., Stich, D., Ferreira, A.~M., \& Morales, J., 2015.
\newblock Extended fault inversion with random slipmaps: a resolution test for the 2012 m w 7.6 nicoya, costa rica earthquake, {\it Geophysical Journal International\/}, {\bf 202}(3), 1505--1521.

\bibitem[Lu et~al.(2022)Lu, Haiman, \& Zorrilla~Matilla]{lu2022simultaneously}
Lu, T., Haiman, Z., \& Zorrilla~Matilla, J.~M., 2022.
\newblock Simultaneously constraining cosmology and baryonic physics via deep learning from weak lensing, {\it Monthly Notices of the Royal Astronomical Society\/}, {\bf 511}(1), 1518--1528.

\bibitem[Lueckmann et~al.(2019)Lueckmann, Bassetto, Karaletsos, \& Macke]{lueckmann2019likelihood}
Lueckmann, J.-M., Bassetto, G., Karaletsos, T., \& Macke, J.~H., 2019.
\newblock Likelihood-free inference with emulator networks, in {\em Symposium on Advances in Approximate Bayesian Inference\/}, pp. 32--53, PMLR.

\bibitem[Lueckmann et~al.(2021)Lueckmann, Boelts, Greenberg, Goncalves, \& Macke]{lueckmann2021benchmarking}
Lueckmann, J.-M., Boelts, J., Greenberg, D., Goncalves, P., \& Macke, J., 2021.
\newblock Benchmarking simulation-based inference, in {\em International conference on artificial intelligence and statistics\/}, pp. 343--351, PMLR.

\bibitem[Masfara et~al.(2022)Masfara, Cullison, \& Weemstra]{masfara2022efficient}
Masfara, L. O.~M., Cullison, T., \& Weemstra, C., 2022.
\newblock An efficient probabilistic workflow for estimating induced earthquake parameters in 3d heterogeneous media, {\it Solid Earth\/}, {\bf 13}(8), 1309--1325.

\bibitem[Matos et~al.(2015)Matos, Silveira, Matias, Caldeira, Ribeiro, Dias, Kr{\"u}ger, \& dos Santos]{matos2015upper}
Matos, C., Silveira, G., Matias, L., Caldeira, R., Ribeiro, M.~L., Dias, N.~A., Kr{\"u}ger, F., \& dos Santos, T.~B., 2015.
\newblock Upper crustal structure of madeira island revealed from ambient noise tomography, {\it Journal of Volcanology and Geothermal Research\/}, {\bf 298}, 136--145.

\bibitem[McBrearty \& Beroza(2022)]{mcbrearty2022earthquake}
McBrearty, I.~W. \& Beroza, G.~C., 2022.
\newblock Earthquake location and magnitude estimation with graph neural networks, in {\em 2022 IEEE international conference on image processing (ICIP)\/}, pp. 3858--3862, IEEE.

\bibitem[McBrearty \& Beroza(2023)]{mcbrearty2023earthquake}
McBrearty, I.~W. \& Beroza, G.~C., 2023.
\newblock Earthquake phase association with graph neural networks, {\it Bulletin of the Seismological Society of America\/}, {\bf 113}(2), 524--547.

\bibitem[McNamara \& Buland(2004)]{mcnamara2004ambient}
McNamara, D.~E. \& Buland, R.~P., 2004.
\newblock Ambient noise levels in the continental united states, {\it Bulletin of the seismological society of America\/}, {\bf 94}(4), 1517--1527.

\bibitem[Molnar \& Lyon-Caent(1989)]{molnar1989fault}
Molnar, P. \& Lyon-Caent, H., 1989.
\newblock Fault plane solutions of earthquakes and active tectonics of the tibetan plateau and its margins, {\it Geophysical Journal International\/}, {\bf 99}(1), 123--153.

\bibitem[Mosegaard \& Tarantola(1995)]{mosegaard1995monte}
Mosegaard, K. \& Tarantola, A., 1995.
\newblock Monte carlo sampling of solutions to inverse problems, {\it Journal of Geophysical Research: Solid Earth\/}, {\bf 100}(B7), 12431--12447.

\bibitem[Musta{\'c} \& Tkal{\v{c}}i{\'c}(2016)]{Musta2016}
Musta{\'c}, M. \& Tkal{\v{c}}i{\'c}, H., 2016.
\newblock Point source moment tensor inversion through a bayesian hierarchical model, {\it Geophysical Journal International\/}, {\bf 204}(1), 311--323.

\bibitem[Musta{\'c} et~al.(2018)Musta{\'c}, Tkal{\v{c}}i{\'c}, \& Burky]{mustac2018variability}
Musta{\'c}, M., Tkal{\v{c}}i{\'c}, H., \& Burky, A.~L., 2018.
\newblock The variability and interpretation of earthquake source mechanisms in the geysers geothermal field from a bayesian standpoint based on the choice of a noise model, {\it Journal of Geophysical Research: Solid Earth\/}, {\bf 123}(1), 513--532.

\bibitem[Münchmeyer et~al.(2021)Münchmeyer, Bindi, Leser, \& Tilmann]{Münchmeyer21}
Münchmeyer, J., Bindi, D., Leser, U., \& Tilmann, F., 2021.
\newblock Earthquake magnitude and location estimation from real time seismic waveforms with a transformer network a preprint.

\bibitem[Nakata \& Fichtner(2019)]{nakata2019seismic}
Nakata, G. \& Fichtner, 2019.
\newblock {\it Visualization of the Seismic Ambient Noise Spectrum\/}, p. 1–29, Cambridge University Press.

\bibitem[Neal(1993)]{neal1993probabilistic}
Neal, R.~M., 1993.
\newblock Probabilistic inference using markov chain monte carlo methods.

\bibitem[Nissen-Meyer et~al.(2014)Nissen-Meyer, van Driel, St{\"a}hler, Hosseini, Hempel, Auer, Colombi, \& Fournier]{nissen2014axisem}
Nissen-Meyer, T., van Driel, M., St{\"a}hler, S.~C., Hosseini, K., Hempel, S., Auer, L., Colombi, A., \& Fournier, A., 2014.
\newblock Axisem: broadband 3-d seismic wavefields in axisymmetric media, {\it Solid Earth\/}, {\bf 5}(1), 425--445.

\bibitem[Nissen-Meyer et~al.(2016)Nissen-Meyer, Fournier, van Driel, St{\"a}hler, Hempel, Hosseini, Ampuero, Chaljub, Colombi, Dahlen, Komatitsch, Nolet, \& Tromp]{axisem}
Nissen-Meyer, T., Fournier, A., van Driel, M., St{\"a}hler, S., Hempel, S., Hosseini, K., Ampuero, J.-P., Chaljub, E., Colombi, A., Dahlen, F., Komatitsch, D., Nolet, G., \& Tromp, J., 2016.
\newblock Axisem v1.3 [software].

\bibitem[Nooshiri et~al.(2022)Nooshiri, Bean, Dahm, Grigoli, Kristj{\'a}nsd{\'o}ttir, Obermann, \& Wiemer]{nooshiri2022multibranch}
Nooshiri, N., Bean, C.~J., Dahm, T., Grigoli, F., Kristj{\'a}nsd{\'o}ttir, S., Obermann, A., \& Wiemer, S., 2022.
\newblock A multibranch, multitarget neural network for rapid point-source inversion in a microseismic environment: examples from the hengill geothermal field, iceland, {\it Geophysical Journal International\/}, {\bf 229}(2), 999--1016.

\bibitem[Oglesby \& Mai(2012)]{oglesby2012fault}
Oglesby, D.~D. \& Mai, P.~M., 2012.
\newblock Fault geometry, rupture dynamics and ground motion from potential earthquakes on the north anatolian fault under the sea of marmara, {\it Geophysical Journal International\/}, {\bf 188}(3), 1071--1087.

\bibitem[Olsen(2000)]{olsen2000site}
Olsen, K., 2000.
\newblock Site amplification in the los angeles basin from three-dimensional modeling of ground motion, {\it Bulletin of the Seismological Society of America\/}, {\bf 90}(6B), S77--S94.

\bibitem[Ousadou et~al.(2024)Ousadou, Ayadi, \& Bezzeghoud]{ousadou2024catalogue}
Ousadou, F., Ayadi, A., \& Bezzeghoud, M., 2024.
\newblock Catalogue of source mechanisms and overview of present-day stress fields in the western region of the africa--eurasia plate boundary, {\it Frontiers in Earth Science\/}, {\bf 12}, 1366156.

\bibitem[Papamakarios \& Murray(2016)]{papamakarios2016fast}
Papamakarios, G. \& Murray, I., 2016.
\newblock Fast $\varepsilon$-free inference of simulation models with bayesian conditional density estimation, {\it Advances in neural information processing systems\/}, {\bf 29}.

\bibitem[Papamakarios et~al.(2017)Papamakarios, Pavlakou, \& Murray]{Papamakarios_MAF_2017}
Papamakarios, G., Pavlakou, T., \& Murray, I., 2017.
\newblock Masked autoregressive flow for density estimation.

\bibitem[Papamakarios et~al.(2019{\natexlab{a}})Papamakarios, Sterratt, \& Murray]{papamakarios2019sequential}
Papamakarios, G., Sterratt, D., \& Murray, I., 2019{\natexlab{a}}.
\newblock Sequential neural likelihood: Fast likelihood-free inference with autoregressive flows, in {\em The 22nd international conference on artificial intelligence and statistics\/}, pp. 837--848, PMLR.

\bibitem[Papamakarios et~al.(2019{\natexlab{b}})Papamakarios, Sterratt, \& Murray]{Papamakarios2019}
Papamakarios, G., Sterratt, D.~C., \& Murray, I., 2019{\natexlab{b}}.
\newblock Sequential neural likelihood: Fast likelihood-free inference with autoregressive flows.

\bibitem[Papamakarios et~al.(2021)Papamakarios, Nalisnick, Rezende, Mohamed, \& Lakshminarayanan]{Papamakarios_review21}
Papamakarios, G., Nalisnick, E., Rezende, D.~J., Mohamed, S., \& Lakshminarayanan, B., 2021.
\newblock Normalizing flows for probabilistic modeling and inference, {\it J. Mach. Learn. Res.\/}, {\bf 22}(1).

\bibitem[Pasyanos et~al.(2014)Pasyanos, Masters, Laske, \& Ma]{pasyanos2014litho1}
Pasyanos, M.~E., Masters, T.~G., Laske, G., \& Ma, Z., 2014.
\newblock Litho1. 0: An updated crust and lithospheric model of the earth, {\it Journal of Geophysical Research: Solid Earth\/}, {\bf 119}(3), 2153--2173.

\bibitem[Peterson(1993)]{peterson1993observations}
Peterson, J.~R., 1993.
\newblock Observations and modeling of seismic background noise, Tech. rep., US Geological Survey.

\bibitem[Phạm(2024)]{pham2024gradient}
Phạm, T.-S., 2024.
\newblock Gradient-based joint inversion of point-source moment tensor and station-specific time-shifts, {\it Geophysical Journal International\/}, {\bf 238}(2), 783--793.

\bibitem[Phạm et~al.(2024)Phạm, Tkal{\v{c}}i{\'c}, Hu, \& Kim]{pham2024towards}
Phạm, T.-S., Tkal{\v{c}}i{\'c}, H., Hu, J., \& Kim, S., 2024.
\newblock Towards a new standard for seismic moment tensor inversion containing 3-d earth structure uncertainty, {\it Geophysical Journal International\/}, {\bf 238}(3), 1840--1853.

\bibitem[Piras et~al.(2023)Piras, {Spurio Mancini}, Ferreira, Joachimi, \& Hobson]{Piras2023}
Piras, D., {Spurio Mancini}, A., Ferreira, A. M.~G., Joachimi, B., \& Hobson, M.~P., 2023.
\newblock Towards fast machine-learning-assisted bayesian posterior inference of microseismic event location and source mechanism, {\it Geophys. J. Int\/}, {\bf 232}, 1219--1235.

\bibitem[Prelogovi{\'c} \& Mesinger(2024)]{prelogovic2024informative}
Prelogovi{\'c}, D. \& Mesinger, A., 2024.
\newblock How informative are summaries of the cosmic 21 cm signal?, {\it Astronomy \& Astrophysics\/}, {\bf 688}, A199.

\bibitem[Prieto et~al.(2012)Prieto, Beroza, Barrett, L{\'o}pez, \& Florez]{prieto2012earthquake}
Prieto, G.~A., Beroza, G.~C., Barrett, S.~A., L{\'o}pez, G.~A., \& Florez, M., 2012.
\newblock Earthquake nests as natural laboratories for the study of intermediate-depth earthquake mechanics, {\it Tectonophysics\/}, {\bf 570}, 42--56.

\bibitem[Pugh et~al.(2016)Pugh, White, \& Christie]{pugh2016bayesian}
Pugh, D., White, R., \& Christie, P., 2016.
\newblock A bayesian method for microseismic source inversion, {\it Geophysical Journal International\/}, {\bf 206}(2), 1009--1038.

\bibitem[Rezende \& Mohamed(2015)]{Rezende_15}
Rezende, D.~J. \& Mohamed, S., 2015.
\newblock Variational inference with normalizing flows.

\bibitem[Ritsema et~al.(2011)Ritsema, Deuss, Van~Heijst, \& Woodhouse]{ritsema2011s40rts}
Ritsema, J., Deuss, A., Van~Heijst, H., \& Woodhouse, J., 2011.
\newblock S40rts: a degree-40 shear-velocity model for the mantle from new rayleigh wave dispersion, teleseismic traveltime and normal-mode splitting function measurements, {\it Geophysical Journal International\/}, {\bf 184}(3), 1223--1236.

\bibitem[Sambridge et~al.(2006)Sambridge, Gallagher, Jackson, \& Rickwood]{sambridge2006trans}
Sambridge, M., Gallagher, K., Jackson, A., \& Rickwood, P., 2006.
\newblock Trans-dimensional inverse problems, model comparison and the evidence, {\it Geophysical Journal International\/}, {\bf 167}(2), 528--542.

\bibitem[Sambridge et~al.(2013)Sambridge, Bodin, Gallagher, \& Tkal{\v{c}}i{\'c}]{sambridge2013transdimensional}
Sambridge, M., Bodin, T., Gallagher, K., \& Tkal{\v{c}}i{\'c}, H., 2013.
\newblock Transdimensional inference in the geosciences, {\it Philosophical Transactions of the Royal Society A: Mathematical, Physical and Engineering Sciences\/}, {\bf 371}(1984), 20110547.

\bibitem[Sambridge et~al.(2022)Sambridge, Jackson, \& Valentine]{sambridge2022geophysical}
Sambridge, M., Jackson, A., \& Valentine, A.~P., 2022.
\newblock Geophysical inversion and optimal transport, {\it Geophysical Journal International\/}, {\bf 231}(1), 172--198.

\bibitem[Sawade et~al.(2022)Sawade, Beller, Lei, \& Tromp]{sawade2022global}
Sawade, L., Beller, S., Lei, W., \& Tromp, J., 2022.
\newblock Global centroid moment tensor solutions in a heterogeneous earth: the cmt3d catalogue, {\it Geophysical Journal International\/}, {\bf 231}(3), 1727--1738.

\bibitem[Sharma et~al.(2024)Sharma, Dai, \& Seljak]{sharma2024comparative}
Sharma, D., Dai, B., \& Seljak, U., 2024.
\newblock A comparative study of cosmological constraints from weak lensing using convolutional neural networks, {\it Journal of Cosmology and Astroparticle Physics\/}, {\bf 2024}(08), 010.

\bibitem[Simutė et~al.(2023)Simutė, Boehm, Krischer, Gokhberg, Vallée, \& Fichtner]{HMCSimute23}
Simutė, S., Boehm, C., Krischer, L., Gokhberg, A., Vallée, M., \& Fichtner, A., 2023.
\newblock Bayesian seismic source inversion with a 3-d earth model of the japanese islands, {\it Journal of Geophysical Research: Solid Earth\/}, {\bf 128}, e2022JB024231.

\bibitem[Smith et~al.(2022)Smith, Ross, Azizzadenesheli, \& Muir]{smith2022hyposvi}
Smith, J.~D., Ross, Z.~E., Azizzadenesheli, K., \& Muir, J.~B., 2022.
\newblock Hyposvi: Hypocentre inversion with stein variational inference and physics informed neural networks, {\it Geophysical Journal International\/}, {\bf 228}(1), 698--710.

\bibitem[{Spurio Mancini} et~al.(2021){Spurio Mancini}, Piras, Ferreira, Hobson, \& Joachimi]{Mancini2021}
{Spurio Mancini}, A., Piras, D., Ferreira, A. M.~G., Hobson, M.~P., \& Joachimi, B., 2021.
\newblock Accelerating bayesian microseismic event location with deep learning, {\it Solid Earth\/}, {\bf 12}, 1683--1705.

\bibitem[{Spurio Mancini} et~al.(2022){Spurio Mancini}, Docherty, Price, \& McEwen]{mancini2022bayesian}
{Spurio Mancini}, A., Docherty, M., Price, M., \& McEwen, J., 2022.
\newblock Bayesian model comparison for simulation-based inference.

\bibitem[Spurio~Mancini et~al.(2022)Spurio~Mancini, Piras, Alsing, Joachimi, \& Hobson]{spurio2022cosmopower}
Spurio~Mancini, A., Piras, D., Alsing, J., Joachimi, B., \& Hobson, M.~P., 2022.
\newblock Cosmopower: emulating cosmological power spectra for accelerated bayesian inference from next-generation surveys, {\it Monthly Notices of the Royal Astronomical Society\/}, {\bf 511}(2), 1771--1788.

\bibitem[St{\"a}hler et~al.(2016)St{\"a}hler, Sigloch, Hosseini, Crawford, Barruol, Schmidt-Aursch, Tsekhmistrenko, Scholz, Mazzullo, \& Deen]{stahler2016performance}
St{\"a}hler, S.~C., Sigloch, K., Hosseini, K., Crawford, W.~C., Barruol, G., Schmidt-Aursch, M.~C., Tsekhmistrenko, M., Scholz, J.-R., Mazzullo, A., \& Deen, M., 2016.
\newblock Performance report of the rhum-rum ocean bottom seismometer network around la r{\'e}union, western indian ocean, {\it Advances in Geosciences\/}, {\bf 41}, 43--63.

\bibitem[Stähler \& Sigloch(2014)]{Stahler2016_a}
Stähler, S.~C. \& Sigloch, K., 2014.
\newblock Fully probabilistic seismic source inversion-part 1: Efficient parameterisation, {\it Solid Earth\/}, {\bf 5}, 1055--1069.

\bibitem[Stähler \& Sigloch(2016)]{Stahler2016_b}
Stähler, S.~C. \& Sigloch, K., 2016.
\newblock Fully probabilistic seismic source inversion-part 2: Modelling errors and station covariances, {\it Solid Earth\/}, {\bf 7}, 1521--1536.

\bibitem[Tape \& Tape(2012)]{tape2012geometric}
Tape, W. \& Tape, C., 2012.
\newblock A geometric setting for moment tensors, {\it Geophysical Journal International\/}, {\bf 190}(1), 476--498.

\bibitem[Tape \& Tape(2013)]{tape2013classical}
Tape, W. \& Tape, C., 2013.
\newblock The classical model for moment tensors, {\it Geophysical Journal International\/}, {\bf 195}(3), 1701--1720.

\bibitem[Tape \& Tape(2015)]{Tape2015}
Tape, W. \& Tape, C., 2015.
\newblock A uniform parametrization of moment tensors, {\it Geophysical Journal International\/}, {\bf 202}, 2074--2081.

\bibitem[Tarantola(2005)]{tarantola2005inverse}
Tarantola, A., 2005.
\newblock {\it Inverse problem theory and methods for model parameter estimation\/}, SIAM.

\bibitem[Tejero-Cantero et~al.(2020)Tejero-Cantero, Boelts, Deistler, Lueckmann, Durkan, Gonçalves, Greenberg, \& Macke]{tejero-cantero2020sbi}
Tejero-Cantero, A., Boelts, J., Deistler, M., Lueckmann, J.-M., Durkan, C., Gonçalves, P.~J., Greenberg, D.~S., \& Macke, J.~H., 2020.
\newblock sbi: A toolkit for simulation-based inference, {\it Journal of Open Source Software\/}, {\bf 5}(52), 2505.

\bibitem[Tilmann et~al.(2020)Tilmann, Sadeghisorkhani, \& Mauerberger]{tilmann2020another}
Tilmann, F., Sadeghisorkhani, H., \& Mauerberger, A., 2020.
\newblock Another look at the treatment of data uncertainty in markov chain monte carlo inversion and other probabilistic methods, {\it Geophysical Journal International\/}, {\bf 222}(1), 388--405.

\bibitem[Tsekhmistrenko et~al.(2024)Tsekhmistrenko, Ferreira, Miranda, Baranbooei, Cabieces, Carapuço, Corela, Duarte, Ferreira, Geissler, Harris, Hicks, Hosseini, Ke, Krüger, Lange, Loureiro, Makus, Marignier, Neres, Ramos, Rein, Saoulis, Schlaphorst, Schmidt-Aursch, \& Tilmann]{tsekhmistrenko2024performance}
Tsekhmistrenko, M., Ferreira, A. M.~G., Miranda, M., Baranbooei, S., Cabieces, R., Carapuço, M., Corela, C., Duarte, J.~L., Ferreira, H., Geissler, W.~H., Harris, K., Hicks, S.~P., Hosseini, K., Ke, K.-Y., Krüger, F., Lange, D., Loureiro, A., Makus, P., Marignier, A., Neres, M., Ramos, L., Rein, T., Saoulis, A., Schlaphorst, D., Schmidt-Aursch, M., \& Tilmann, F., 2024.
\newblock Performance of the {UPFLOW} large ocean bottom seismometer array in the azores-madeira-canary islands region, mid-atlantic ocean, {\it Seismica (in prep)\/}.

\bibitem[Uria et~al.(2016)Uria, C{\^o}t{\'e}, Gregor, Murray, \& Larochelle]{uria2016neural}
Uria, B., C{\^o}t{\'e}, M.-A., Gregor, K., Murray, I., \& Larochelle, H., 2016.
\newblock Neural autoregressive distribution estimation, {\it The Journal of Machine Learning Research\/}, {\bf 17}(1), 7184--7220.

\bibitem[USGS(2024)]{usgs-2024}
USGS, 2024.
\newblock {USGS Earthquake Catalogue}.

\bibitem[Vack{\'a}{\v{r}} et~al.(2017)Vack{\'a}{\v{r}}, Burj{\'a}nek, Gallovi{\v{c}}, Zahradn{\'\i}k, \& Clinton]{vackavr2017bayesian}
Vack{\'a}{\v{r}}, J., Burj{\'a}nek, J., Gallovi{\v{c}}, F., Zahradn{\'\i}k, J., \& Clinton, J., 2017.
\newblock Bayesian {ISOLA}: new tool for automated centroid moment tensor inversion, {\it Geophysical Journal International\/}, {\bf 210}(2), 693--705.

\bibitem[Valentine \& Sambridge(2023)]{valentine2023emerging}
Valentine, A.~P. \& Sambridge, M., 2023.
\newblock Emerging directions in geophysical inversion, {\it Applications of Data Assimilation and Inverse Problems in the Earth Sciences\/}, {\bf 5}(9).

\bibitem[Valentine \& Trampert(2012)]{valentine2012assessing}
Valentine, A.~P. \& Trampert, J., 2012.
\newblock Assessing the uncertainties on seismic source parameters: Towards realistic error estimates for centroid-moment-tensor determinations, {\it Physics of the Earth and Planetary Interiors\/}, {\bf 210}, 36--49.

\bibitem[Vall{\'e}e et~al.(2011)Vall{\'e}e, Charl{\'e}ty, Ferreira, Delouis, \& Vergoz]{vallee2011scardec}
Vall{\'e}e, M., Charl{\'e}ty, J., Ferreira, A.~M., Delouis, B., \& Vergoz, J., 2011.
\newblock Scardec: a new technique for the rapid determination of seismic moment magnitude, focal mechanism and source time functions for large earthquakes using body-wave deconvolution, {\it Geophysical Journal International\/}, {\bf 184}(1), 338--358.

\bibitem[van Driel et~al.(2015)van Driel, Krischer, St\"ahler, Hosseini, \& Nissen-Meyer]{instaseis_paper}
van Driel, M., Krischer, L., St\"ahler, S.~C., Hosseini, K., \& Nissen-Meyer, T., 2015.
\newblock Instaseis: instant global seismograms based on a broadband waveform database, {\it Solid Earth\/}, {\bf 6}(2), 701--717.

\bibitem[Vasist et~al.(2023)Vasist, Rozet, Absil, Molli{\`e}re, Nasedkin, \& Louppe]{vasist2023neural}
Vasist, M., Rozet, F., Absil, O., Molli{\`e}re, P., Nasedkin, E., \& Louppe, G., 2023.
\newblock Neural posterior estimation for exoplanetary atmospheric retrieval, {\it Astronomy \& Astrophysics\/}, {\bf 672}, A147.

\bibitem[Vasyura-Bathke et~al.(2020)Vasyura-Bathke, Dettmer, Steinberg, Heimann, Isken, Zielke, Mai, Sudhaus, \& J{\'o}nsson]{vasyura2020bayesian}
Vasyura-Bathke, H., Dettmer, J., Steinberg, A., Heimann, S., Isken, M.~P., Zielke, O., Mai, P.~M., Sudhaus, H., \& J{\'o}nsson, S., 2020.
\newblock The bayesian earthquake analysis tool, {\it Seismological Research Letters\/}, {\bf 91}(2A), 1003--1018.

\bibitem[Vasyura-Bathke et~al.(2021)Vasyura-Bathke, Dettmer, Dutta, Mai, \& Jonsson]{vasyura2021accounting}
Vasyura-Bathke, H., Dettmer, J., Dutta, R., Mai, P.~M., \& Jonsson, S., 2021.
\newblock Accounting for theory errors with empirical bayesian noise models in nonlinear centroid moment tensor estimation, {\it Geophysical Journal International\/}, {\bf 225}(2), 1412--1431.

\bibitem[Vavry{\v{c}}uk et~al.(2017)Vavry{\v{c}}uk, Adamov{\'a}, Doubravov{\'a}, \& Jakoubkov{\'a}]{vavryvcuk2017moment}
Vavry{\v{c}}uk, V., Adamov{\'a}, P., Doubravov{\'a}, J., \& Jakoubkov{\'a}, H., 2017.
\newblock Moment tensor inversion based on the principal component analysis of waveforms: Method and application to microearthquakes in west bohemia, czech republic, {\it Seismological Research Letters\/}, {\bf 88}(5), 1303--1315.

\bibitem[W{\'e}ber(2006)]{weber2006probabilistic}
W{\'e}ber, Z., 2006.
\newblock Probabilistic local waveform inversion for moment tensor and hypocentral location, {\it Geophysical Journal International\/}, {\bf 165}(2), 607--621.

\bibitem[Weston et~al.(2011)Weston, Ferreira, \& Funning]{weston2011global}
Weston, J., Ferreira, A., \& Funning, G., 2011.
\newblock Global compilation of interferometric synthetic aperture radar earthquake source models: 1. comparisons with seismic catalogs, {\it Journal of Geophysical Research: Solid Earth\/}, {\bf 116}(B8).

\bibitem[Weston et~al.(2012)Weston, Ferreira, \& Funning]{weston2012systematic}
Weston, J., Ferreira, A.~M., \& Funning, G.~J., 2012.
\newblock Systematic comparisons of earthquake source models determined using insar and seismic data, {\it Tectonophysics\/}, {\bf 532}, 61--81.

\bibitem[Wiens(2001)]{wiens2001seismological}
Wiens, D.~A., 2001.
\newblock Seismological constraints on the mechanism of deep earthquakes: Temperature dependence of deep earthquake source properties, {\it Physics of the Earth and Planetary Interiors\/}, {\bf 127}(1-4), 145--163.

\bibitem[Wu \& Xiao(2012)]{wu2012covariance}
Wu, W.~B. \& Xiao, H., 2012.
\newblock Covariance matrix estimation in time series, in {\em Handbook of Statistics\/}, vol.~30, pp. 187--209, Elsevier.

\bibitem[Yagi \& Fukahata(2008)]{yagi2008importance}
Yagi, Y. \& Fukahata, Y., 2008.
\newblock Importance of covariance components in inversion analyses of densely sampled observed data: an application to waveform data inversion for seismic source processes, {\it Geophysical Journal International\/}, {\bf 175}(1), 215--221.

\bibitem[Yu et~al.(2018)Yu, Vavry{\v{c}}uk, Adamov{\'a}, \& Bohnhoff]{yu2018moment}
Yu, C., Vavry{\v{c}}uk, V., Adamov{\'a}, P., \& Bohnhoff, M., 2018.
\newblock Moment tensors of induced microearthquakes in the geysers geothermal reservoir from broadband seismic recordings: implications for faulting regime, stress tensor, and fluid pressure, {\it Journal of Geophysical Research: Solid Earth\/}, {\bf 123}(10), 8748--8766.

\bibitem[Zha et~al.(2013)Zha, Webb, \& Menke]{zha2013determining}
Zha, Y., Webb, S.~C., \& Menke, W., 2013.
\newblock Determining the orientations of ocean bottom seismometers using ambient noise correlation, {\it Geophysical Research Letters\/}, {\bf 40}(14), 3585--3590.

\bibitem[Zhang et~al.(2022)Zhang, Liu, Feng, Wang, \& Zhu]{zhang2022loc}
Zhang, M., Liu, M., Feng, T., Wang, R., \& Zhu, W., 2022.
\newblock Loc-flow: An end-to-end machine learning-based high-precision earthquake location workflow, {\it Seismological Society of America\/}, {\bf 93}(5), 2426--2438.

\end{thebibliography}

\section*{Supporting Information}

The Supporting Information for this study contains additional figures and explanations that provide further details on the methodology, results, and discussions presented in the main manuscript. Below is a brief description of each figure in the Supporting Information:

\refstepcounter{supfig}
\noindent
\textbf{Figure S1:} Visualisation of the Tests of Accuracy with Random Points (TARP) method for producing empirical coverage estimates.\label{SI:1}

\refstepcounter{supfig}
\noindent
\textbf{Figure S2:} Diagram demonstrating how overconfident posteriors are detected by TARP, and how the resulting distribution of observed credibility levels can be interpreted in the empirical coverage plot. \label{SI:2}

\refstepcounter{supfig}
\noindent
\textbf{Figure S3:} Same as in Fig. S\ref{SI:2}, dealing with biased posteriors. \label{SI:3}

\refstepcounter{supfig}
\noindent
\textbf{Figure S4:} Empirical coverage tests of a range of SBI experiments, contrasting results when using only land or OBS stations, as well as with different covariance parametrisations. \label{SI:4}

\refstepcounter{supfig}
\noindent
\textbf{Figure S5:} Same as in Fig. S\ref{SI:5}, but using Gaussian likelihood-based inversions for the different configurations. \label{SI:5}

\refstepcounter{supfig}
\noindent
\textbf{Figure S6:} Empirical coverage tests that suggest Gaussian likelihood-based inversions underestimate the uncertainty intervals by up to a factor of $3$.\label{SI:6}

\refstepcounter{supfig}
\noindent
\textbf{Figure S7:} Diagram of the prior-constraining procedure, using iterative least squares to find the best fitting model parameters and constraining the prior about this point. \label{SI:7}

\refstepcounter{supfig}
\noindent
\textbf{Figure S8:} The effects of the prior-constraining procedure on the empirical coverage performance of SBI. \label{SI:8}

\refstepcounter{supfig}
\noindent
\textbf{Figure S9:} The degradation of compression quality as a function of the constrained prior size for the Azores event studied in the manuscript. \label{SI:9}

\refstepcounter{supfig}
\noindent
\textbf{Figure S10:} Same as in Fig. S\ref{SI:9} for the Madeira event. \label{SI:10}

\refstepcounter{supfig}
\noindent
The full Supporting Information file is available online, linked with this article.

\label{lastpage}

\end{document}


\maketitle

\section{Empirical Coverage Testing}
\label{sec:empirical_coverage_testing}
Empirical coverage testing of an inference method can be performed by running many artificial inversions and checking how well the posterior samples match the artificial (and therefore known) input model parameters. A method that produces well-calibrated posteriors should have a uniform distribution of observed credibility regions (i.e. at what quantile of the posterior do the true model parameters fall). Concretely, this means that the same proportion of events should fall within e.g. every $5\%$ interval, i.e. from $[0,5]\%$, $[5,10]\%$, ..., $[95,100]\%$.  By repeatedly performing inference, an empirical distribution of credibility levels can be computed and compared to the ideal distribution. This allows practitioners to identify issues with the inference method \citep{hermans2021,lueckmann2021benchmarking}. 

\begin{figure*}
\includegraphics[width=0.8\textwidth]{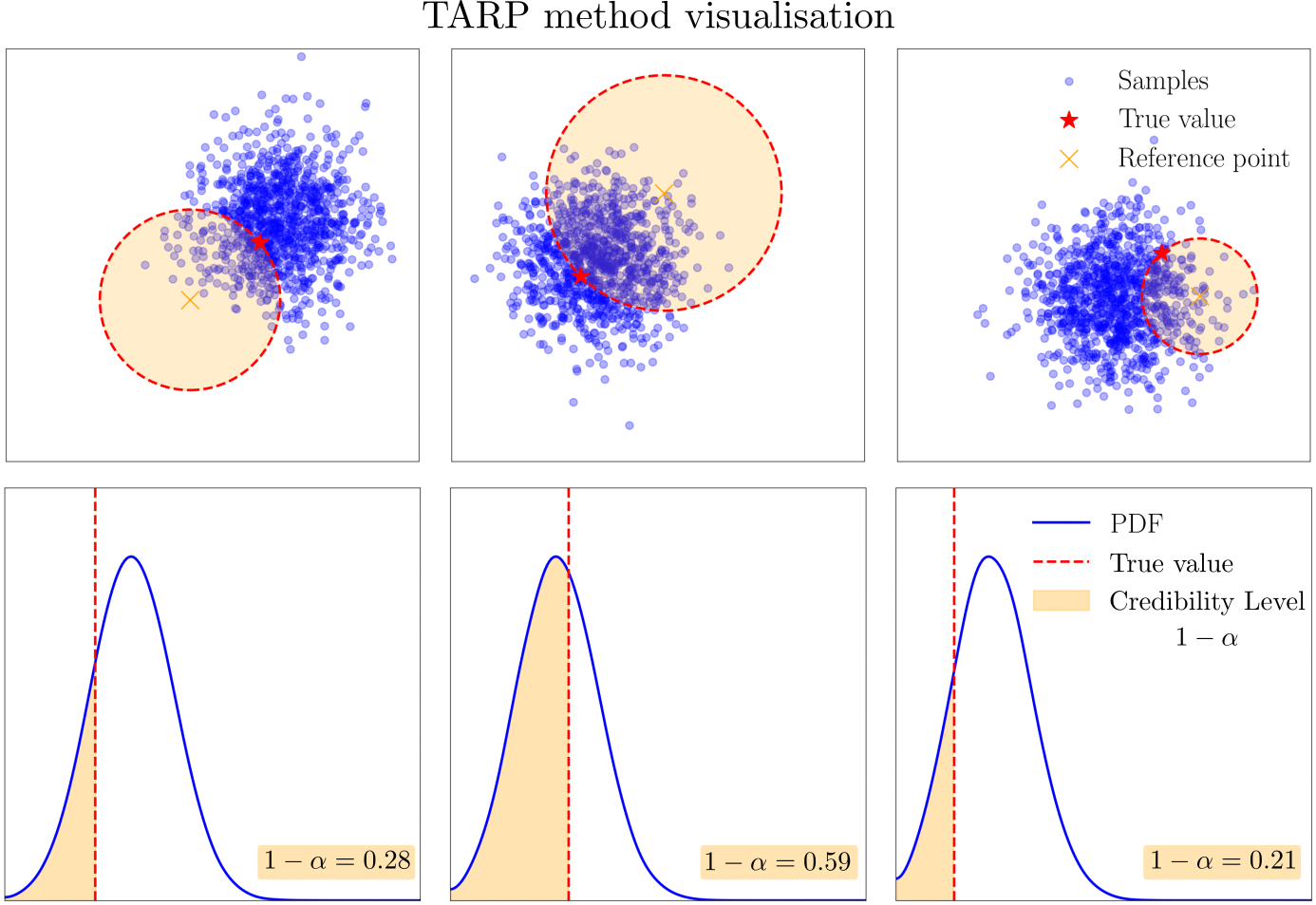}
\caption{Top row: three examples of artificial inversions, showing the posterior samples (blue scatter) and known input model parameters (red stars). TARP generates random reference points (yellow crosses) and computes the fraction of samples that fall between the reference point and the true solution. Bottom row: the probability density function (PDF) of samples as a function of distance from the reference point. The cumulative fraction of samples covered by each ball estimates the credibility interval $1 - \alpha$ for each event. This process can then be repeated to check the posterior calibration.}
\label{fig:tarp_cartoon}
\end{figure*}

Tests of Accuracy with Random Points (TARP) can be used to estimate expected coverage curves in an efficient manner \citep{lemos2023sampling}. We present a simplified 2D visualisation to demonstrate how TARP works in Fig. \ref{fig:tarp_cartoon}.

\begin{figure*}
\includegraphics[width=0.8\textwidth]{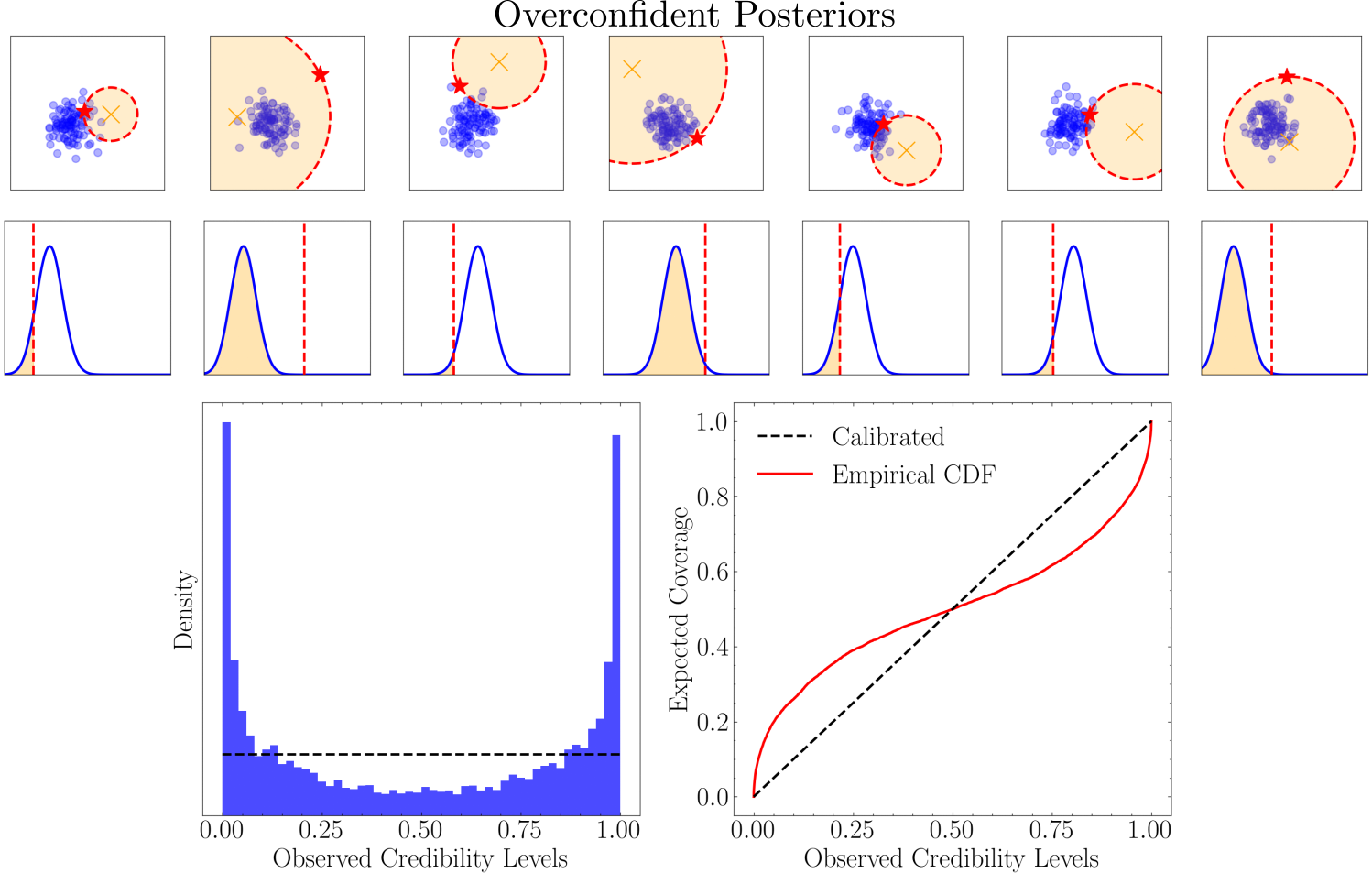}
\caption{Top row: same as in Fig. \ref{fig:tarp_cartoon}, showing artificial inversions for an inference method that produces overconfident posteriors. Posterior overconfidence (very low sample spread) exaggerated for visual effect. Bottom left: the distribution of observed credibility levels from repeated inversions with a method that produces overconfidence. Under the TARP method, this leads to clustering of credibility levels near $0$ and $1$. Bottom right: the distribution of credibility levels can be converted into an expected coverage plot by computing the cumulative distribution function (CDF).}
\label{fig:tarp_cartoon_over}
\end{figure*}

Overconfident posteriors that underestimate the uncertainties often have the true value well outside the bulk of the posterior (see e.g. Fig. 7 in the main text). This means that in a coverage test, the observed distribution of credibility levels is skewed from the ideal case. Specifically, for an overconfident posterior, we expect a greater proportion of true model parameters to fall well outside the central quantiles. This is demonstrated in Fig. \ref{fig:tarp_cartoon_over}.

\begin{figure*}
\includegraphics[width=0.8\textwidth]{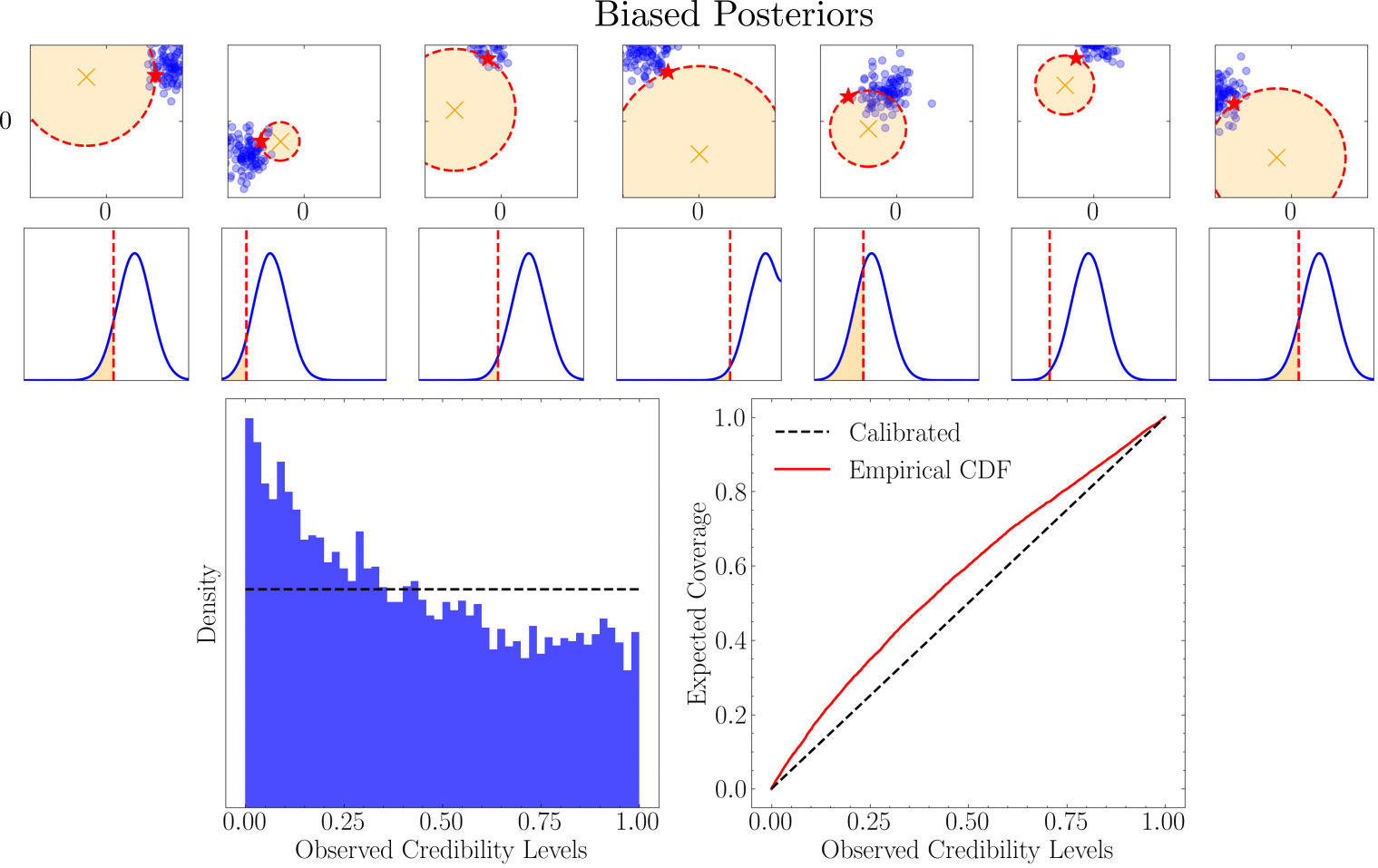}
\caption{Top row: same as in Fig. \ref{fig:tarp_cartoon}, showing repeated artificial inversions for an inference method that produces biased posteriors. Bias exaggerated for visual effect. In this case, the posteriors have a bias that over-estimate the magnitude of each model parameter. Bottom left: this bias lead to a left-hand skew of the credibility level distribution towards $0$. Bottom right: the distribution of credibility levels can be converted into an expected coverage plot by computing the cumulative distribution function (CDF). }
\label{fig:tarp_cartoon_bias}
\end{figure*}

Inference methods can also produce biased posteriors, which may produce biased maximum \textit{a posteriori} estimates of the model parameters, or have parameter-dependent biases. One example of this in the context of moment tensor inversion could be a bias that over-estimates the magnitude of the scalar moment $M_0$. A 2D visualisation of this behaviour, and the resulting empirical credibility level distribution, is demonstrated in Fig. \ref{fig:tarp_cartoon_bias}.


\section{Calibration Tests}
\begin{figure*}
\includegraphics[width=\textwidth]{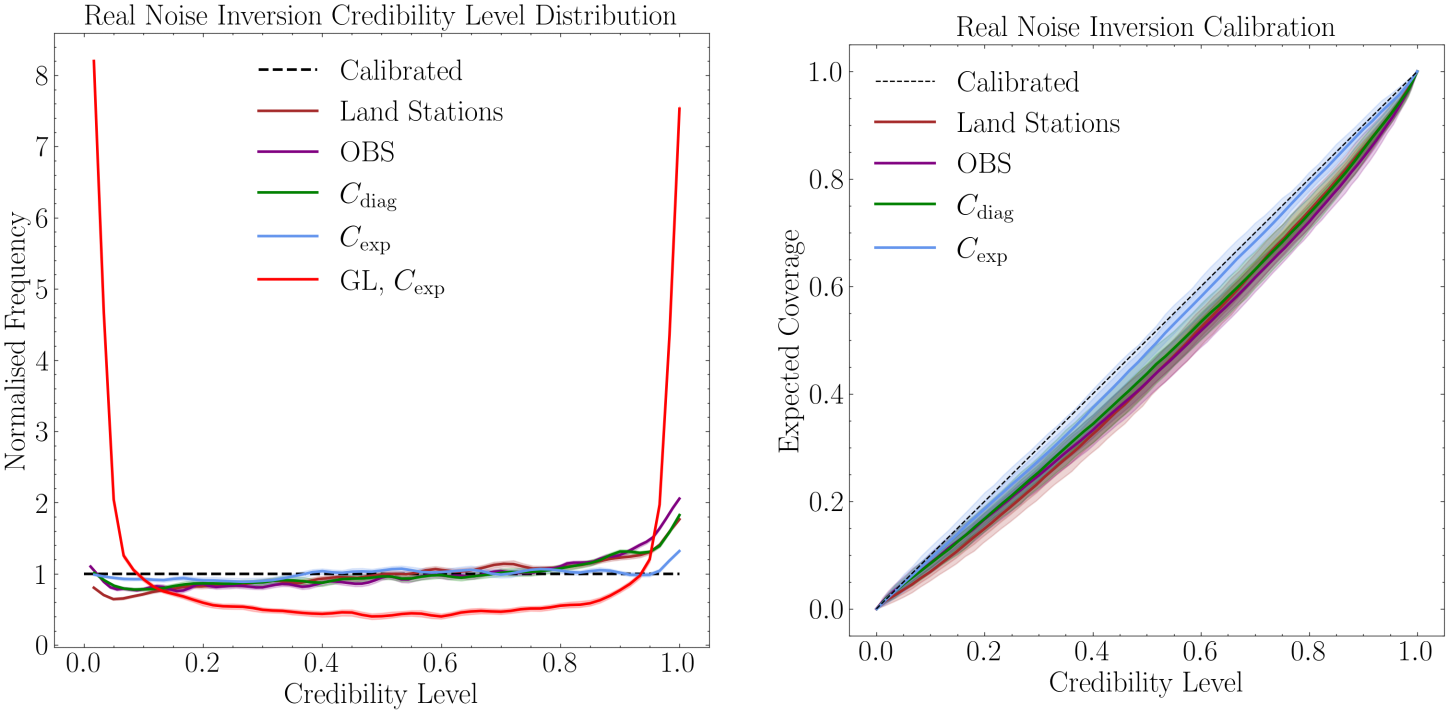}   
\caption{Left: SBI calibration tests showing the credibility interval distributions, for a range of experimental configurations: only land stations, only OBS stations, and the two different covariance parametrisations used for compression. The Gaussian likelihood (GL, red) calibration test is shown for reference. Right: expected coverage plots for experiments repeating the synthetic real noise example.  All SBI experiments lead to relatively good calibration, with a mild bias indicated by the right-skewed credibility level distribution. There is also a hint of mild overconfidence, which has been reported in the SBI literature.}
\label{fig:calsbi}
\end{figure*}

We apply the techniques introduced in Section \ref{sec:empirical_coverage_testing} to explore the performance of the inference methods introduced in the main text. We begin with a range SBI empirical calibration tests, all performed using artificial source inversions with real noise as in the main manuscript. We run 600 independent artificial inversions to compute the empirical coverage for each setup, and present the results in Fig. \ref{fig:calsbi}. In the credibility level distribution plot, we include a comparison with the Gaussian likelihood method to demonstrate the significant improvement in performance  of all SBI configurations compared to standard likelihood-based inference. Fig. \ref{fig:calsbi} shows that SBI performs well across the board, though there are still some issues of mild bias and overconfidence. Mild overconfidence is well-documented in the SBI literature \cite{hermans2021, delaunoy2022towards}, resulting from modelling errors of the neural density estimators (NDEs) at the tails of the distribution. 

One potential source of bias in the SBI case is that the NDEs struggle to model the posterior over the large prior hypercube in the synthetic example ($M_{np} \sim \mathcal{U}[-3, 3] \times 10^{14} ~ \text{Nm}$). Though this should be mitigated by increasing the training dataset size, we were not able to completely remove the bias by running more simulations. Future work could explore improved NDE architectures to see whether the bias is caused by a bottleneck in the modelling abilities of the NDE. Another potential cause of bias is the noise rescaling step (explained in Section 4.4) performed to produce real data noise samples that roughly matched the event covariance $\mathbf{C}_E$. This was done to produce a more realistic likelihood given an estimate of the prevailing seismic noise background of the event, $\hat{p}(\mathbf{D} \! \mid \! \mathbf{m}; \hat{\mathbf{C}}_\text{E})$. However, clearly this is a simplifying assumption as the character of the noise (and thus the ``true'' likelihood $p(\mathbf{D} \! \mid \! \mathbf{m}; \hat{\mathbf{C}}_\text{E})$) will change at different noise levels; this could therefore introduce some bias into the inversions.  We do not explore this further.

\begin{figure*}
\includegraphics[width=\textwidth]{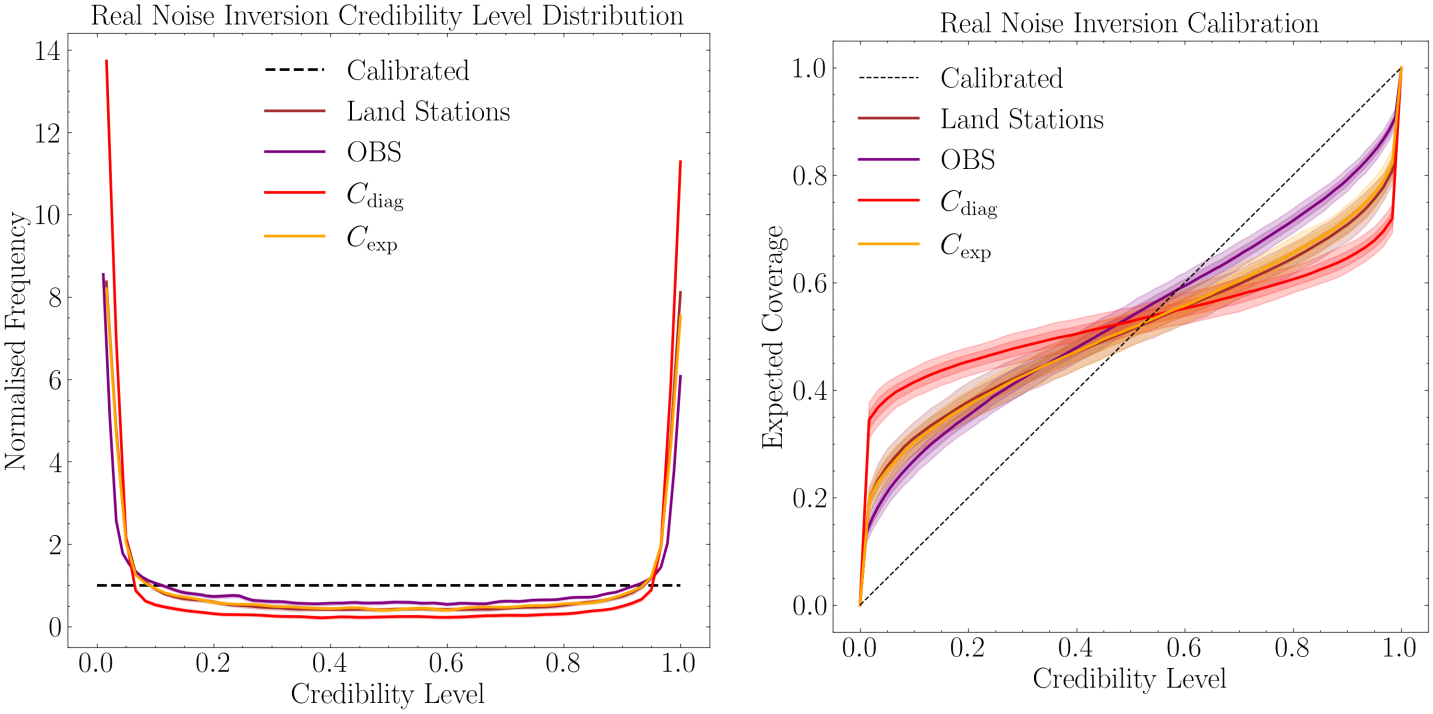}   
\caption{Left: same as in Fig. \ref{fig:calsbi} but with calibration tests showing the credibility interval distributions for the Gaussian likelihood approach. Right: expected coverage plots for the various setups repeating the synthetic real noise example.  All Gaussian likelihood experiments lead to highly overconfident posteriors. Slight asymmetry is also seen in the credibility level distributions, indicated a mild bias as well as the severe overconfidence.}
\label{fig:glcal}
\end{figure*}

We then applied TARP to the Gaussian likelihood inversion method, with results shown in Fig. \ref{fig:glcal}. As explored in the main paper, the diagonal covariance $\mathbf{C}_\text{diag}$ performs substantially worse than the exponentially tapering parametrisation of the covariance $\mathbf{C}_\text{exp}$. We also explore different subsets of the station configuration from Fig. 5 in the main manuscript, with station locations and noise taken from (i) only land stations and (ii) only OBS stations. 

Both sets of stations led to highly overconfident posteriors that significantly underestimate the uncertainty intervals on the moment tensor components. This finding suggests that overconfidence caused by a Gaussian likelihood is not driven by land seismometer or OBS-specific effects, and instead is more universally a result of the characteristics of the seismic noise we are analysing. Interestingly, we found that using the land station array led to slightly \textit{worse} overconfidence than using an OBS only array. We believe that this is driven by the specific geometry of the land station configuration from Fig. 5 in the main manuscript. Since there are several land stations that are very close to the fixed earthquake location, these stations have high sensitivity to the random noise instances that strongly (anti-)correlates with the sensitivity kernels (see Fig. 8 in the main manuscript), leading to more substantial errors in the MLE that a Gaussian likelihood excludes. We found that the high proportion of outliers in the Gaussian likelihood coverage tests (shown at the tails of Fig. \ref{fig:glcal}) are driven by the stations closest to the seismic source.  On the other hand, the OBS stations all have more similar distances to the source, meaning that most high MLE errors are driven by noise at several stations, reducing the rate of occurrence of these outlying events.

\begin{figure*}
\includegraphics[width=\textwidth]{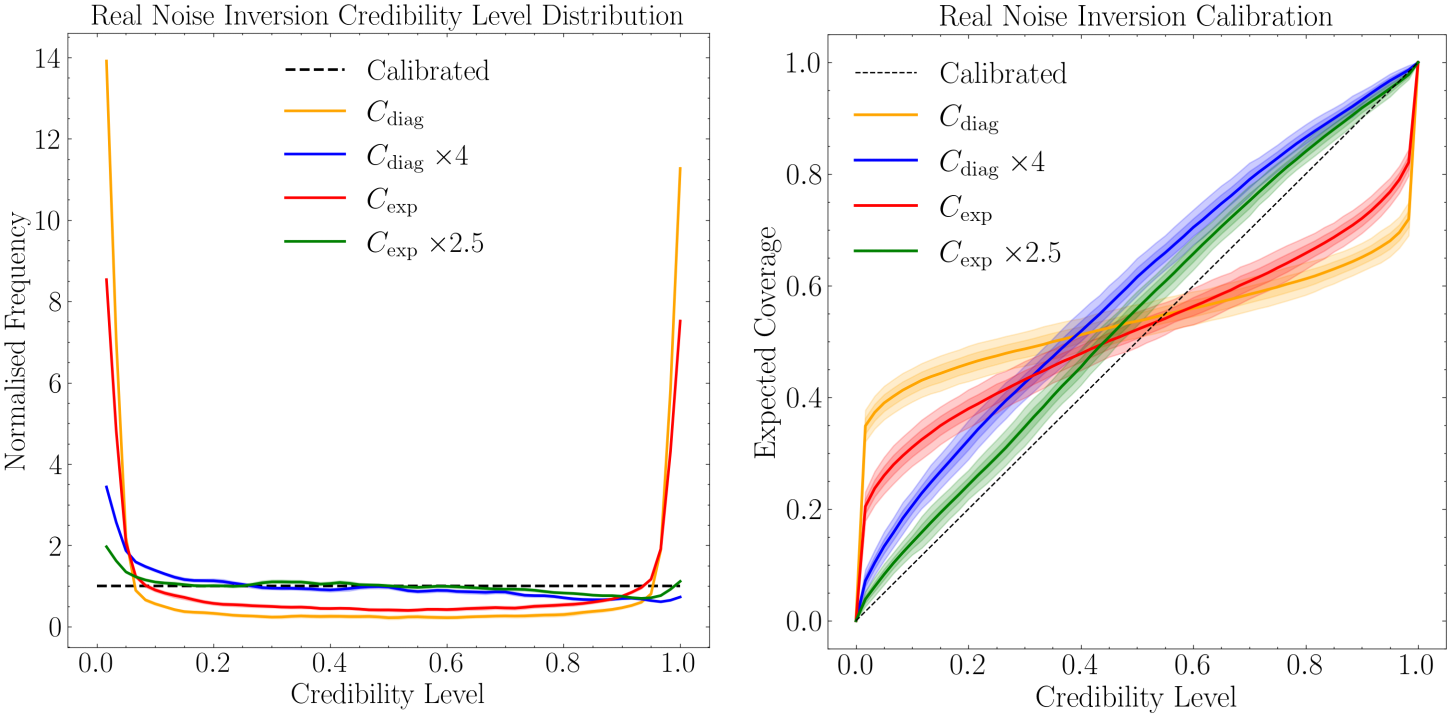}   
\caption{Left: same as in Fig. \ref{fig:calsbi}, now showing the credibility interval distributions for the Gaussian likelihood approach where the posterior samples have been rescaled by a empirically chosen factor (shown as $\times 2.5$, $\times 4$ ). Right: corresponding expected coverage plots. The rescaled posteriors still show minor upticks in the credibility level distribution at 0 and 1, indicating a persisting mild overconfidence. The rescaled posteriors also highlight a bias in the Gaussian likelihood approach (shown by asymmetry in the credibility level distribution), which existed in the original Gaussian likelihood posteriors but was overshadowed by the overconfidence issues. }
\label{fig:unskewed}
\end{figure*}

Finally, we explored the degree to which the Gaussian likelihood-based inference procedure underestimated the scale of the uncertainties. Here, we utilise a simple heuristic by na\"ively rescaling the inferred posteriors. This is done by applying the following transformation to the posterior samples $\mathbf{m}_i$:
\begin{equation}
\mathbf{m}_i' = \widehat{\mathbf{m}}_\text{MAP} + f \times (\mathbf{m}_i - \widehat{\mathbf{m}}_\text{MAP}) 
\end{equation}
where $\widehat{\mathbf{m}}_\text{MAP}$ is estimated by computing the mean of samples $\mathbf{m}_i$, and $f$ is an arbitrary scaling factor. This approach can only correct the absolute scale of the uncertainties, and cannot address biases. We explored a range of values of $f$ and performed empirical coverage tests on the resulting sample ensembles $\mathbf{m}_i'$. An example of the resulting credibility level distribution is given in Fig. \ref{fig:unskewed}.

We found that for covariance parametrisations $\mathbf{C}_\text{diag}$ and $\mathbf{C}_\text{exp}$, posterior overconfidence could be substantially corrected using factors $f=4$ and $f=2.5$ respectively. Even then, Fig. \ref{fig:unskewed} indicates that the rescaled sample ensembles still suffer from minor issues with overconfidence. We found that for $\mathbf{C}_\text{exp}$, the overconfidence was removed at around $f=3$, though at this stage a significant bias dominated the errors in the rescaled posteriors. We conclude that the best covariance parametrisation explored in this work, $\mathbf{C}_\text{exp}$, underestimates the uncertainty in the posterior by up to a factor of 3. 

To be clear, this rescaling approach should not be used to somehow ``correct'' the incorrect posteriors produced by the Gaussian likelihood-based inference. This is just an illustrative procedure to explore the scale and extent of the overconfidence.

\section{Prior Constraining Step}
\label{sec:s_prior_constrain}

Since score compression breaks down when significant non-linear effects are introduced, we must restrict the dataset generation step to regions where the compression is valid. In order to ensure we accurately cover as much of the posterior region as possible, we expand about the maximum likelihood model parameters $\mathbf{m}_*$. This is achieved by performing damped iterative least squares with an estimate of the likelihood to find $\mathbf{m}_*$. Then, using the Fisher information as an estimate of the Gaussian posterior, we can use the per-parameter posterior standard deviations $\boldsymbol{\sigma}$ to constrain the prior. We reproduce the constraining step from the main text in the manuscript here:
\begin{equation}
    p_\text{Fisher}(\mathbf{m}; N) = \mathcal{U}\left[\mathbf{m_*} - N \boldsymbol{\sigma} , \mathbf{m_*} + N\boldsymbol{\sigma}\right],
\label{eq:fisher_prior_S}
\end{equation}
where the posterior covariance estimate $\mathbf{\hat{C}} = \mathbf{F}^{-1}$ is used to construct the per-parameter uncertainties, $\boldsymbol{\sigma}^2 = \text{diag}(\mathbf{F}^{-1})$. 

We then perform empirical coverage tests to check to what extent this step introduces biases into the SBI procedure. We repeat the artificial moment tensor inversion setup from the main manuscript and perform 600 inversions with a range of prior constraints. Note that since the moment tensor inversion is linear, this test explicitly checks the extent to which constraining the prior with $p_\text{Fisher}(\mathbf{m}; N)$ biases the resulting posteriors. We are not exploring how non-linearity effects degrade the compression yet. 

The results for several empirical coverage tests with different dataset sizes and prior sizes $N$ are given in Fig. \ref{fig:sbifisher}. We clearly see an introduction of overconfidence into the posteriors across all test configurations. This is perhaps unsurprising, as constraining the prior will inevitably mean that some of the rare events will be excluded from the prior volume. We also see some bias in the event inversion, which could be caused by errors bias in the MLE $\mathbf{m}_*$ used to constrain the prior. This could also be related to the fact that rescaling the uniform prior changes the prior density in each dimension, further biasing the posterior estimate. However, the $10,000$ simulation test with $N=15$ mitigates against some of these problems, leading to an inference procedure which performs substantially better than a Gaussian likelihood inversion.   

\begin{figure*}
\includegraphics[width=\textwidth]{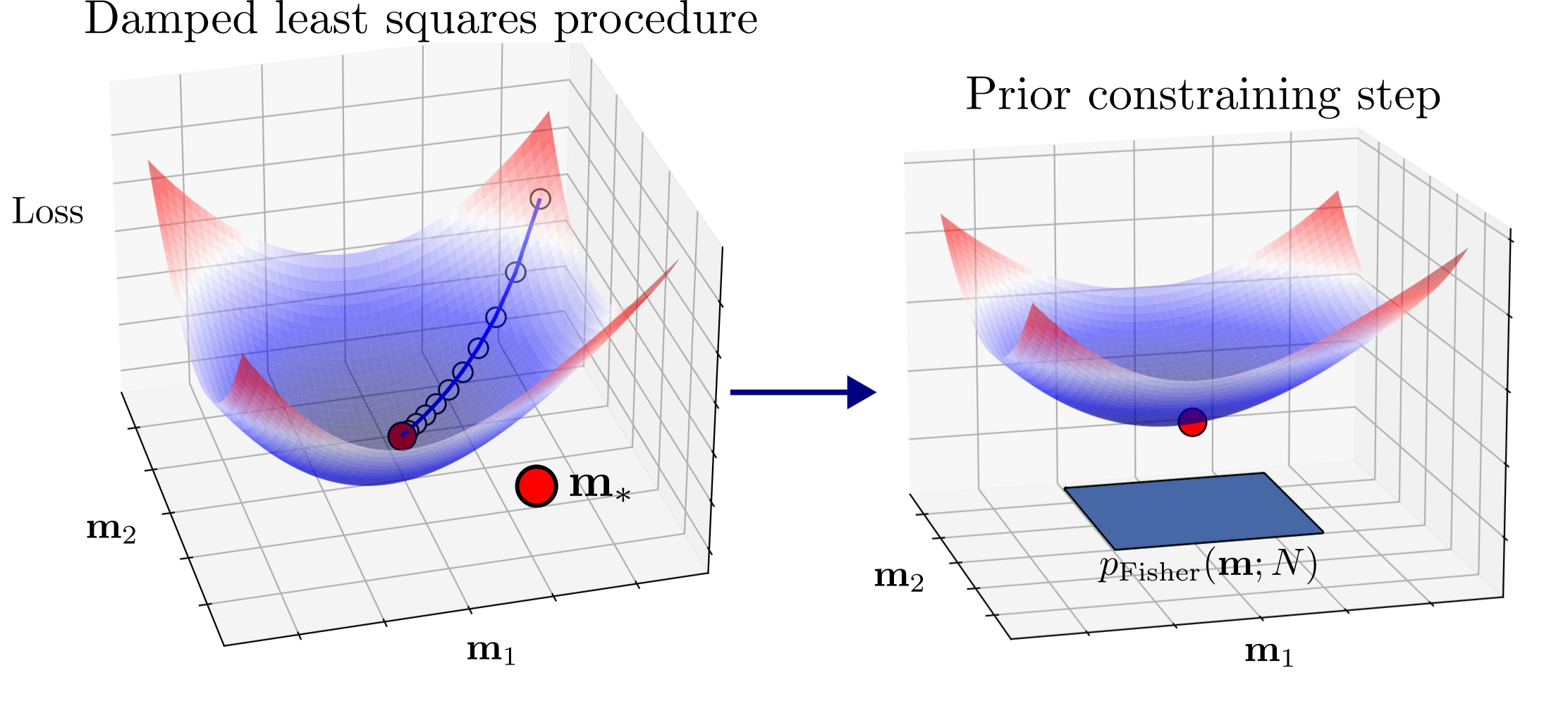}   
\caption{Diagram of the prior-constraining step to ensure good compression performance in non-linear inversion settings. Left: iterative least squares is performed until convergence. Right: the local estimate of the Fisher information $\mathbf{F}_*$ is used to construct a prior around the minimum $p_\text{Fisher}(\mathbf{m};N)$. $N$ varies the size of the prior as a factor of the estimated posterior variances $\text{diag} \: \mathbf{F}^{-1}$.}
\label{fig:priorconst}
\end{figure*}

\begin{figure*}
\includegraphics[width=0.9\textwidth]{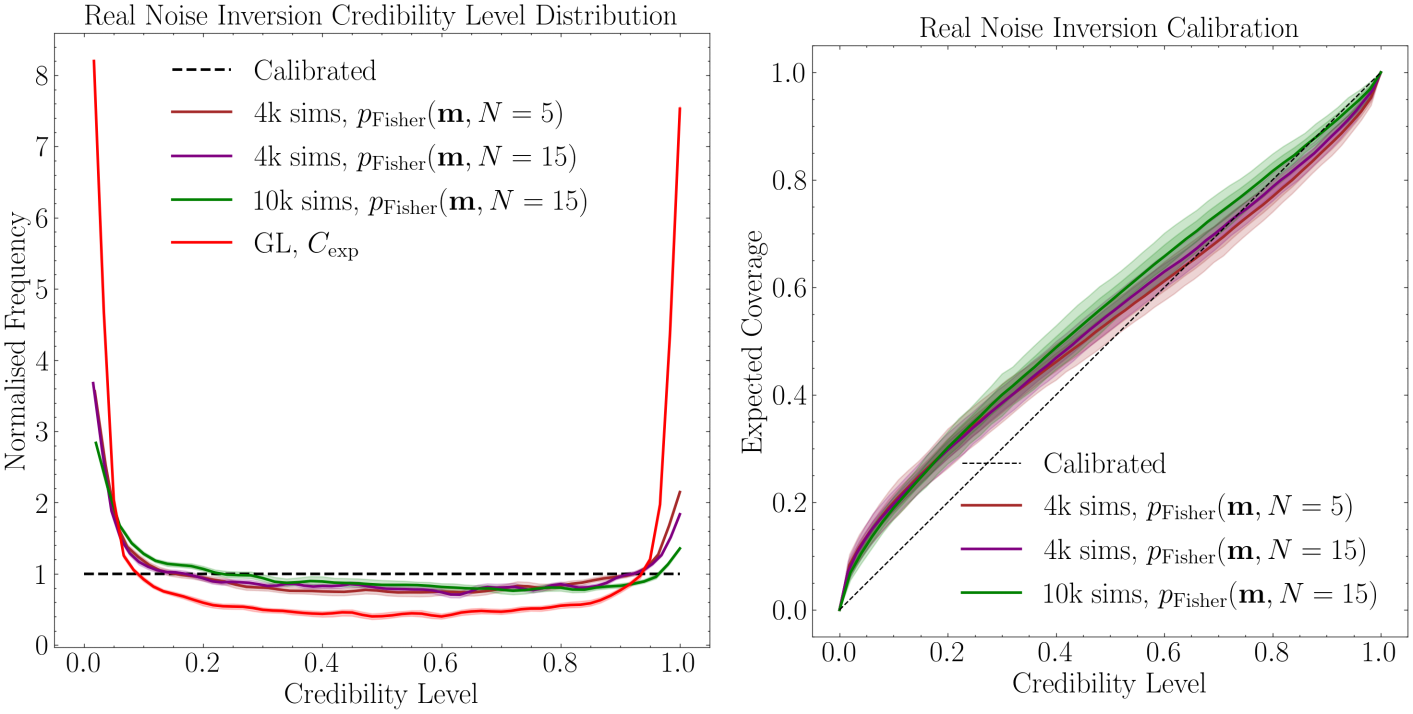}   
\caption{Same as in Figs \ref{fig:calsbi}--\ref{fig:unskewed} but for SBI using our constrained prior method $p_\text{Fisher}$ (for more details see Section 4.5 in the main text and Section \ref{sec:s_prior_constrain}). We experiment with a range of dataset sizes and prior sizes $N$. This step appears to introduce a stronger over-confidence and bias than in Fig. \ref{fig:calsbi}. Increasing the prior volume (increased $N$) and the number of simulations both mitigate against these issues but do not solve them completely.}
\label{fig:sbifisher}
\end{figure*}

\section{Real Event Compression Quality}

We explore how the prior-constraining step $p_\text{Fisher}(\mathbf{m};N)$ affects real event inversions, where the prior is restricted to a uniform hypercube as in Eq. \ref{eq:fisher_prior_S} centred on the MLE $\mathbf{m}_*$. As $N$ increases, we find a trade-off between producing an uninformative prior that supports the true posterior, and worsening compression quality as non-linear effects begin to dominate. 

We compute a relative moment tensor error between the simulated moment tensor and the compressed estimate, $\left|\mathbf{t_M} - \mathbf{m_M}\right|/M_0$, for a range of increasing prior volume factors $N$. Note that since $\mathbf{t}$ is a maximum likelihood estimate of $\mathbf{m}$ under our compression scheme, the error would be zero if there were no noise and the compression was exact. We use this as a metric to explore how the compression quality worsens. 

\begin{figure*}
\includegraphics[width=\textwidth]{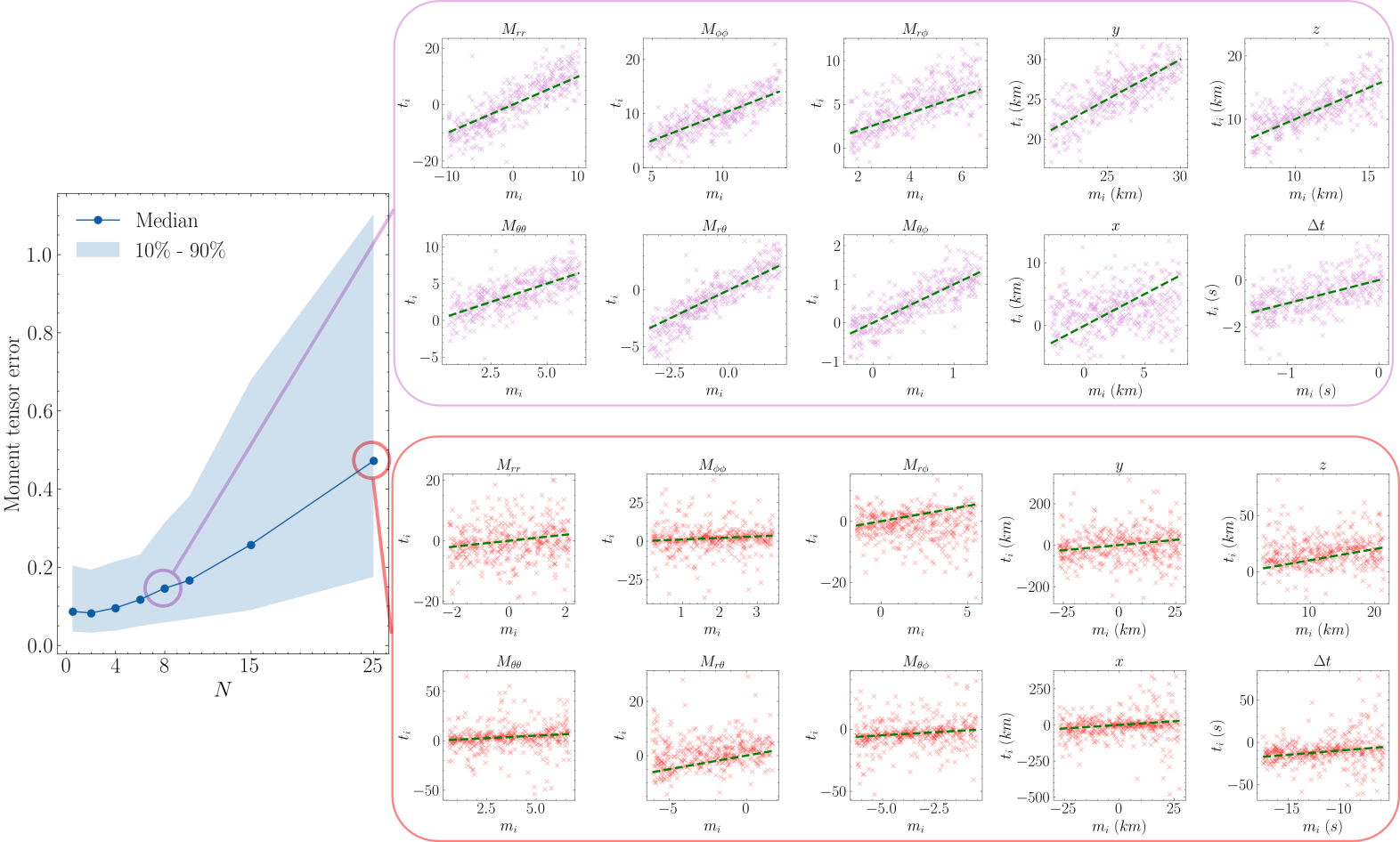}   
\caption{Left: the $M_W \: 4.8  $ Azores 13/01/2022 earthquake moment tensor compression errors as a function of the prior constraining parameter $N$. Increasing compression error is caused by non-linearity effects as the prior volume is increased. Right: compression quality across all model parameters at $N=8$ and $N=25$. Dashed green line represents ideal compression, and the scatter shows compression results for 400 simulated events, which would be used to train a model of the posterior. All moment tensor components are in units of $\times 10^{15} ~\text{Nm}$. $m_i$ and $t_i$ denote the elements of the model parameter vector $\mathbf{m}$ and compressed statistic vector $\mathbf{t}$, respectively (see Section 3.5 of the main text for details).}
\label{fig:north_compression}
\end{figure*}

We investigate compression quality for the $M_W \: 4.8  $ Azores earthquake on 13/01/2022 for a range of values of $N$, varying from $N=0.5$ to $N=25$. Results are shown in Fig. \ref{fig:north_compression}. At $N=0.5$, the score compression is near optimal since the linearity assumption holds well. This ensures that the moment tensor errors are driven by noise. The increase in errors as $N$ is increased is caused by the non-linearity effects that degrade score compression, as simulated model parameters $\mathbf{m}_i$ move further from the expansion point $\mathbf{m_*}$. We find that up until $N=8$, compression errors remain relatively low. This means that the scatter in Fig. \ref{fig:north_compression} closely follows the linear line of ``perfect'' parameter recovery. As the compression quality worsens, the compression estimates $\mathbf{t} = \widehat{\mathbf{m}}$ accrue larger errors, leading to larger variance in $\mathbf{m}$ for a given observation $\mathbf{t}_\text{obs}$. This provides a heuristic to see how the posterior quality (and constraining power) of SBI worsens as the compression quality degrades.

\begin{figure*}
\includegraphics[width=\textwidth]{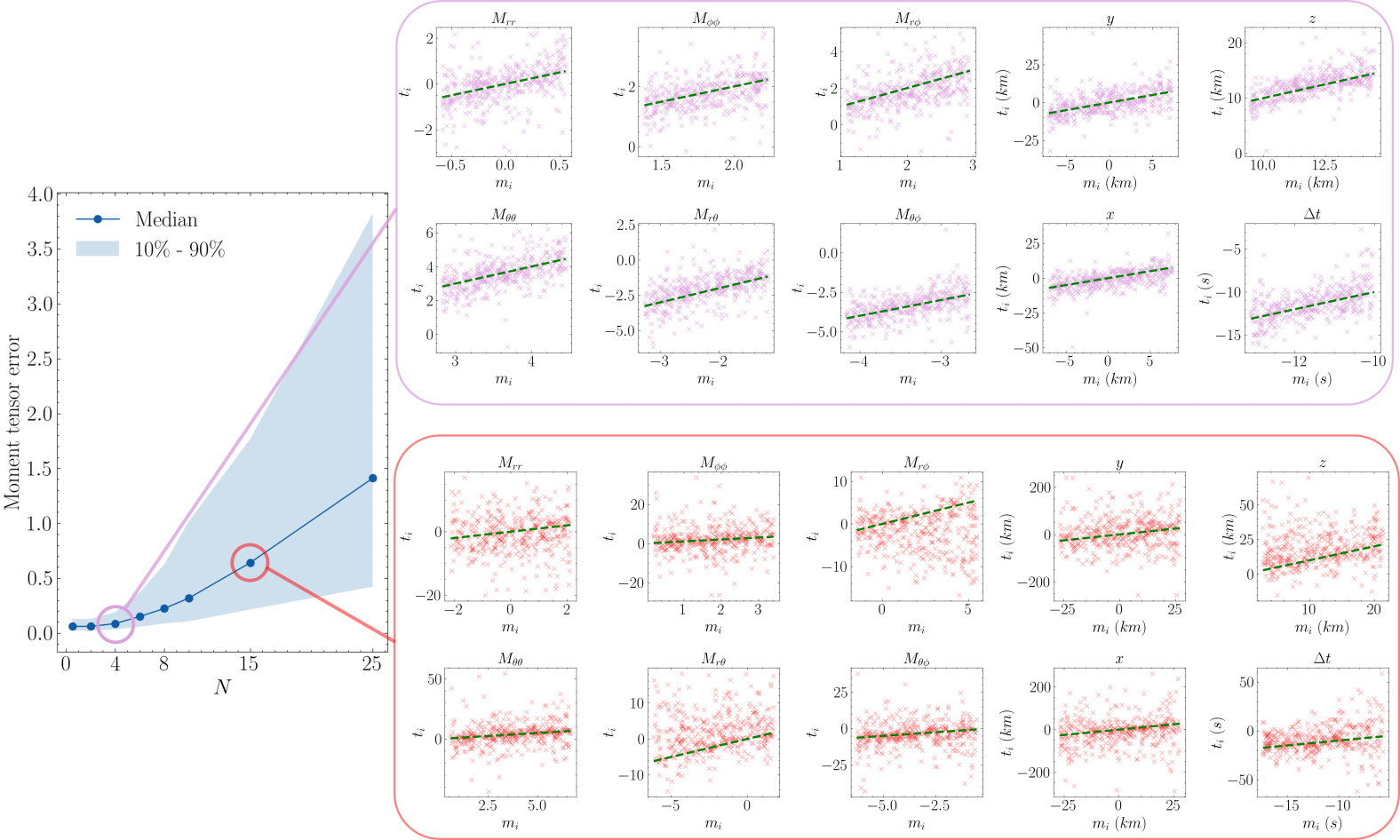}   
\caption{Left: same as in Fig. \ref{fig:north_compression} but for the $M_W \: 4.9$ Madeira earthquake that occurred on 16/02/2022. Errors increase much more rapidly than in Fig. \ref{fig:north_compression}. Right: compression quality across all model parameters at $N=4$ and $N=15$. Dashed green line represents ideal compression, and the scatter shows compression results for 400 simulated events. All moment tensor components are in units of $\times 10^{16}~ \text{Nm}$.}
\label{fig:madeira_compression}
\end{figure*}

Next, we repeat the analysis for the $M_W \: 4.9  $ earthquake near Madeira island that occurred on 16/02/2022, shown in Fig. \ref{fig:madeira_compression}. Here we find that the compression quality degrades much faster as $N$ increases (note the difference in the y-axis scale compared to Fig. \ref{fig:north_compression}). The compression errors remain relatively low up to $N=4$, staying near the optimal quality in the linear regime. We therefore use this value of $N$ to produce the posteriors presented in the main manuscript.

\newpage

\newcommand{\newblock}{} 
\bibliographystyle{gji}
\bibliography{bibliography}